\documentclass[a4paper,10pt]{article}
\usepackage[utf8]{inputenc}
\usepackage{graphicx,tabularx,amsmath,amsthm,mathtools,amsfonts,bm,amssymb,setspace,color,float,caption,subcaption}
\usepackage[pdfpagemode=UseNone,pdfstartview={XYZ null null null}]{hyperref}
\usepackage[title]{appendix}
\usepackage[longnamesfirst,numbers,square]{natbib}
\usepackage[margin=0.5in]{geometry} % Smaller margi
\usepackage{color}

\DeclareMathOperator{\sech}{sech}
\DeclareMathOperator{\cosech}{cosech}
\newcommand{\eps}{\varepsilon}

\newcommand{\R}{\mathbb{R}}

\newcommand{\C}{\mathbb{C}}

\newcommand{\LL}{\mathcal{L}}
\newcommand{\BB}{\mathcal{B}}
\newcommand{\CC}{\mathcal{C}}

\newcommand{\WW}{\mathcal{W}}
\newcommand{\HH}{\mathcal{H}}
\newcommand{\FF}{\mathcal{F}}

\newcommand{\bWW}{\bm{\mathcal{W}}}
\newcommand{\bFF}{\bm{\mathcal{F}}}
\newcommand{\bGG}{\bm{\mathcal{G}}}

\newcommand{\bu}{\bm{u}}
\newcommand{\bv}{\bm{v}}

\newcommand{\bF}{\bm{F}}
\newcommand{\bphi}{\bm{\phi}}
\newcommand{\bPhi}{\bm{\Phi}}
\newcommand{\bpsi}{\bm{\psi}}

\newcommand{\orderOne}{\mathcal{O}\left(\varepsilon\right)}
\newcommand{\orderTwo}{\mathcal{O}\left(\varepsilon^2\right)}
\newcommand{\orderThree}{\mathcal{O}\left(\varepsilon^3\right)}
\newcommand{\orderFour}{\mathcal{O}\left(\varepsilon^4\right)}

\newtheorem{theorem}{Theorem}[section]

\newtheorem{lemma}[theorem]{Lemma}

\title{Weakly Nonlinear Theory for Oscillatory Dynamics in a One-Dimensional PDE-ODE Model of Membrane Dynamics Coupled by a Bulk Diffusion Field}

\author{Fr\'ed\'eric Paquin-Lefebvre \thanks{Dept. of Mathematics, UBC, Vancouver, Canada. (corresponding author {\tt paquinl@math.ubc.ca})} \and Wayne Nagata \thanks{Dept. of Mathematics, UBC, Vancouver, Canada. {\tt nagata@math.ubc.ca}} \and Michael J. Ward \thanks{Dept. of Mathematics, UBC, Vancouver, Canada. {\tt ward@math.ubc.ca}}}

\begin{document}

\maketitle

\begin{abstract}
We study the dynamics of systems consisting of two spatially segregated ODE compartments coupled through a one-dimensional bulk diffusion field. For this coupled PDE-ODE system, we first employ a multi-scale asymptotic expansion to derive amplitude equations near codimension-one Hopf bifurcation points for both in-phase and anti-phase synchronization modes. The resulting normal form equations pertain to any vector nonlinearity restricted to the ODE compartments. In our first example, we apply our weakly nonlinear theory to a coupled PDE-ODE system with Sel'kov membrane kinetics, and show that the symmetric steady state undergoes supercritical Hopf bifurcations as the coupling strength and the diffusivity vary. We then consider the PDE diffusive coupling of two Lorenz oscillators. It is shown that this coupling mechanism can have a stabilizing effect, characterized by a significant increase in the Rayleigh number required for a Hopf bifurcation. Within the chaotic regime, we can distinguish between synchronous chaos, where both the left and right oscillators are in-phase, and chaotic states characterized by the absence of synchrony. Finally, we compute the largest Lyapunov exponent associated with a linearization around the synchronous manifold that only considers odd perturbations. This allows us to predict the transition to synchronous chaos as the coupling strength and the diffusivity increase.
\end{abstract}

% \begin{keywords}
% Hopf bifurcation, weakly nonlinear theory, multi-scale expansion, in-phase/anti-phase oscillations, Lyapunov exponents, synchronous chaos.
% \end{keywords}
% 
% \begin{AMS}
% 37N30, 37N25, 65P30, 35B36, 35B35
% \end{AMS}

\section{Introduction}\label{sec:intro} 

We investigate, through a weakly nonlinear analysis, the oscillatory dynamics in a class of one-dimensional coupled PDE-ODE models. The class of models considered allows us to study the collective synchronization of two dynamically active compartments, modeled by systems of nonlinear ODEs, that are indirectly coupled via the diffusion of some spatially extended variable in a 1-D bulk interval. In particular, this modeling paradigm has been used in the study of intracellular polarization and oscillations in fission yeast, where each compartment represents the opposite tips of an elongated rod-shaped cell (cf.~\cite{xu2018}, \cite{xu1}). Pattern formation behavior and linear stability analyses of coupled 1-D membrane-bulk PDE-ODE systems have been analyzed in other specific contexts (cf.~\cite{gomez2007}, \cite{gou2015}, \cite{gou2016}, \cite{gou2017}, \cite{lawley}), and in multi-dimensional domains in \cite{gou2016QS} and \cite{kuttler2006}, where they have been employed to study intercellular communication and the related concepts of quorum and diffusion sensing. Quasi-steady versions of the coupled membrane-bulk models, whereby the membrane is at steady state and contributes only nonlinear flux source terms, have been used to model spatial effects in gene regulatory networks (cf.~\cite{hess1}, \cite{hess2}, \cite{hess3}) and cascades in biological signal transduction (cf.~\cite{levy1}, \cite{levy2}).

To formulate our 1-D model, we assume that some spatially extended bulk variable $C(x,t)$ undergoes linear diffusion and decay with rate constants $D$ and $k$ within an interval of length $2L$,
\begin{equation}\label{eq:bulk}
C_t = D C_{xx} - k C\,, \qquad 0 < x < 2L\,, \qquad t > 0\,.
\end{equation}
We impose the following linear Robin-type boundary conditions to model the exchange between the bulk and the compartments:
\begin{equation}\label{eq:BC}
-D C_x(0,t) = \kappa (e_1^T\bu(t) - C(0,t))\,, \quad DC_x(2L,t) = \kappa (e_1^T\bv(t) - C(2L,t))\,,
\end{equation}
where $e_1 = (1, 0, \ldots,0)^T \in \, \R^n$. Here, $\bu(t),\,\bv(t) \in \R^n$ denote the variables in the left and right local compartments, of which only the first component is released within the 1-D bulk {region}. In this model, the leakage parameter $\kappa$ controls the permeability {of the compartments at each endpoint.} Furthermore, letting $\bFF(\cdot) \in \, \R^n$ be the nonlinear vector function modeling each oscillator, {which we assume to be identical}, and {denoting $\beta$ as the coupling strength, we impose} that the {ODE systems}
\begin{equation}\label{eq:kinetics}
\dfrac{d \bu}{dt} = \bFF(\bu) + \beta(C(0,t) - e_1^T\bu)e_1\,, \quad \dfrac{d \bv}{dt} = \bFF(\bv) + \beta(C(2L,t) - e_1^T\bv)e_1\,,
\end{equation}
govern the dynamics in each compartment. The coupled PDE-ODE system \eqref{eq:bulk}--\eqref{eq:kinetics} given here is in dimensionless form. The geometry for this 1-D model can be viewed as a long rectangular strip separating two vertical 1-D membranes, where there is assumed to be no transverse solution dependence.

There is a rather wide literature investigating the dynamics of
diffusively coupled oscillators, where the coupling usually consists
of the discrete Laplacian acting on a lattice of several oscillators
with periodic boundary conditions, or through some other discretely
coupled network. Examples of such coupled ODE systems include discrete
chains of bistable kinetics, such as the Lorenz or Fitzhugh-Nagumo
systems, in which the formation of propagating fronts was studied in
\cite{balmforth2005}, \cite{pazo2003} and \cite{josic2000}, and the
well-known Kuramoto-type oscillator models as surveyed in
\cite{kura}. However, relatively few studies have considered spatially
segregated oscillators that are indirectly coupled via a PDE bulk
diffusion field.

For our PDE-ODE coupled system, our primary goal is to derive
amplitude equations, or normal forms, near Hopf bifurcation points for
either the in-phase or anti-phase synchronization modes, while
allowing for an arbitrary, but identical, vector nonlinearity in each
ODE compartment. This rather general framework will provide us with
explicit formulae for the normal form coefficients that can easily be
evaluated numerically in order to classify whether Hopf bifurcations
are sub- or supercritical and also to detect possible criticality
switches indicated by sign changes. Our weakly nonlinear theory is
given in \S \ref{sec:weakly_nonlinear_theory}, where for calculational
efficiency we employ a multi-scale asymptotic expansion to derive the
two distinct normal forms. {The work presented} {
  here extends the weakly nonlinear stability analysis from \S 6 of
  \cite{gou2015}, which focused on synchronous oscillations in a class
  of coupled PDE-ODE models with nonlinear boundary conditions and a
  single active species in each compartment.}

{In \cite{gou2017} it was shown that the coupling of two
  conditional biochemical oscillators, each of them at a quiescent
  state when isolated, via a diffusive chemical signal could lead to
  robust in-phase and anti-phase oscillatory dynamics. The study
  employed the two-component Sel'kov kinetics, originally used to
  model glycolysis oscillations \cite{selkov1968}.} {As an
  extension of this previous work}, in \S \ref{sec:selkov} we apply
our weakly nonlinear theory to a coupled PDE-ODE model with Sel'kov
kinetics and find a rather wide parameter regime for which the
base-state can lose stability to a supercritical Hopf bifurcation. Our
weakly nonlinear results are validated against numerical bifurcation
results and time-dependent numerical simulations.

Then, in \S \ref{sec:lorenz} we assume that the dynamics in each
compartment is governed by identical Lorenz ODE
oscillators. {For an isolated Lorenz ODE system, we recall
  that increasing the Rayleigh number (the typical bifurcation
  parameter) causes a number of bifurcations including pitchfork,
  homoclinic and Hopf, and ultimately chaotic oscillations
  \cite{sparrow1982}. Here, we would like to determine how this
  cascade of bifurcations is affected by the PDE bulk diffusive
  coupling of two identical Lorenz ODE systems.} Our analysis will
show that such a coupling mechanism causes a significant increase in
the critical Rayleigh number where the Hopf bifurcation occurs,
suggesting a delay in the appearance of chaotic oscillations. For this
problem we also consider the case where the bulk domain is well-mixed
and spatially homogeneous, corresponding to the infinite bulk
diffusion limit and for which the coupled PDE-ODE system is reduced to
a single system of globally coupled ODEs, as shown in Appendix
\ref{sec:well_mixed}.

Finally, for both finite and infinite diffusion cases, we predict in \S \ref{subsec:synchronous_chaos} the transition to synchronous chaos as the diffusivity $D$ and the coupling strength $\beta$ are increased. {Here, synchronous chaos is defined as sensitivity to initial conditions along an invariant synchronous manifold where both Lorenz oscillators are completely in phase. Our
  predictions are based on the computation of the largest Lyapunov
  exponent of an appropriate non-autonomous linearization of our
  coupled PDE-ODE system \eqref{eq:bulk}--\eqref{eq:kinetics},
  obtained by only selecting transverse perturbations to the
  synchronous manifold.} {We remark that the master stability
  functions}, {for determining the stability of the
  synchronous state of a network of oscillators with an arbitrary
  discrete coupling function, are similarly obtained
  (cf.~\cite{carroll1998}). Furthermore, our results are consistent
  with the discrete case} {in the sense that}
{complete synchronization of two interacting chaotic
  oscillators occurs when the coupling is strong enough to suppress
  chaotic instabilities
  (cf.~\cite{yamada1983,pikovsky1984,carroll2015}).}

In \S \ref{sec:discussion}, we conclude by briefly summarizing our main results and by suggesting a few open problems that warrant further investigation.

%%%%%%%%%%%%%%%%%%%%%%%

\section{Weakly nonlinear theory for 1-D coupled PDE-ODE systems}\label{sec:weakly_nonlinear_theory}

\subsection{{Symmetric steady state and linear stability analysis}}
We first rewrite the coupled PDE-ODE system
\eqref{eq:bulk}-\eqref{eq:kinetics} as an evolution equation {in
the form}
\begin{equation}\label{eq:evolution}
 \dot{W} = \bF(W) = \begin{pmatrix}
                     D C_{xx} - k C \\
                     \bFF(\bu) + \beta(C|_{x=0} - e_1^T\bu)e_1 \\
                     \bFF(\bv) + \beta(C|_{x=2L} - e_1^T\bv)e_1
                    \end{pmatrix}.
\end{equation}
Here, $\bF$ is a nonlinear functional acting on $\bm{\WW}$, defined as
the space of vector functions whose components satisfy the appropriate
linear Robin-type boundary conditions:
\begin{equation}\label{eq:function_space}
\bWW = \left. \left\{ 
W = \begin{pmatrix}
C(x) \\
\bu \\
\bv
\end{pmatrix} \right|
\begin{matrix}
-D C_x|_{x=0} = \kappa \left( e_1^T\bu - C|_{x=0} \right) \\
D C_x|_{x=2L} = \kappa \left( e_1^T\bv - C|_{x=2L} \right)
\end{matrix}
\right\}.
\end{equation}
A symmetric steady state for \eqref{eq:evolution} is given by
\begin{equation}\label{eq:symmetric_steady_state}
W_e = \begin{pmatrix}
       (1-p_0)\dfrac{\cosh(\omega (L - x))}{\cosh(\omega L)} e_1^T\bu_e \\
       \bu_e \\
       \bu_e
      \end{pmatrix}, \quad \omega = \sqrt{\dfrac{k}{D}}, \quad p_0 = \frac{D \omega \tanh(\omega L)}{D \omega \tanh(\omega L) + \kappa},
\end{equation}
where $\bu_e$ is a solution of a nonlinear algebraic system of equations
\begin{equation}\label{eq:ss_eqn}
 \bFF(\bu_e) - \beta p_0 E \bu_e = 0, \quad E \equiv e_1e_1^T,
\end{equation}
and where $E$ is an $n \times n$ rank-one matrix.

Next, we consider the linear stability of a symmetric steady state by introducing a perturbation of the form
\begin{equation}\label{eq:perturbation}
  W(x,t) = W_e(x) + \WW(x) e^{\lambda t}, \qquad \WW(x) =
  \begin{pmatrix} \eta(x) \\ \bphi \\ \bpsi \end{pmatrix}\,.
\end{equation}
Substitution of \eqref{eq:perturbation} within \eqref{eq:evolution}
yields, after expanding and collecting coefficients of
$e^{\lambda t}$, the following nonstandard eigenvalue problem:
\begin{equation}\label{eq:linearized_operator}
\lambda \WW = \LL(\WW) \quad \text{with} \quad \LL(\WW) \equiv
\begin{pmatrix}
 D \eta_{xx} - k\eta \\
 J_e \bphi + \beta (\eta(0) - e_1^T \bphi) e_1 \\
 J_e \bpsi + \beta (\eta(2L) - e_1^T \bpsi) e_1
\end{pmatrix}.
\end{equation}
Here, $J_e$ is the Jacobian matrix of the nonlinear vector function evaluated at a steady state $\bu_e$, while $\LL$ is the linearized operator acting on the function space defined in \eqref{eq:function_space}. The eigenfunction $\WW(x)$ therefore satisfies the same boundary conditions, given by
\begin{equation}
-D \eta_x(0) = \kappa \left( e_1^T\bphi - \eta(0) \right), \quad D \eta_x(2L) = \kappa \left( e_1^T\bpsi - \eta(2L) \right).
\end{equation}
We can write the solution in the bulk as a linear combination of an even and an odd eigenfunction in the form
\begin{equation}\label{eq:eta}
\eta(x) = \dfrac{1-p_+(\lambda)}{2} e_1^T (\bphi + \bpsi) \frac{\cosh(\Omega (L-x))}{\cosh(\Omega L)} + \dfrac{1-p_-(\lambda)}{2} e_1^T (\bphi - \bpsi) \frac{\sinh(\Omega (L-x))}{\sinh(\Omega L)}.
\end{equation}
Here, $p_+(\lambda)$, $p_-(\lambda)$ and $\Omega$ are each defined by
\begin{equation}
p_+(\lambda) = \dfrac{D \Omega \tanh(\Omega L)}{D \Omega \tanh(\Omega L) + \kappa}, \quad p_-(\lambda) = \dfrac{D \Omega \coth(\Omega L)}{D \Omega \coth(\Omega L) + \kappa}, \quad \Omega = \sqrt{\frac{k + \lambda}{D}},
\end{equation}
where we take the principal branch for $\Omega$ if $\lambda$ is complex. The eigenvectors $\bphi$ and $\bpsi$ in \eqref{eq:eta} satisfy the homogeneous linear system of equations given by
\begin{equation}
 \begin{pmatrix}
  J_e - \lambda I - \beta \left( \frac{p_+(\lambda) + p_-(\lambda)}{2} \right) E & \beta \left( \frac{p_-(\lambda) - p_+(\lambda)}{2}\right) E \\
  \beta \left( \frac{p_-(\lambda) - p_+(\lambda)}{2}\right) E & J_e - \lambda I - \beta \left( \frac{p_+(\lambda) + p_-(\lambda)}{2} \right) E 
 \end{pmatrix} 
 \begin{pmatrix}
  \bphi \\
  \bpsi
 \end{pmatrix} =
 \begin{pmatrix}
  \bm{0} \\  
  \bm{0}
 \end{pmatrix}.
\end{equation}

From symmetry considerations and since every {perturbation} can be written as the sum of an even part with $\bphi = \bpsi$, and an odd part with $\bphi = -\bpsi$, this system can be reduced to $n$ equations. Letting $\bphi_+$ and $\bphi_-$ denote the even and odd eigenvectors, we can readily establish a reduced homogeneous linear system for each case as
\begin{equation}\label{eq:lin_hom}
\bPhi_\pm(\lambda) \bphi_\pm = \left[J_e - \lambda I - \beta p_\pm(\lambda) E \right] \bphi_\pm = \bm{0}.
\end{equation}
In this way, the eigenvalue parameter $\lambda$ must satisfy the transcendental equation
\begin{equation}\label{eq:transcendental}
\det\left[ \bPhi_\pm(\lambda) \right] = 0
\end{equation}
in order for the system to admit a non-trivial solution $\bphi_\pm \neq \bm{0}$. Finally, the eigenfunctions $\WW_\pm$ for both even and odd cases are defined by
\begin{equation}\label{eq:eigenfunctions}
\WW_+ = \begin{pmatrix} (1-p_+(\lambda))\frac{\cosh(\Omega(L-x))}{\cosh(\Omega L)} e_1^T \bphi_+ \\ \bphi_+ \\ \bphi_+ \end{pmatrix}, \quad
\WW_- = \begin{pmatrix} (1-p_-(\lambda))\frac{\sinh(\Omega(L-x))}{\sinh(\Omega L)} e_1^T \bphi_- \\ \bphi_- \\ - \bphi_- \end{pmatrix}.
\end{equation}

\subsection{{Adjoint linear operator and inner product}}

The imposition of a solvability condition in the multi-scale asymptotic expansion presented below requires the appropriate formulation of an adjoint linear operator $\LL^\star$ defined by
\begin{equation}\label{eq:adjoint_operator}
\LL^\star (W^\star) = 
\begin{pmatrix}
 D C_{xx}^\star - kC^\star \\
 J_e^T \bu^\star + (\kappa C^\star|_{x=0} - \beta e_1^T \bu^\star) e_1 \\
 J_e^T \bv^\star + (\kappa C^\star|_{x=2L} - \beta e_1^T \bv^\star) e_1
\end{pmatrix},
\end{equation}
which acts on the space $\bWW^\star$ of vector functions satisfying the adjoint boundary conditions,
\begin{equation}
  -D C_x^\star|_{x=0} = \beta e_1^T\bu^\star - \kappa C^\star|_{x=0},
  \quad D C_x^\star|_{x=2L} = \beta e_1^T\bv^\star - \kappa C^\star|_{x=2L}.
\end{equation}
For any $W \in \bWW$ and $W^\star \in \bWW^\star$, we have 
\begin{equation}\label{eq:adjoint_prop}
\langle W^\star,\LL W \rangle = \langle \LL^\star W^\star, W \rangle,
\end{equation}
where the inner product in \eqref{eq:adjoint_prop} is defined by
\begin{equation}\label{eq:inner_product}
\langle W^\star,W \rangle = \int_0^{2L} \overline{C^\star}C \, dx + \overline{\bu^\star}^T\bu + \overline{\bv^\star}^T\bv.
\end{equation}

Next, upon calculating the even and the odd adjoint eigenfunctions we obtain
\begin{equation}\label{eq:adjoint_eigenfunctions}
 \WW_+^\star = 
 \begin{pmatrix}
  \frac{\beta}{\kappa}(1 - \overline{p_+(\lambda)})\frac{\cosh(\overline{\Omega}(L-x))}{\cosh(\overline{\Omega} L)} e_1^T\bphi_+^\star \\
  \bphi_+^\star \\
  \bphi_+^\star
 \end{pmatrix}, \quad
 \WW_-^\star = 
 \begin{pmatrix}
  \frac{\beta}{\kappa}(1 - \overline{p_-(\lambda)})\frac{\sinh(\overline{\Omega}(L-x))}{\sinh(\overline{\Omega} L)} e_1^T\bphi_-^\star \\
  \bphi_-^\star \\
  - \bphi_-^\star
 \end{pmatrix}\,,
\end{equation}
where $\bphi_\pm^\star$ satisfies the conjugate transpose of the system \eqref{eq:lin_hom},
\begin{equation}
 \left[\overline{\bPhi_\pm(\lambda)} \right]^T \bphi_\pm^\star = \bm{0}.
\end{equation}
From the definitions \eqref{eq:eigenfunctions}, \eqref{eq:adjoint_eigenfunctions} and \eqref{eq:inner_product}, we can verify that the eigenfunctions and their adjoints form an orthogonal set, which can be normalized for convenience as
\begin{equation}
  \langle \WW_+^\star, \WW_- \rangle = \langle \WW_-^\star, \WW_+ \rangle = 0,
  \qquad
  \langle \WW_+^\star, \WW_+ \rangle = \langle \WW_-^\star, \WW_- \rangle = 1,
\end{equation}
and that the following properties hold:
\begin{equation}
  \LL(\WW_\pm) = \lambda \WW_\pm, \quad
  \LL^\star(\WW_\pm^\star) = \overline{\lambda} \WW_\pm^\star.
\end{equation}
 
\subsection{Multi-scale expansion}

Let $\mu = (\beta, D)^T$ be a vector of bifurcation parameters. As usual, a slow time-scale $\tau = \eps^2 t$ with $\eps \ll 1$ is introduced. Using the same scaling, we perturb the vector of bifurcation parameters to yield
\begin{equation}\label{eq:detuning}
\mu = \mu_0 + \eps^2\mu_1\,, \quad \mbox{where} \quad {\mu_0=\begin{pmatrix} \beta_0 \\ D_0 \end{pmatrix} \quad \text{and} \quad \mu_1=\begin{pmatrix} \beta_1 \\ D_1 \end{pmatrix}} \quad {\mbox{with}} \quad \|\mu_1\| = 1\,.
\end{equation}
Here $\mu_0$ is the bifurcation point, while $\mu_1$ is a unit vector indicating the direction of the bifurcation. We then expand the state variable in a regular asymptotic power series around a symmetric steady state as
\begin{equation}\label{eq:asymptotic}
 W(x,t,\tau) = W_e(x) + \eps W_1(x,t,\tau) + \eps^2 W_2(x,t,\tau) + \eps^3 W_3(x,t,\tau) + \orderFour.
\end{equation}
Next, by inserting \eqref{eq:detuning} and \eqref{eq:asymptotic} into \eqref{eq:evolution}, and collecting powers of $\eps$, we obtain that
\begin{equation}\label{eq:perturbed_equation}
\begin{split}
& \eps \partial_t W_1 + \eps^2\partial_t W_2 + \eps^3(\partial_t W_3 + \partial_\tau W_1) = \\
& \eps \LL(\mu_0;W_1) + \eps^2 \left( \LL(\mu_0;W_2) + \BB(W_1,W_1) + 
\begin{pmatrix} \omega^2 C_e D_1 \\ - p_0 E \bu_e \beta_1 \\ - p_0 E \bu_e \beta_1 \end{pmatrix} \right) + \\
& \eps^3 \left( \LL(\mu_0;W_3) + 2\BB(W_1,W_2) + \CC(W_1,W_1,W_1) +
\begin{pmatrix}
 \frac{1}{D_0}(\partial_t + k)C_1 D_1 \\
 (C_1|_{x=0}e_1 - E\bu_1) \beta_1 \\
 (C_1|_{x=2L}e_1 - E\bv_1) \beta_1
\end{pmatrix}\right),
\end{split}
\end{equation}
and that the perturbed boundary conditions satisfy
\begin{equation}\label{eq:perturbed_BC}
\begin{split}
& \sum_{j=1}^3 \eps^j\left(\partial_x C_j + \frac{\kappa}{D_0}\left(e_1^T\bu_j - C_j\right) \right) = \left(\eps^2 p_0e_1^T\bu_e + \eps^3\left(e_1^T\bu_1 - C_1\right)\right) \frac{\kappa}{D_0^2}D_1, \quad x = 0, \\
& \sum_{j=1}^3 \eps^j\left(\partial_x C_j - \frac{\kappa}{D_0}\left(e_1^T\bv_j - C_j\right) \right) = \left(-\eps^2 p_0e_1^T\bu_e - \eps^3\left(e_1^T\bv_1 - C_1\right)\right) \frac{\kappa}{D_0^2}D_1, \quad x = 2L.
\end{split}
\end{equation}
Finally, we precisely define the multilinear forms $\BB(\cdot,\cdot)$ and $\CC(\cdot, \cdot, \cdot)$ in \eqref{eq:perturbed_equation} as
\begin{equation}
 \BB(W_j,W_k) = \begin{pmatrix} 0 \\ B(\bu_j,\bu_k) \\ B(\bv_j,\bv_k) \end{pmatrix}, \quad \CC(W_j,W_k,W_l) = \begin{pmatrix} 0 \\ C(\bu_j,\bu_k,\bu_l) \\ C(\bv_j,\bv_k,\bv_l) \end{pmatrix},
\end{equation}
where the non-trivial components satisfy
\begin{equation}
 B(\bu_j,\bu_k) = \frac{1}{2}(I \otimes \bu_k^T) H_e \bu_j, \quad C(\bu_j,\bu_k,\bu_l) = \frac{1}{6}(I \otimes \bu_l^T) T_e (\bu_j \otimes \bu_k).
\end{equation}
Here, $I \in \, \R^{n \times n}$ and the matrices $H_e$ and $T_e$ can be defined as
\begin{equation}\label{eq:hessian}
H_e = \begin{bmatrix} \HH(\FF_1) \\ \vdots \\ \HH(\FF_n) \end{bmatrix} \in \R^{n^2 \times n}, \quad 
T_e = \begin{bmatrix}
       \HH\left(\frac{\partial \FF_1}{\partial u_1}\right) & \ldots & \HH\left(\frac{\partial \FF_1}{\partial u_n}\right) \\
       \vdots & \ddots & \vdots \\
       \HH\left(\frac{\partial \FF_n}{\partial u_1}\right) & \ldots & \HH\left(\frac{\partial \FF_n}{\partial u_n}\right)
      \end{bmatrix} \in \R^{n^2 \times n^2},
\end{equation}
where $\HH(\cdot)$ corresponds to the Hessian operator that acts on a scalar function of $n$ variables and returns an $n \times n$ matrix with all the possible second-order derivatives. As usual, all the partial derivatives in \eqref{eq:hessian} are evaluated at a steady state $\bu_e$.

From \eqref{eq:perturbed_equation} and \eqref{eq:perturbed_BC}, we can derive a sequence of problems for each power of $\eps$. By collecting terms at $\orderOne$, we obtain the linearized system evaluated at the bifurcation point,
\begin{equation}\label{eq:order_one}
\partial_t W_1 = \LL(\mu_0; W_1), \quad
\begin{cases}
 \partial_x C_1 + \frac{\kappa}{D_0}\left(e_1^T\bu_1 - C_1\right) = 0, & x = 0, \\
 \partial_x C_1 - \frac{\kappa}{D_0}\left(e_1^T\bv_1 - C_1\right) = 0, & x = 2L.
\end{cases}
\end{equation}
The solution to \eqref{eq:order_one} depends on the spatial mode considered. In what follows, we treat the even $(+)$ and the odd $(-)$ modes simultaneously, although we only consider codimension-one Hopf bifurcations. We denote $\{i\lambda_I^\pm,-i\lambda_I^\pm \}$ as the set of critical eigenvalues and $A_\pm(\tau)$ as an unknown complex amplitude depending on the slow time-scale. Then, we can write $W_1$ as
\begin{equation}\label{eq:W_1}
 W_1 = \WW_\pm A_\pm(\tau) e^{i\lambda_I^\pm t} + \overline{\WW_\pm} \overline{A_\pm(\tau)} e^{-i\lambda_I^\pm t},
\end{equation}
where the eigenfunctions are evaluated at $\mu_0$ and
$\lambda = i\lambda_I^\pm$. Our goal is to derive an evolution
equation for $A_\pm(\tau)$.

Repeating a similar procedure at $\orderTwo$, we obtain
\begin{equation}
 \partial_t W_2 = \LL(\mu_0;W_2) + \BB(W_1,W_1) + 
 \begin{pmatrix} \omega^2 C_e D_1 \\ - p_0 E \bu_e \beta_1 \\ - p_0 E \bu_e
   \beta_1 \end{pmatrix}\,,
\end{equation}
together with the appropriate boundary conditions
\begin{equation}
\begin{split}
& \partial_x C_2 + \frac{\kappa}{D_0}\left(e_1^T\bu_2 - C_2\right) = \frac{\kappa p_0}{D_0^2}e_1^T\bu_e D_1, \quad x = 0, \\
& \partial_x C_2 - \frac{\kappa}{D_0}\left(e_1^T\bv_2 - C_2\right) = -\frac{\kappa p_0}{D_0^2}e_1^T\bu_e D_1, \quad x = 2L.
\end{split}
\end{equation}
By inserting \eqref{eq:W_1} within the bilinear form, we obtain the
following quadratic terms,
\begin{equation}
 \BB(W_1,W_1) = A_\pm^2 \BB(\WW_\pm,\WW_\pm) e^{2i\lambda_I^\pm t} + |A_\pm|^2 2 \BB(\WW_\pm,\overline{\WW_\pm}) + \overline{A_\pm}^2 \BB(\overline{\WW_\pm},\overline{\WW_\pm}) e^{-2i\lambda_I^\pm t}.
\end{equation}
{This expression justifies a decomposition for $W_2$ in the form}
\begin{equation}
  W_2 = W_{0000} + A_+^2 W_{2000} e^{2i\lambda_I^+ t} + |A_+|^2 W_{1100} +
  \overline{A_+}^2 W_{0200} e^{-2i\lambda_I^+ t}
\end{equation}
for the even mode, {together with}
\begin{equation}
  W_2 = W_{0000} + A_-^2 W_{0020} e^{2i\lambda_I^- t} + |A_-|^2 W_{0011} +
  \overline{A_-}^2 W_{0002} e^{-2i\lambda_I^- t}
\end{equation}
for the odd mode. Explicit solutions for the coefficients $W_{jklm}$ and a brief outline of their computation are given in Appendix \ref{sec:Wjklm}.

\subsection{Solvability condition and amplitude equations}

Upon collecting terms of $\orderThree$ in
\eqref{eq:perturbed_equation} and \eqref{eq:perturbed_BC}, we obtain
that
\begin{equation}\label{eq:order_three}
 \partial_t W_3 - \LL(\mu_0;W_3) = -\partial_\tau W_1 + 2\BB(W_1,W_2) + \CC(W_1,W_1,W_1) +
\begin{pmatrix}
  \frac{\partial_tC_1 + kC_1}{D_0}D_1\\
 (C_1|_{x=0} - e_1^T\bu_1 )e_1 \beta_1 \\
 (C_1|_{x=2L} - e_1^T\bv_1 )e_1\beta_1
\end{pmatrix},
\end{equation}
{together with the following boundary conditions:}
\begin{equation}
\begin{split}
& D_0\partial_x C_3 + \kappa\left(e_1^T\bu_3 - C_3\right) = \left(e_1^T\bu_1 - C_1\right) \frac{\kappa}{D_0}D_1, \quad x = 0, \\
& D_0\partial_x C_3 - \kappa\left(e_1^T\bv_3 - C_3\right) = -\left(e_1^T\bv_1 - C_1\right) \frac{\kappa}{D_0}D_1, \quad x = 2L.
\end{split} 
\end{equation}
As usual when applying multi-scale expansion methods to oscillatory
problems, we suppose that the solution at $\orderThree$ is given by
the harmonic oscillator as
\begin{equation}\label{eq:ansatz_3}
 W_3 = U_\pm(\tau) e^{i\lambda_I^\pm t} + \overline{U_\pm(\tau)} e^{-i\lambda_I^\pm t}, \quad U_\pm(\tau) = \begin{pmatrix} C_\pm(x,\tau) \\ \bu_\pm(\tau) \\ \bv_\pm(\tau) \end{pmatrix},
\end{equation}
where the temporal frequency corresponds to the imaginary part of the critical eigenvalue of the spatial mode considered.

Upon inserting \eqref{eq:ansatz_3} into \eqref{eq:order_three}, and collecting the coefficients of $e^{i\lambda_I^+ t}$, we obtain that
\begin{equation}\label{eq:U_plus}
\begin{split}
& i\lambda_I^+ U_+ - \LL(\mu_0;U_+) = - \WW_+ \frac{d A_+}{d\tau} + \left( 2\BB(\WW_+,W_{0000}) + \begin{pmatrix} \left(\Omega_I^+\right)^2\eta_+(x) D_1 \\ -p_+(i\lambda_I^+)E\bphi_+ \beta_1 \\ -p_+(i\lambda_I^+)E\bphi_+ \beta_1 \end{pmatrix}\right) A_+ \\
& + \left(2\BB(\WW_+,W_{1100}) + 2\BB(\overline{\WW_+},W_{2000}) + 3\CC(\WW_+,\WW_+,\overline{\WW_+})\right)|A_+|^2A_+
\end{split}
\end{equation}
for the even mode, with the boundary conditions given by
\begin{equation}\label{eq:BC_plus}
\begin{split}
& D_0\partial_x C_+ + \kappa\left(e_1^T\bu_+ - C_+\right) =  \frac{\kappa}{D_0}p_+(i\lambda_I^+) e_1^T\bphi_+ D_1 A_+, \quad x = 0, \\
& D_0\partial_x C_+ - \kappa\left(e_1^T\bv_+ - C_+\right) = -\frac{\kappa}{D_0}p_+(i\lambda_I^+) e_1^T\bphi_+ D_1 A_+, \quad x = 2L. 
\end{split} 
\end{equation}
{Alternatively, for the odd mode, we obtain that}
\begin{equation}\label{eq:U_minus}
\begin{split}
& i\lambda_I^- U_- - \LL(\mu_0;U_-) = - \WW_- \frac{d A_-}{d\tau} + \left( 2\BB(\WW_-,W_{0000}) + \begin{pmatrix} \left(\Omega_I^-\right)^2\eta_-(x) D_1 \\ -p_-(i\lambda_I^-)E\bphi_- \beta_1 \\ p_-(i\lambda_I^-)E\bphi_- \beta_1 \end{pmatrix}\right) A_- \\
& + \left(2\BB(\WW_-,W_{0011}) + 2\BB(\overline{\WW_-},W_{0020}) + 3\CC(\WW_-,\WW_-,\overline{\WW_-})\right)|A_-|^2A_-
\end{split}
\end{equation}
with the boundary conditions given by
\begin{equation}\label{eq:BC_minus}
\begin{split}
& D_0\partial_x C_- + \kappa\left(e_1^T\bu_- - C_-\right) =  \frac{\kappa}{D_0}p_-(i\lambda_I^-) e_1^T\bphi_- D_1 A_-, \quad x = 0, \\
& D_0\partial_x C_- - \kappa\left(e_1^T\bv_- - C_-\right) = \frac{\kappa}{D_0}p_-(i\lambda_I^-) e_1^T\bphi_- D_1 A_-, \quad x = 2L. 
\end{split} 
\end{equation}

We now derive a solvability condition for the systems \eqref{eq:U_plus} and \eqref{eq:U_minus} subject to the boundary conditions \eqref{eq:BC_plus} and \eqref{eq:BC_minus}, respectively.

\begin{lemma}[Solvability condition]\label{lemma:solvability_condition}
Let $\lambda \, \in \, \C$ be an eigenvalue of the linearized operator $\LL$ {defined in \eqref{eq:linearized_operator}}, and let us consider the linear inhomogeneous system
\begin{equation}\label{eq:lemma_eqn}
 \lambda U - \LL(U) = \bGG\,,
\end{equation}
{where $\bGG$ is some generic right-hand side and $U \equiv (C(x), \bu, \bv)^T$ satisfies the following inhomogeneous boundary conditions:}
\begin{equation}\label{eq:lemma_bc}
- D\partial_x C|_{x=0} - \kappa\left(e_1^T\bu - C|_{x=0}\right) = \gamma, \quad D\partial_x C|_{x=2L} - \kappa\left(e_1^T\bv - C|_{x=2L}\right) = \xi.
\end{equation}
Then, a necessary and sufficient condition for \eqref{eq:lemma_eqn} and \eqref{eq:lemma_bc} to have a solution $U$ is that
\begin{equation}\label{fin:solve}
\langle \WW^\star , \bGG \rangle + \overline{\eta^\star(0)}\gamma + \overline{\eta^\star(2L)}\xi = 0\,,
\end{equation}
where $\WW^\star = (\eta^\star(x),\bphi^\star,\bpsi^\star)^T$ is an eigenfunction of the adjoint linearized operator {defined in \eqref{eq:adjoint_operator}}, satisfying $\LL^\star(\WW^\star) = \overline{\lambda} \WW^\star$.

\begin{proof}
The Fredholm alternative theorem guarantees the existence of a solution to \eqref{eq:lemma_eqn} and \eqref{eq:lemma_bc} if and only if the inhomogeneous terms are orthogonal to $\ker(\overline{\lambda} I - \LL^\star)$. Hence, upon taking the inner product with the adjoint eigenfunction $\WW^\star$, we obtain that
\begin{equation}\label{eq:proof_1}
0 = \langle \WW^\star, \bGG \rangle - \langle \WW^\star,  \lambda U - \LL(U) \rangle = \langle \WW^\star, \bGG \rangle - \lambda \langle \WW^\star, U \rangle + \langle \WW^\star, \LL(U) \rangle\,.
\end{equation}
Next, we integrate by parts using the definition of the inner product and further derive that
\begin{equation}\label{eq:proof_2}
\langle \WW^\star, \LL(U) \rangle = \langle \LL^\star(\WW^\star), U \rangle + \overline{\eta^\star(0)}\gamma + \overline{\eta^\star(2L)}\xi = \lambda \langle \WW^\star, U \rangle + \overline{\eta^\star(0)}\gamma + \overline{\eta^\star(2L)}\xi \,.
\end{equation}
The result \eqref{fin:solve} is readily obtained after substituting \eqref{eq:proof_2} back into \eqref{eq:proof_1}.
\end{proof}
\end{lemma}

As a direct application of Lemma \ref{lemma:solvability_condition}, we now obtain the desired amplitude equations. For the even mode, we have that
\begin{equation}\label{eq:amplitude_even}
\frac{d A_+}{d\tau} = g_{1000}^T \mu_1 A_+ + g_{2100} |A_+|^2 A_+\,,
\end{equation}
while similarly for the odd mode we have
\begin{equation}\label{eq:amplitude_odd}
\frac{d A_-}{d\tau} = g_{0010}^T \mu_1 A_- + g_{0021} |A_-|^2 A_-\,.
\end{equation}
The coefficients $g_{2100},\, g_{0021} \in \, \C$ of the cubic terms in these amplitude equations are given by
\begin{subequations}
\begin{align}
g_{2100} &= \langle \WW_+^\star, 2\BB(\WW_+,W_{1100}) + 2\BB(\overline{\WW_+},W_{2000}) + 3 \CC(\WW_+,\WW_+,\overline{W_+}) \rangle\,, \\
g_{0021} &= \langle \WW_-^\star, 2\BB(\WW_-,W_{0011}) + 2\BB(\overline{\WW_-},W_{0020}) + 3 \CC(\WW_-,\WW_-,\overline{W_-}) \rangle\,,
\end{align}
\end{subequations}
while the vector coefficients $g_{1000},\, g_{0010} \in \, \C^2$ satisfy
\begin{subequations}
\begin{align}
 &g_{1000} = \overline{\bphi_+^\star}^T E \bphi_+ \bigg(\frac{\beta_0}{\kappa}(1-p_+(i\lambda_I^+))^2 \Omega_I^+ \left( \tanh(\Omega_I^+L) + \Omega_I^+L \sech^2(\Omega_I^+ L) \right)\xi_2 \\ 
 & - 2p_+(i\lambda_I^+) \xi_1 + 2\frac{\beta_0}{D_0}(p_+(i\lambda_I^+) - 1)p_+(i\lambda_I^+) \xi_2 \bigg) + 4 \overline{\bphi_+^\star}^T B(\bphi_+,[\bPhi_+(0)]^{-1}E\bu_e)\alpha\,, \nonumber \\
 &g_{0010} = \overline{\bphi_-^\star}^T E \bphi_- \bigg(\frac{\beta_0}{\kappa}(1-p_-(i\lambda_I^-))^2 \Omega_I^- \left( \coth(\Omega_I^-L) - \Omega_I^-L \cosech^2(\Omega_I^- L) \right)\xi_2 \\
 & - 2p_-(i\lambda_I^-) \xi_1 + 2\frac{\beta_0}{D_0}(p_-(i\lambda_I^-) - 1)p_-(i\lambda_I^-) \xi_2 \bigg) + 4 \overline{\bphi_-^\star}^T B(\bphi_-,[\bPhi_+(0)]^{-1}E\bu_e)\alpha\,. \nonumber
\end{align}
\end{subequations}
Finally, the following lemma summarizes our asymptotic approximations for the weakly nonlinear oscillations in the vicinity of a Hopf bifurcation point for our PDE-ODE system:

\begin{lemma}[In-phase and anti-phase periodic solutions in the weakly nonlinear regime]\label{lemma:weakly_nonlinear_regime}
Let $g_{2100}, g_{0021} \in \C$ be the cubic term coefficients in \eqref{eq:amplitude_even} and \eqref{eq:amplitude_odd}, and assume that their real part is nonzero, hence excluding degenerate cases. Then, in the limit $\eps \to 0$ with $\eps = \sqrt{\|\mu - \mu_0\|}$ denoting the square-root of the distance {from} the bifurcation point, a leading-order approximate family of in-phase and anti-phase periodic solutions is given by
\begin{equation}\label{eq:periodic_solution}
W_\pm(t) = W_e + \eps \rho_{e\pm} \left[ \WW_\pm e^{i\left(\lambda_I^\pm t + \theta_\pm(0)\right)} + \overline{\WW_\pm} e^{-i\left(\lambda_I^\pm t + \theta_\pm(0)\right)} \right] + \orderTwo\,,
\end{equation}
for any $\theta_\pm(0) \in \R$ and with $\rho_{e\pm}$ defined by 
\begin{equation}
\rho_{e+} = \sqrt{\frac{\|g_{1000}\|}{|g_{2100}|}}, \quad \rho_{e-} = \sqrt{\frac{\|g_{0010}\|}{|g_{0021}|}}.  
\end{equation}
Furthermore, let $\bu_\text{amp}$ denote the amplitude of the bifurcating limit cycle near the Hopf bifurcation point, for both left and right local species. A leading-order approximation for $\bu_\text{amp}$ is given by
\begin{equation}\label{eq:u_amp}
 \bu_{\text{amp}} = \max_{0 \leq t < T_p^\pm} \left\{ \|\bu_\pm(t) - \bu_e\| \right\} = 2\eps \rho_{e\pm}\|\bphi_\pm\| + \orderTwo,
\end{equation}
where the period $T_p^\pm$ of small-amplitude oscillations satisfies
\begin{equation}\label{eq:period}
 T_p^\pm = \frac{2\pi}{\lambda_I^\pm} + \orderTwo.
\end{equation}
Finally, the periodic solution in \eqref{eq:periodic_solution} is asymptotically stable when $\Re(g_{2100}),\,\Re(g_{0021}) < 0$ (supercritical Hopf) and it is unstable for $\Re(g_{2100}),\,\Re(g_{0021}) > 0$ (subcritical Hopf).
\end{lemma}

%%%%%%%%%%%%%%%%%%%%%%%

\section{Diffusive coupling of two identical Sel'kov oscillators}\label{sec:selkov}

{We first recall the full coupled PDE-ODE model, formulated as 
\begin{equation}
\begin{split}
& C_t = D C_{xx} - k C\,, \qquad 0 < x < 2L\,, \qquad t > 0\,, \\
& -D C_x(0,t) = \kappa (e_1^T\bu(t) - C(0,t))\,, \quad DC_x(2L,t) = \kappa (e_1^T\bv(t) - C(2L,t))\,, \\
& \dfrac{d \bu}{dt} = \bFF(\bu) + \beta(C(0,t) - e_1^T\bu)e_1\,, \quad \dfrac{d \bv}{dt} = \bFF(\bv) + \beta(C(2L,t) - e_1^T\bv)e_1\,.
\end{split}
\end{equation}
In this section we consider a two-dimensional nonlinear vector function $\bFF$ that corresponds to the Sel'kov model, given by}
\begin{equation}\label{eq:selkov}
\bFF(\bu) = \begin{pmatrix} A u_2 + u_2u_1^2 - u_1 \\ \epsilon \left[M - (A u_2 + u_2u_1^2) \right]\end{pmatrix}\,, \quad \bu = \begin{pmatrix} u_1 \\ u_2 \end{pmatrix} \in\, \R^2\,,
\end{equation}
where $A$, $M$ and $\epsilon$ are three positive reaction
parameters. Upon solving \eqref{eq:ss_eqn} for a symmetric steady
state, we find a unique solution given by
\begin{equation}
\bu_e = \left( \frac{M}{1+\beta p_0}\,, \frac{M(1+\beta p_0)^2}{A(1+\beta p_0)^2 + M^2}\right)^T\,,
\end{equation}
where $p_0$ is defined in \eqref{eq:symmetric_steady_state}. We assume that in the absence of coupling ($\beta = 0$), each isolated
compartment is quiescent. This is guaranteed when the Sel'kov parameters satisfy the inequality
\begin{equation}\label{eq:cond_oscillator}
\epsilon > \frac{M^2 - A}{(M^2 + A)^2}\,.
\end{equation}
As a result, the spatio-temporal oscillations studied below are due to the coupling between the two compartments and the 1-D bulk diffusion field.

\begin{figure}[htbp]
\centering
\begin{subfigure}{.32\linewidth}
\includegraphics[width=\linewidth]{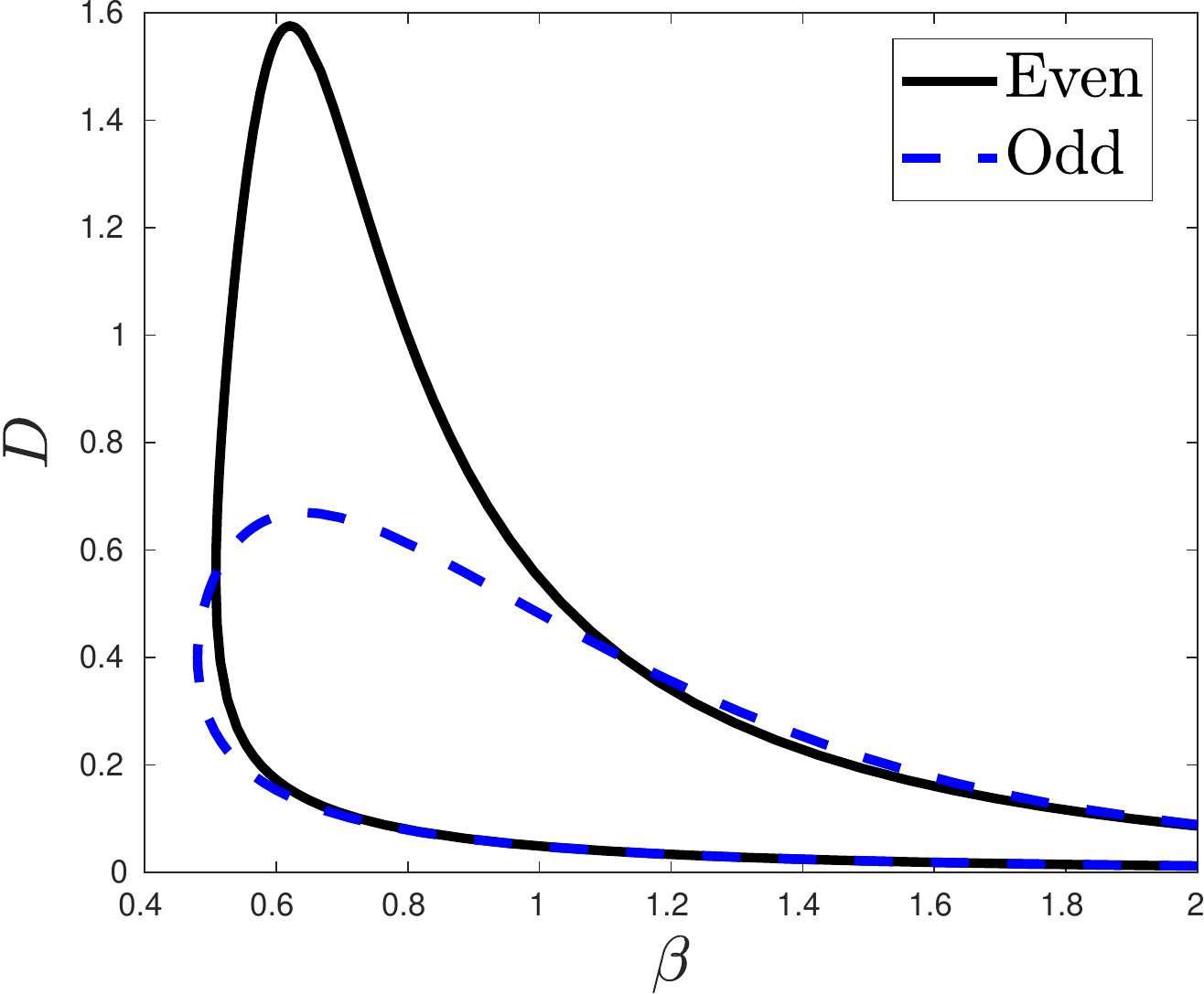}
\end{subfigure}
\begin{subfigure}{.32\linewidth}
\includegraphics[width=\linewidth]{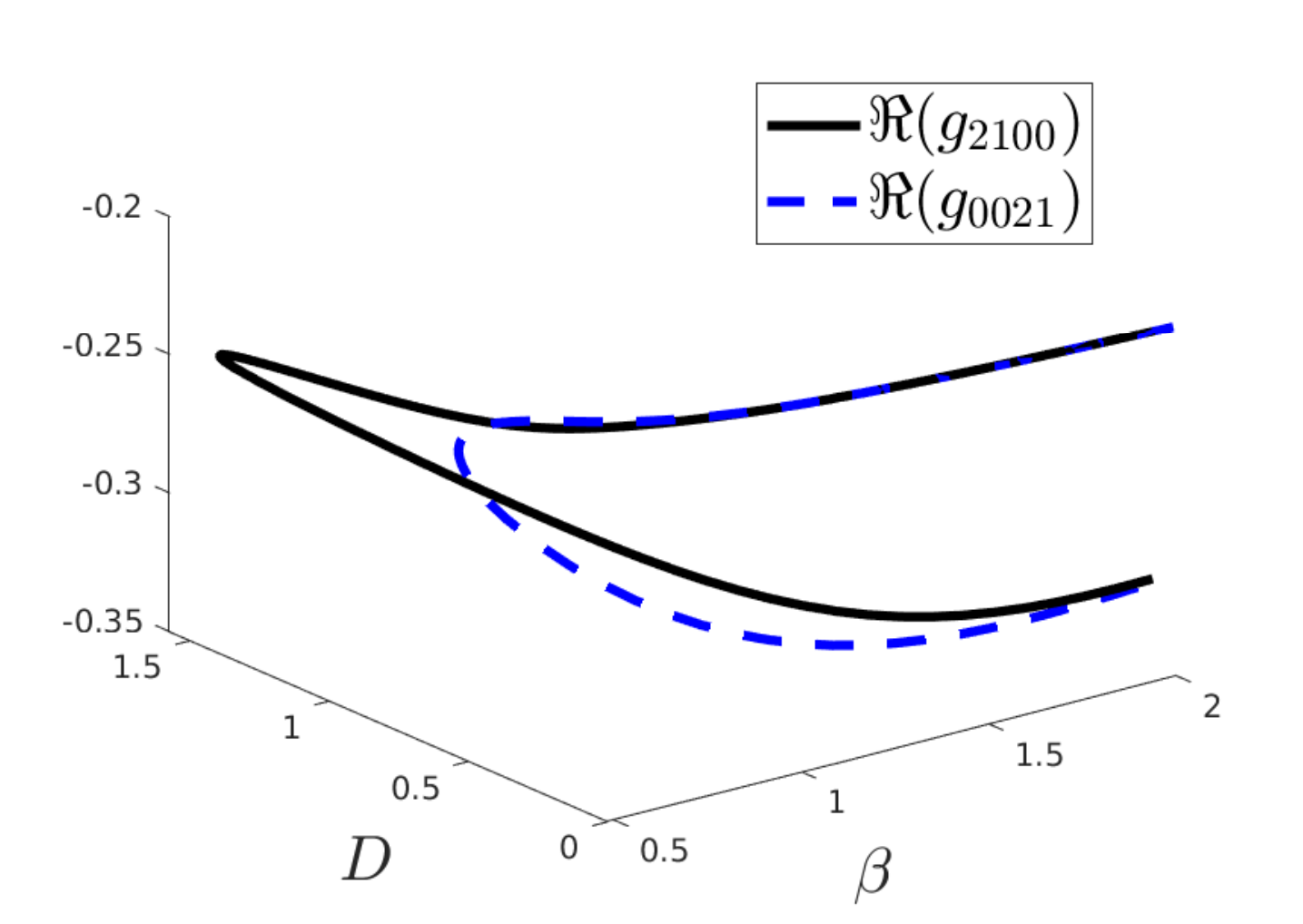} 
\end{subfigure}
\caption{\label{fig:stabDiagSelkov_1} Stability diagrams in the plane of parameters $(\beta,\, D)$ {for the Sel'kov model \eqref{eq:selkov}}. The parameter regime of oscillatory dynamics is located inside the curves. In the right panel, we numerically evaluate the real part of the cubic normal form coefficients in \eqref{eq:amplitude_even} and \eqref{eq:amplitude_odd} over the two stability boundaries. {These coefficients are negative, indicating a supercritical bifurcation.} Parameters values are $L = k = \kappa = 1,\, M = 2,\, A = 0.9$ and $\epsilon = 0.15$.}
\end{figure}

To illustrate the theory developed in \S \ref{sec:weakly_nonlinear_theory}, we choose the parameter values $M=2$, $A=0.9$ and $\epsilon = 0.15$ and numerically solve the eigenvalue relation \eqref{eq:transcendental} in the parameter plane defined by the coupling strength $\beta$ and the diffusion level $D$. The resulting stability diagram is shown in the left panel of Fig.~\ref{fig:stabDiagSelkov_1}, with the black and dashed-blue curves, respectively, corresponding to the in-phase and the anti-phase oscillatory modes. In the right panel, we numerically evaluate the real part of the cubic normal form coefficients in \eqref{eq:amplitude_even} and \eqref{eq:amplitude_odd}. Our numerical computations show that $\Re(g_{2100})$ and $\Re(g_{0021})$ are both negative, which indicate that supercritical Hopf bifurcations can be expected while crossing either the even or the odd Hopf stability boundaries. Hence, we predict the existence of stable weakly nonlinear spatio-temporal oscillations when a single oscillatory mode becomes unstable. This prediction may not hold when the two distinct instabilities coincide, which for instance occurs when $D$ is small.

{We remark that the linear stability phase diagram in the left panel of Fig.~\ref{fig:stabDiagSelkov_1} was previously computed in \cite{gou2017}, where the resulting oscillatory dynamics was studied numerically from PDE simulations and global bifurcation software. The new weakly nonlinear theory developed in this paper establishes that this Hopf bifurcation is supercritical. Finally, in \cite{gou2016DH} a center manifold analysis predicted the presence of unstable mixed-mode oscillations in the vicinity of the codimension-two Hopf bifurcation point at $\mu_0 \approx (0.508,0.556)$.}

\begin{figure}[htbp]
\centering
\begin{subfigure}{0.32\linewidth}
\includegraphics[width=\linewidth]{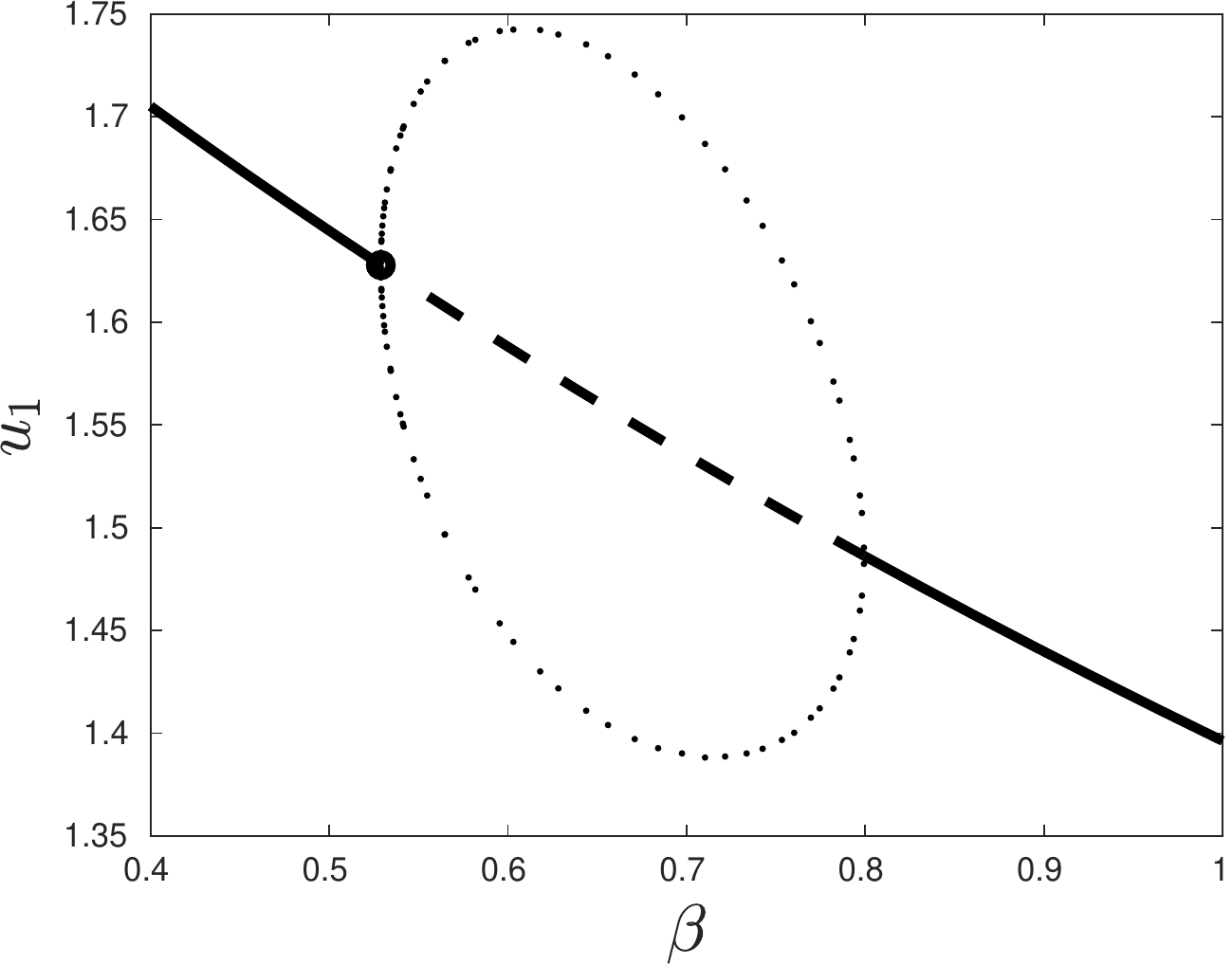}
\caption{$D=1$}
\end{subfigure}
\begin{subfigure}{0.32\linewidth}
\includegraphics[width=\linewidth]{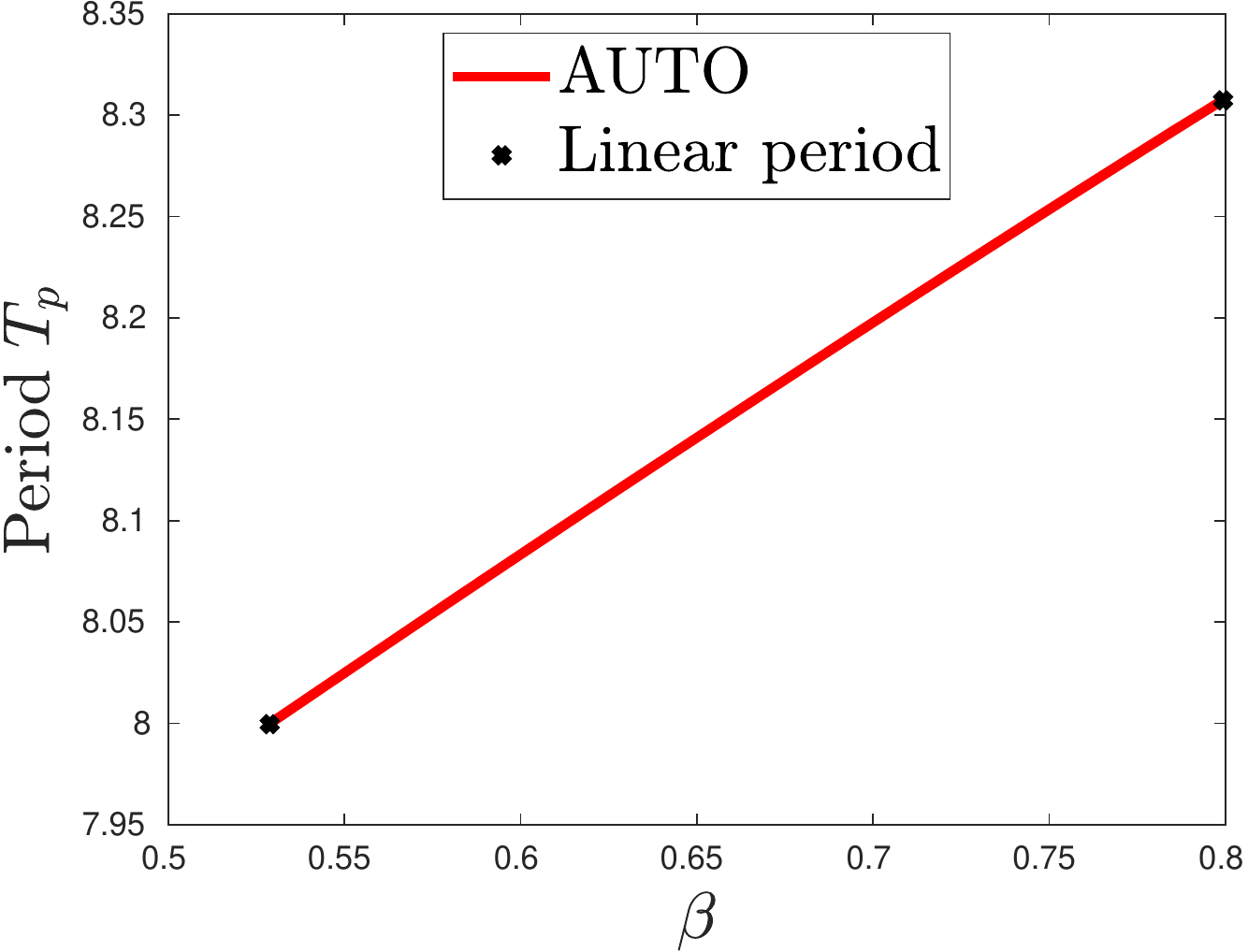}
\caption{$D=1$}
\end{subfigure}
\begin{subfigure}{0.32\linewidth}
\includegraphics[width=\linewidth]{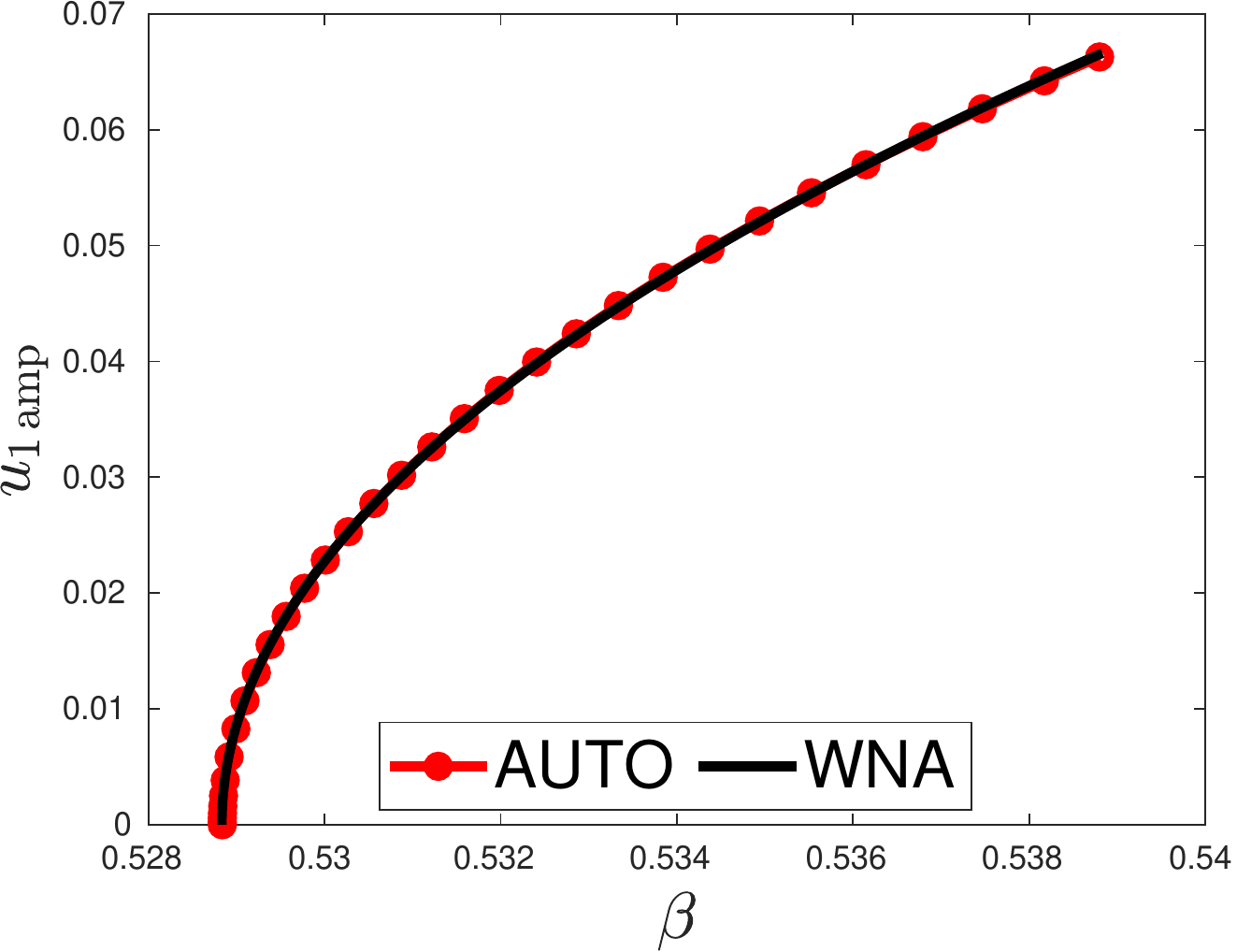}
\caption{$D=1$}
\end{subfigure}

\begin{subfigure}{0.32\linewidth}
\includegraphics[width=\linewidth]{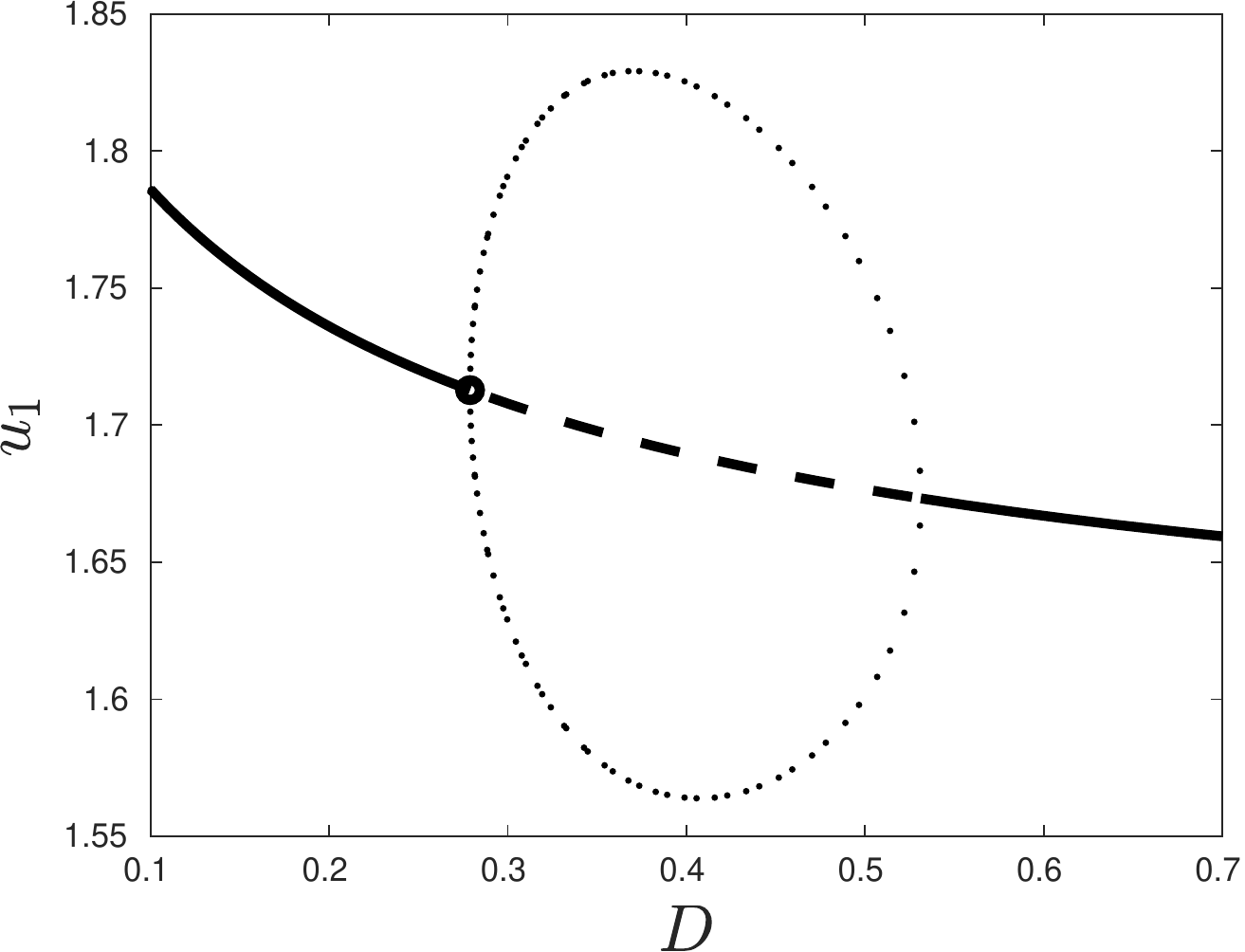}
\caption{$\beta=0.5$}
\end{subfigure}
\begin{subfigure}{0.32\linewidth}
\includegraphics[width=\linewidth]{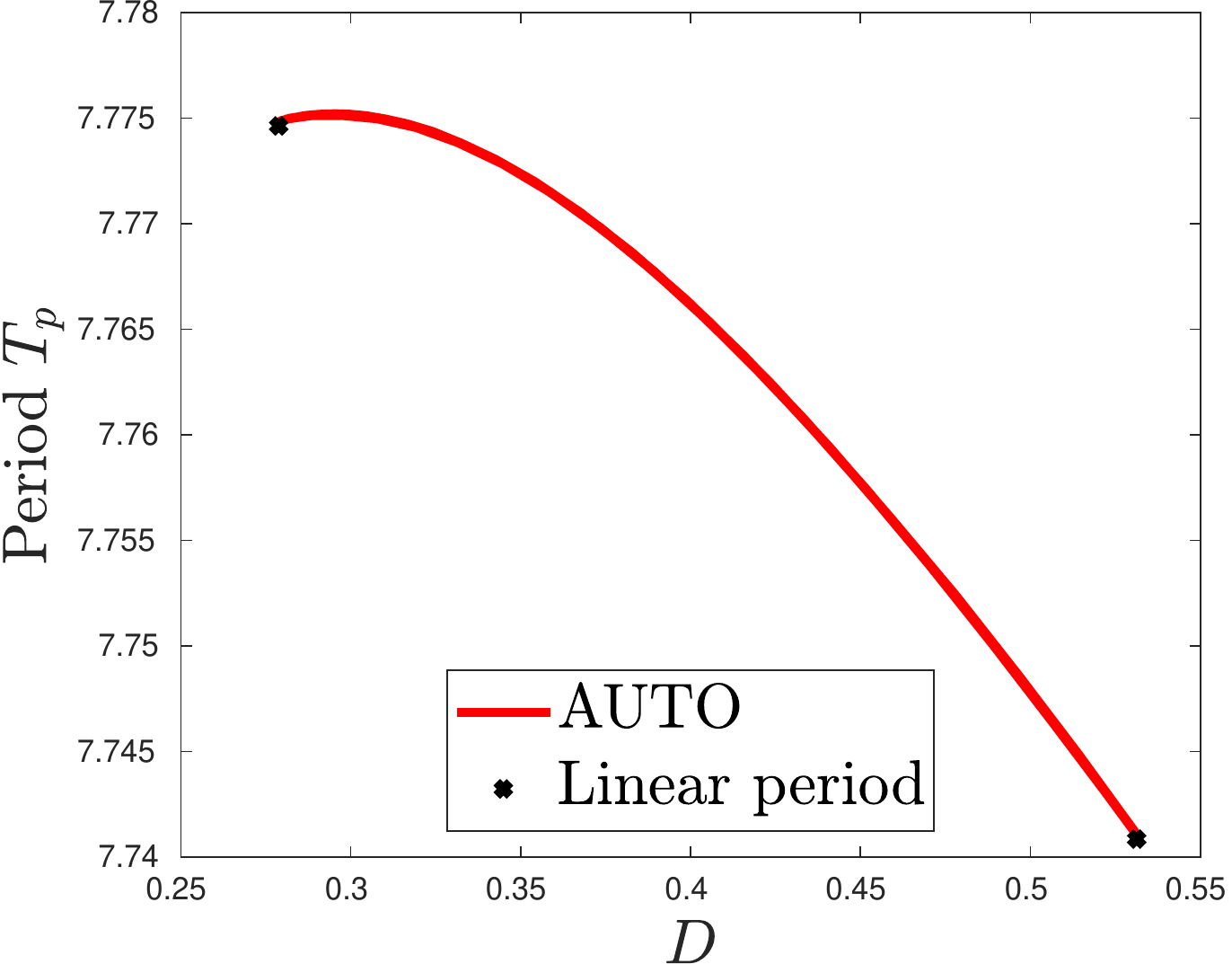}
\caption{$\beta=0.5$}
\end{subfigure}
\begin{subfigure}{0.32\linewidth}
\includegraphics[width=\linewidth]{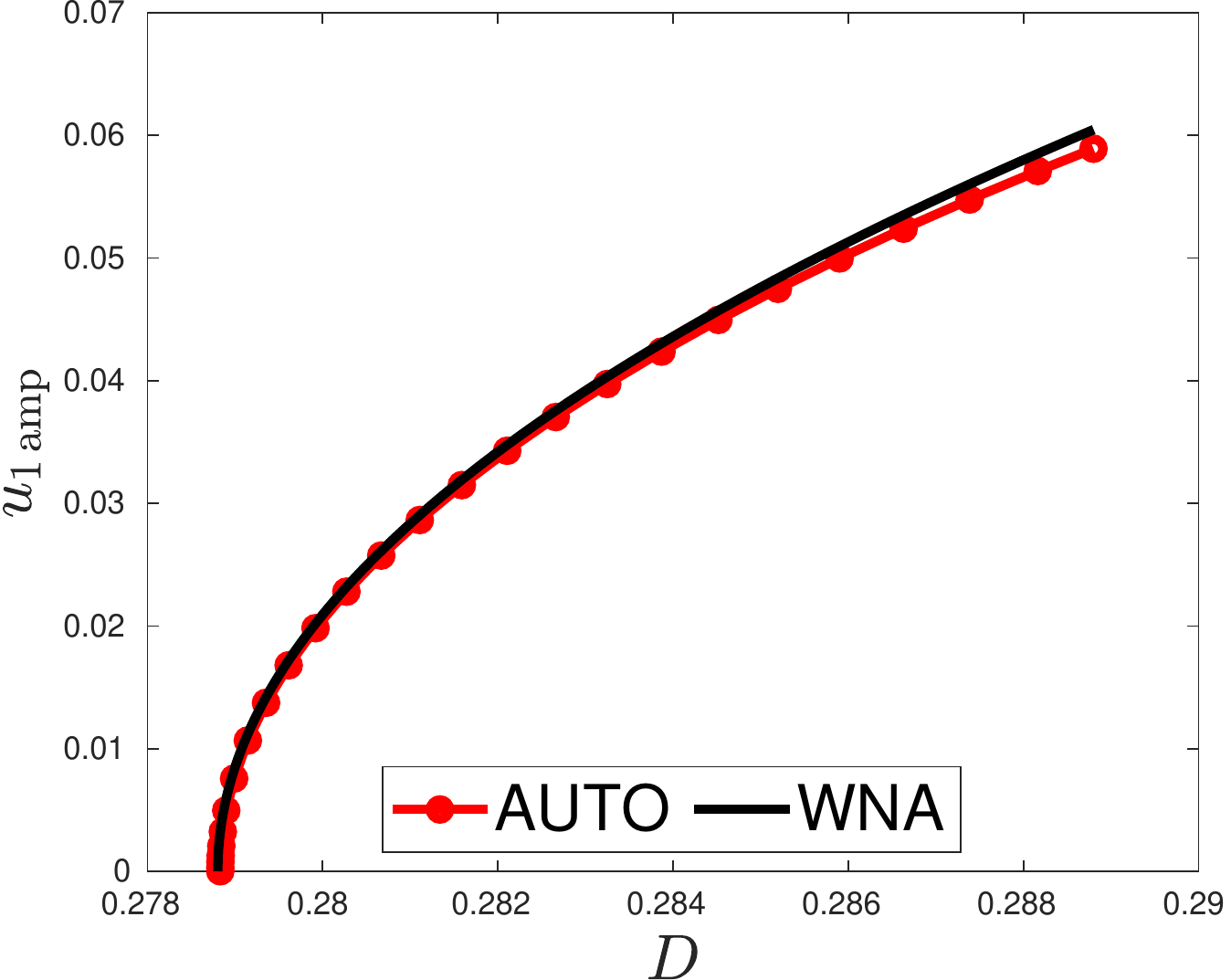}
\caption{$\beta=0.5$}
\end{subfigure}
\caption{\label{fig:slice_1} Stable branches of periodic solutions on the slice $D=1$ (panels (a)-(c)) and on the slice $\beta=0.5$ (panels (d)-(f)) of the stability diagram in Fig.~\ref{fig:stabDiagSelkov_1}. Panels (a), (d): Global branch of periodic solutions. Panels (b), (e): Oscillatory period.  Panels (c), (f): Near the first supercritical Hopf bifurcation along each slice, we compare the numerically computed amplitude (red curves) and the weakly nonlinear prediction as obtained from \eqref{eq:u_amp} with $\eps = 0.1$ (black curves). The 1-D bulk interval is spatially discretized with $N=200$ grid points and other parameter values are the same as in the caption of Fig.~\ref{fig:stabDiagSelkov_1}.}
\end{figure}

\begin{figure}[htbp]
\centering
\begin{subfigure}{0.32\linewidth}
\includegraphics[width=\linewidth]{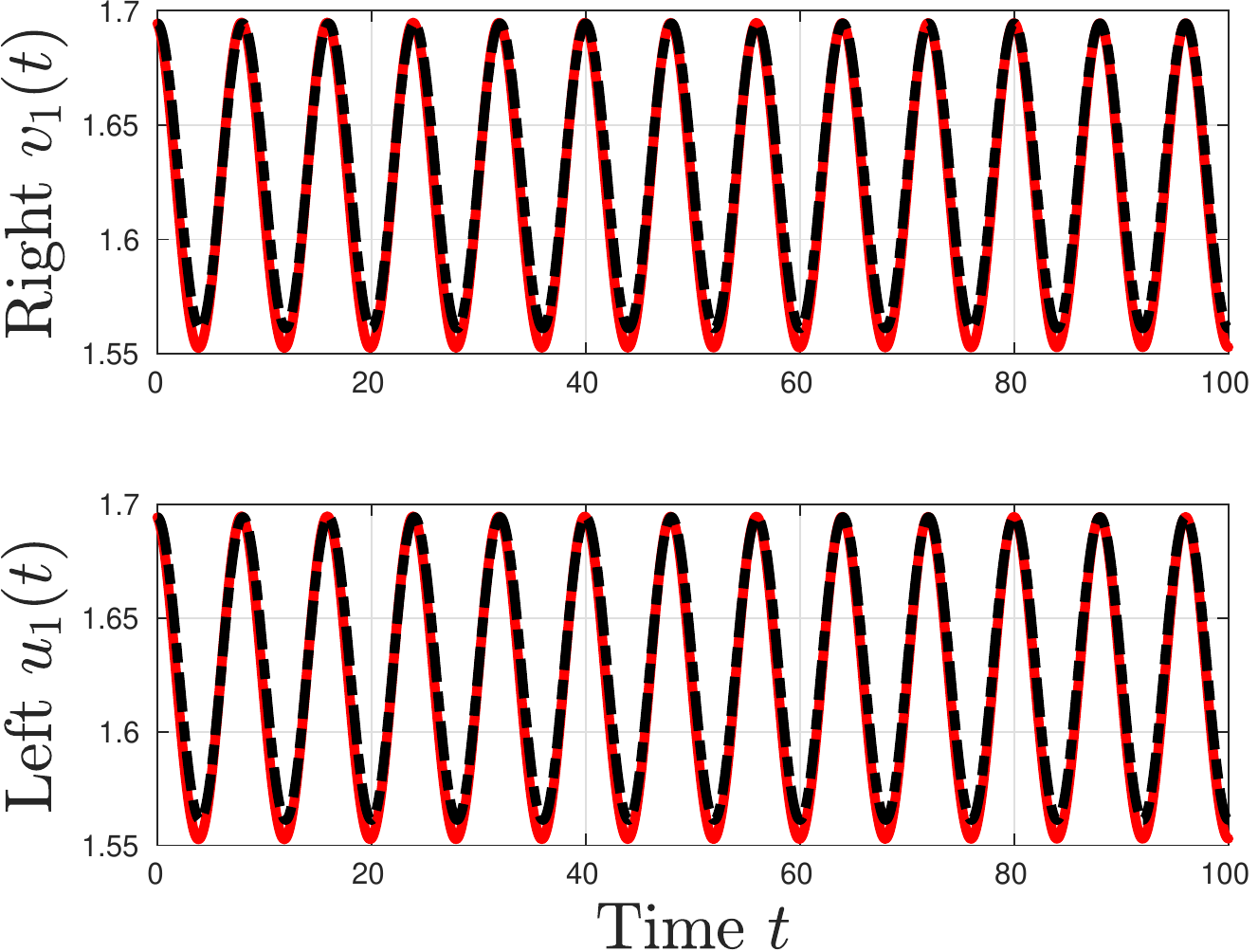}
\caption{$\mu = (0.54, 1)$.}
\end{subfigure}
\begin{subfigure}{0.32\linewidth}
\includegraphics[width=\linewidth]{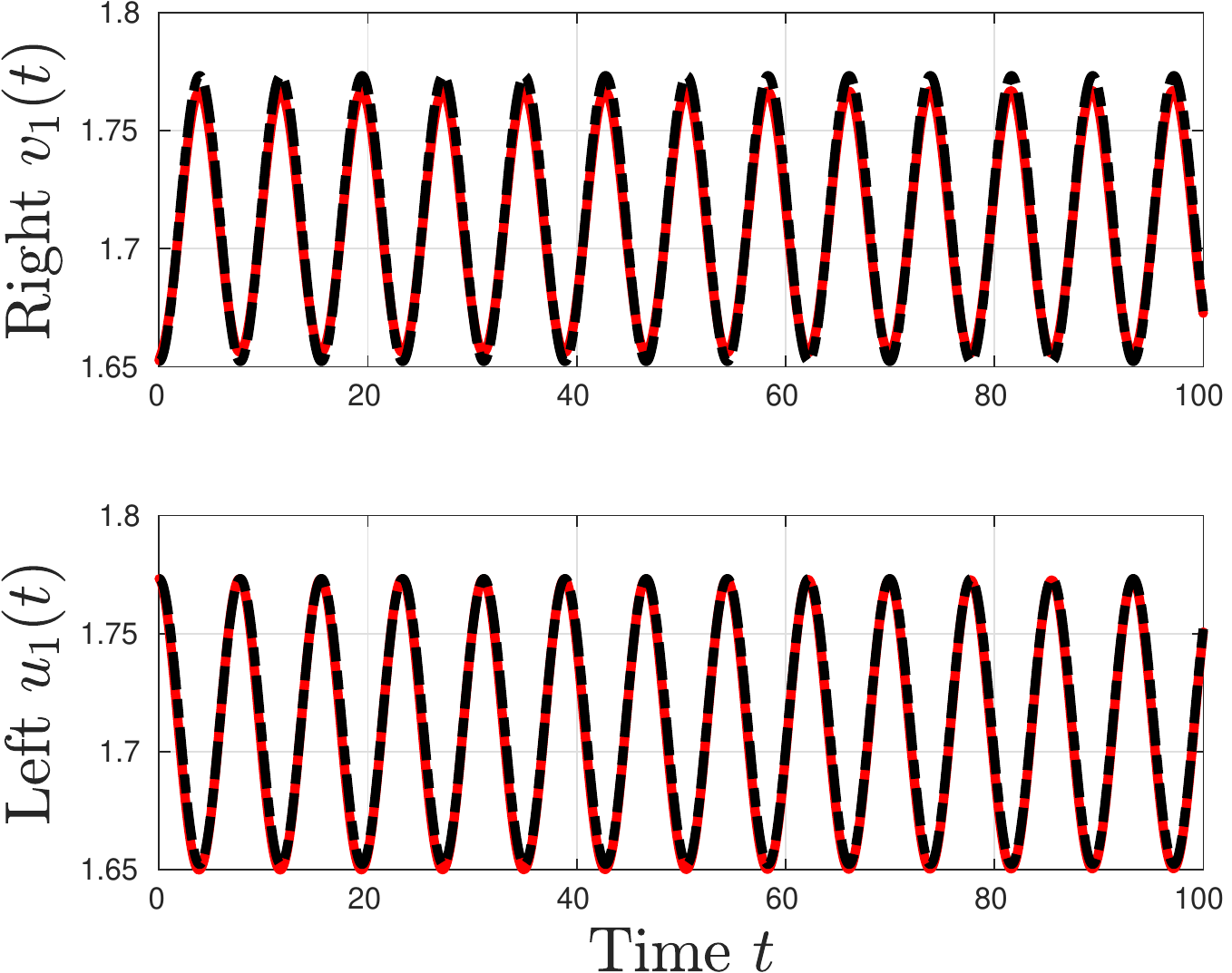}
\caption{$\mu = (0.5, 0.29)$.}
\end{subfigure}
\caption{\label{fig:WNA} In-phase (panel (a)) and anti-phase (panel (b)) oscillations near supercritical Hopf bifurcations, with the red and black-dashed curves respectively corresponding to numerical simulations and to weakly nonlinear periodic solutions (formula \eqref{eq:periodic_solution} with $\eps = 0.1$). The initial conditions for the simulations are given by the weakly nonlinear periodic solutions. The same discretization as in Fig.~\ref{fig:slice_1} is employed.}
\end{figure}

Next, we compare our weakly nonlinear theory against numerical bifurcation results obtained with AUTO (cf.~\cite{doedel2007}) after spatially discretizing \eqref{eq:bulk} with finite differences. In panels (a)-(c) of {Fig.~\ref{fig:slice_1}}, we compute the stable branch of in-phase periodic solutions along the horizontal slice $D=1$, as a function of the coupling strength $\beta$. Near one of the supercritical Hopf bifurcation {points}, we observe in panel (c) a good agreement between the amplitude of the limit cycle computed numerically and as obtained from \eqref{eq:u_amp} with $\eps=0.1$. Qualitatively similar results are shown in panels (d)-(f) of Fig.~\ref{fig:slice_1} for the vertical slice $\beta = 0.5$, {which crosses the boundary} of anti-phase oscillations. Finally in Fig.~\ref{fig:WNA}, and for each oscillatory mode, we give numerically computed time-courses as evolved directly from the solutions in the weakly nonlinear regime (given by \eqref{eq:periodic_solution} with $\eps=0.1$). Such an agreement between the two solutions should also hold for random initial conditions given a sufficiently long integration time and an adjustment of the temporal phase shift.

\begin{figure}[htbp]
\centering
\begin{subfigure}{0.32\linewidth}
\includegraphics[width=\linewidth]{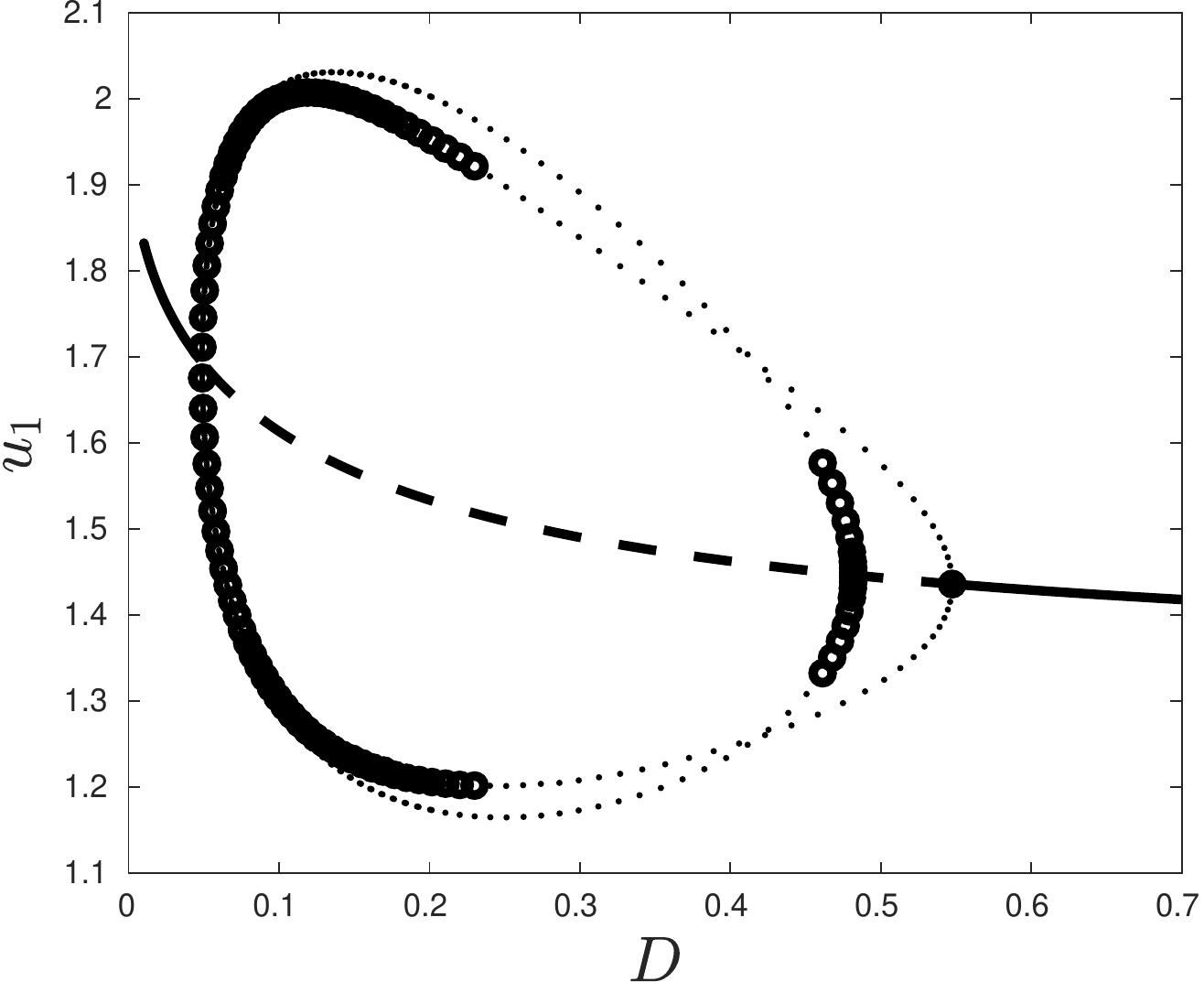}
\caption{$\beta=1$}
\end{subfigure}
\begin{subfigure}{0.32\linewidth}
\includegraphics[width=\linewidth]{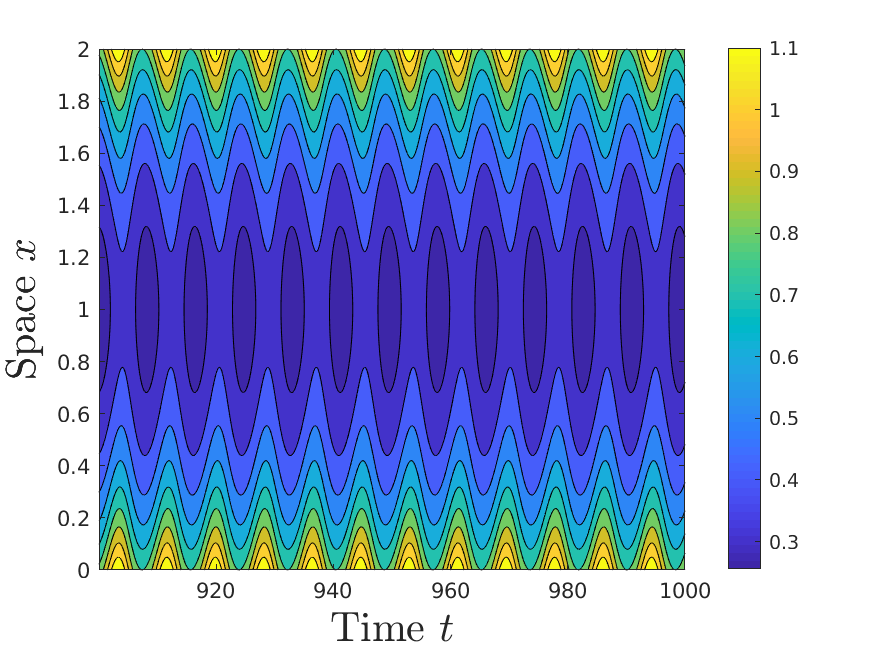}
\caption{$\beta=1,\,D=0.3$}
\end{subfigure}
\begin{subfigure}{0.32\linewidth}
\includegraphics[width=\linewidth]{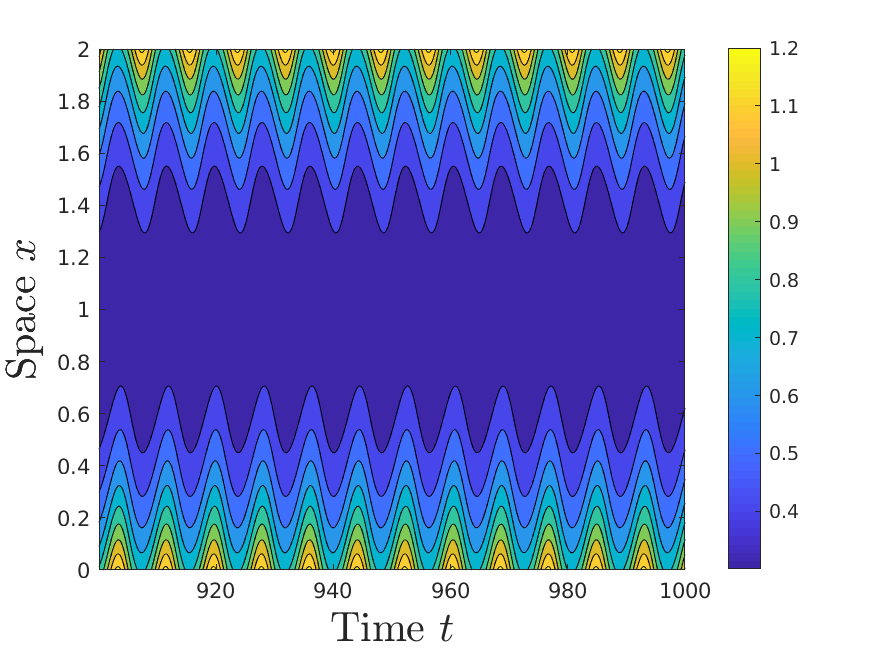}
\caption{$\beta=1,\,D=0.3$}
\end{subfigure}
\caption{\label{fig:slice_2} Interaction of in-phase and anti-phase periodic solutions on the vertical slice $\beta=1$. In panel (a), unstable limit cycles are indicated by open circles while the black dots indicate stable limit cycles. Inner (outer) loops are in-phase (anti-phase) periodic solution branches. Panels (b)-(c): Bistability between in-phase and anti-phase spatio-temporal oscillations, with the spatial variable on the vertical axis and the temporal variable on the horizontal axis. Other parameter values are as in the caption of Fig.~\ref{fig:stabDiagSelkov_1}. Once again, $N=200$ grid points are employed to discretize the 1-D bulk diffusion field.}
\end{figure}

We {conclude} this section with numerical results illustrating the possible bistability between the in-phase and anti-phase oscillations. In Fig.~\ref{fig:slice_2}, we show in panel (a) the global bifurcation diagram on the vertical slice $\beta=1$, where we find an intermediate range of bulk diffusion values $(0.25 < D < 0.45)$ {where both oscillatory modes} are stable. This is confirmed in panels (b) and (c), where numerically computed time-courses are seen to evolve either into in-phase or anti-phase spatio-temporal oscillations, depending on the initial conditions. Here, the boundaries of this bistability parameter range correspond to bifurcations of invariant tori, at which a certain branch of limit cycles switches stability.

%%%%%%%%%%%%%%%%%%%%%%%%%

\section{Diffusive coupling of two identical chaotic Lorenz oscillators}\label{sec:lorenz}

In this section, we consider the diffusive coupling of two identical Lorenz oscillators. We define the nonlinear vector function
$\bFF(\bu)$ for the Lorenz oscillator as
\begin{equation}
\bFF(\bu) = \begin{pmatrix} \sigma (u_2-u_1) \\ -u_1u_3 + r u_1 - u_2 \\ u_1u_2 - bu_3 \end{pmatrix}, \quad \bu = \begin{pmatrix} u_1 \\ u_2 \\ u_3 \end{pmatrix} \in\, \R^3\,,
\end{equation}
where $r$, $\sigma$ and $b$ are the usual Lorenz constants. We take
the classical values $\sigma = 10$ and $b=\frac{8}{3}$, while keeping
$r$, which is proportional to the Rayleigh number, as a bifurcation
parameter. {The general form of the coupled PDE-ODE system
  remains the same as in section \ref{sec:intro}, with the exception
  of the leakage parameter $\kappa$, which we here take to be
  identical to the coupling strength $\beta$.} {In this way, the
  full coupled PDE-ODE model is formulated as} {
\begin{equation}\label{eq:PDE_ODE_lorenz}
\begin{split}
& C_t = D C_{xx} - k C\,, \qquad 0 < x < 2L\,, \qquad t > 0\,, \\
& -D C_x(0,t) = \beta (e_1^T\bu(t) - C(0,t))\,, \quad DC_x(2L,t) = \beta (e_1^T\bv(t) - C(2L,t))\,, \\
& \dfrac{d \bu}{dt} = \bFF(\bu) + \beta(C(0,t) - e_1^T\bu)e_1\,, \quad \dfrac{d \bv}{dt} = \bFF(\bv) + \beta(C(2L,t) - e_1^T\bv)e_1\,.
\end{split}
\end{equation}
With this choice of boundary conditions, the outward flux at each endpoint is identical to the local feedback within the ODEs.}

We will also investigate in this section {the infinite bulk diffusion limit} $(D = \infty)$, corresponding to the well-mixed regime. In this regime, the coupled PDE-ODE system can be reduced to the following globally coupled system of ODEs (see Appendix \ref{sec:well_mixed}):
\begin{equation}\label{eq:well_mixed}
\dfrac{d}{dt}\begin{pmatrix} C_0 \\ \bu \\ \bv \end{pmatrix} = 
\begin{pmatrix}
\frac{\beta}{2L}e_1^T\left(\bu + \bv\right) - \left(k + \frac{\beta}{L}\right)C_0 \\ 
\bFF(\bu) + \beta(C_0 - e_1^T\bu)e_1 \\
\bFF(\bv) + \beta(C_0 - e_1^T\bv)e_1
\end{pmatrix}\,,
\end{equation}
where $C_0(t)$ is the spatially {uniform} bulk variable.

\subsection{{Linear stability analysis}}

Next, we solve for the symmetric steady states of the two coupled Lorenz oscillators, for either a finite or an infinite bulk diffusivity. We find two non-trivial solutions satisfying the steady state equation \eqref{eq:ss_eqn}, given by
\begin{equation}
\bu_e^\pm = \left( \pm \sqrt{\frac{b\left(r- 1 - \frac{\beta}{\sigma} p_0\right)}{1 + \frac{\beta}{\sigma}p_0}}\,, 
\pm \sqrt{b\left(r-1 - \frac{\beta}{\sigma}p_0\right) \left(1 + \frac{\beta}{\sigma}p_0\right)}\,, 
r - 1 - \frac{\beta}{\sigma}p_0\right)^T\,,
\end{equation}
that branch from the origin in a pitchfork bifurcation at the critical value
\begin{equation}
r = 1 + \frac{\beta}{\sigma}p_0 \quad \text{with} 
\quad p_0 = 
\begin{cases}
\frac{D \omega \tanh(\omega L)}{D \omega \tanh(\omega L) + \beta}\,, & D = \mathcal{O}(1)\,, \\
\frac{k}{k + \beta/L}\,, & D = \infty\,.
\end{cases}
\end{equation} 
{By linearity of the diffusive coupling and its Robin boundary conditions, the coupled PDE-ODE formulation preserves the reflection symmetry of the Lorenz system and stability results will be the same for both non-trivial steady states. Hence, we restrict our analysis to the positive non-trivial steady state $\bu_e$, where the superscript $+$ has been dropped to simplify notations.} To determine the linear stability of this steady state, we recall from \eqref{eq:transcendental} that the growth rate $\lambda$ of in-phase and anti-phase perturbations satisfies
\begin{equation}\label{eq:transcendental_beta}
\det\left[J_e - \lambda I - \beta p_\pm(\lambda) E \right] = 0\,,
\end{equation}
where 
\begin{equation}
p_+(\lambda) =
\begin{cases}
\frac{D \Omega \tanh(\Omega L)}{D \Omega \tanh(\Omega L) + \beta}\,, & D = \mathcal{O}(1)\,, \\
\frac{k + \lambda}{k + \lambda + \beta/L}\,, & D = \infty\,,
\end{cases} \quad 
p_-(\lambda) =
\begin{cases}
\frac{D \Omega \coth(\Omega L)}{D \Omega \coth(\Omega L) + \beta}\,, & D = \mathcal{O}(1)\,, \\
1 \,, & D = \infty\,.
\end{cases}
\end{equation}
In the absence of coupling $(\beta=0)$, we recover the usual steady state structure of the Lorenz ODE system. In particular, the non-trivial steady states are well known to lose stability in a subcritical Hopf bifurcation when the Rayleigh number reaches the following critical value:
\begin{equation}\label{eq:uncoupled}
 r_0 = \frac{\sigma(\sigma + b + 3)}{\sigma - b - 1} \approx 24.74\,.
\end{equation}
The corresponding critical frequency is given by $\lambda_I = \sqrt{b(\sigma + r_0)} \approx 9.62$.

\begin{figure}[htbp]
\centering
\begin{subfigure}{0.32\linewidth}
\includegraphics[width=\linewidth]{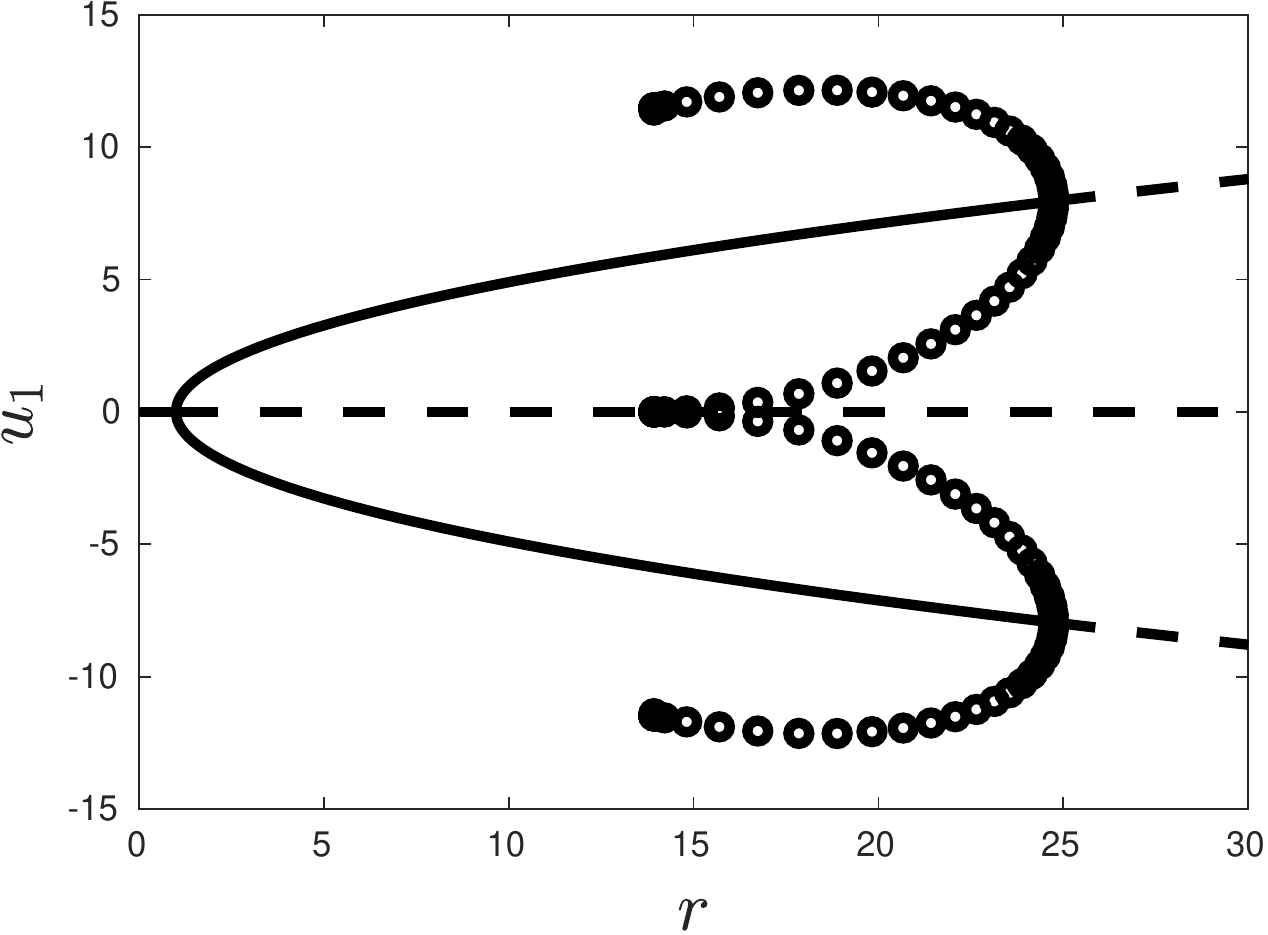}
\end{subfigure}
\begin{subfigure}{0.32\linewidth}
\includegraphics[width=\linewidth]{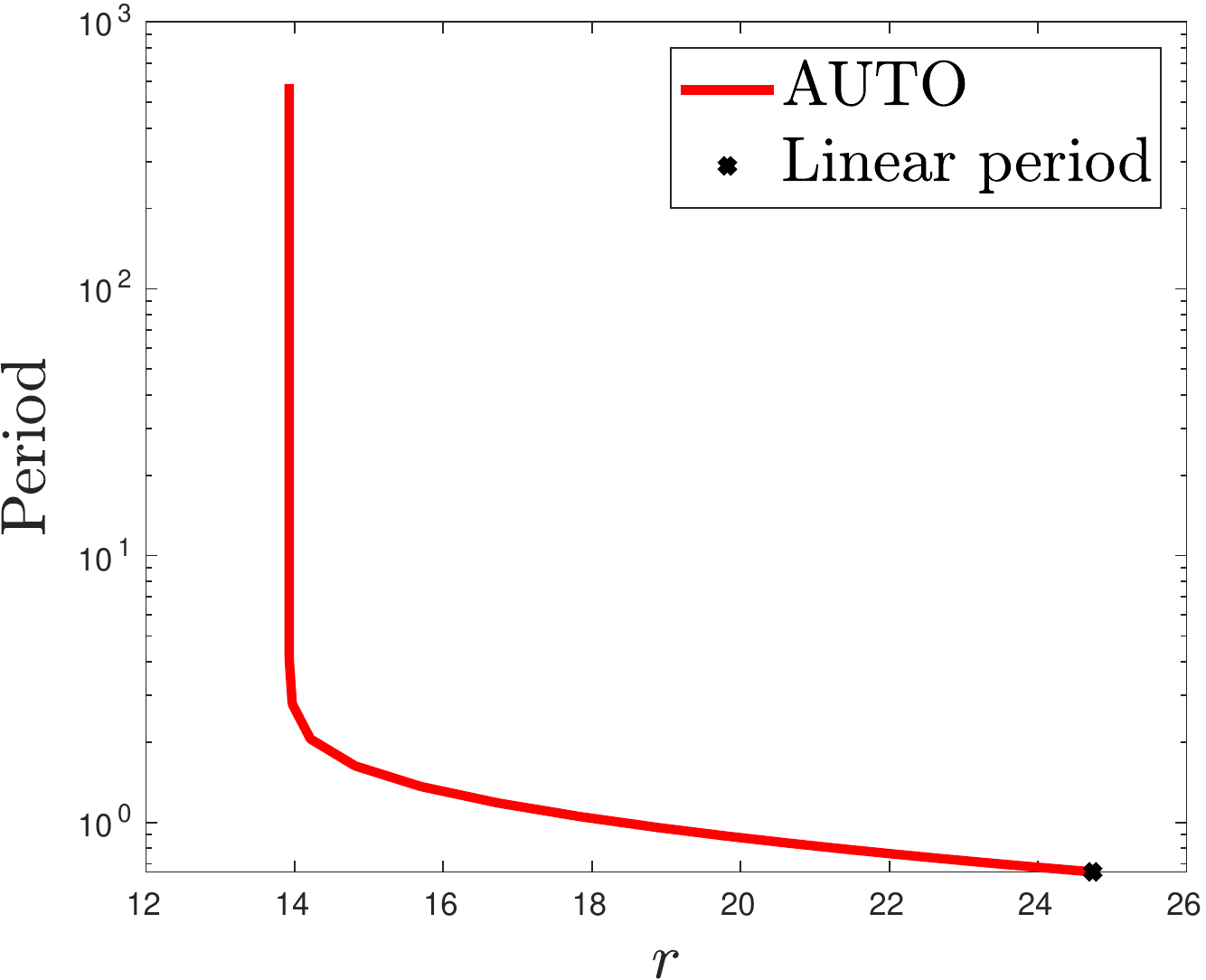}
\end{subfigure}
\caption{\label{fig:lorenz} {For a single Lorenz system with $\sigma = 10$ and $b=8/3$, the non-trivial steady states undergo a subcritical Hopf bifurcation at $r \approx 24.74$. Unstable branches of periodic solutions collide with the origin at $r \approx 13.926$ in a homoclinic bifurcation, with the period approaching infinity as shown in the right panel.}}
\end{figure}

{The succession of bifurcations as the parameter $r$ increases for a single Lorenz ODE is graphically summarized in Fig.~\ref{fig:lorenz}. We recall from \cite{sparrow1982} that the appearance of transient chaos coincides with the homoclinic bifurcation in $r \approx 13.926$, while the onset of attracting chaos is near $r \approx 24.06$, which is slightly before the subcritical Hopf bifurcation point. Hence, there is a small window where chaotic oscillations coexist with the stable non-trivial steady states.}

\subsection{{Weakly nonlinear theory near Hopf stability boundaries}}

In contrast to section \ref{sec:weakly_nonlinear_theory}, where the coupling strength and the bulk diffusivity were employed within the multiple time-scale expansion, here we choose the Rayleigh number {as the bifurcation parameter and}, so we set $\mu \equiv r$ in \eqref{eq:detuning}. Because this parameter does not arise in the boundary conditions, this particular choice simplifies the computation of the linear terms within the amplitude equations \eqref{eq:amplitude_even} and \eqref{eq:amplitude_odd}, which are now defined as
\begin{subequations}
\begin{align}
g_{1000} &= 4 \overline{\bphi_+^\star}^T B\left(\bphi_+,[\bPhi_+(0)]^{-1}\left.\frac{\partial\bFF}{\partial r}\right|_{(r_0;\bu_e)}\right) + \left\langle \WW_+^\star,\left.\frac{\partial\LL}{\partial r}\right|_{(r_0;\WW_+)} \right\rangle, \\
g_{0010} &= 4 \overline{\bphi_-^\star}^T B\left(\bphi_-,[\bPhi_+(0)]^{-1}\left.\frac{\partial\bFF}{\partial r}\right|_{(r_0;\bu_e)}\right) + \left\langle \WW_-^\star,\left.\frac{\partial\LL}{\partial r}\right|_{(r_0;\WW_-)} \right\rangle,
\end{align}
\end{subequations}
We do not perform a separate detailed weakly nonlinear analysis of the PDE-ODE system in the well-mixed regime \eqref{eq:well_mixed}. In fact, the formulae derived in Section \ref{sec:weakly_nonlinear_theory} still apply, provided that we take the appropriate limiting expressions of $p_\pm(\lambda)$ for $D=\infty$.

\begin{figure}[htbp]
\centering
\begin{subfigure}{0.32\linewidth}
\includegraphics[width=\linewidth]{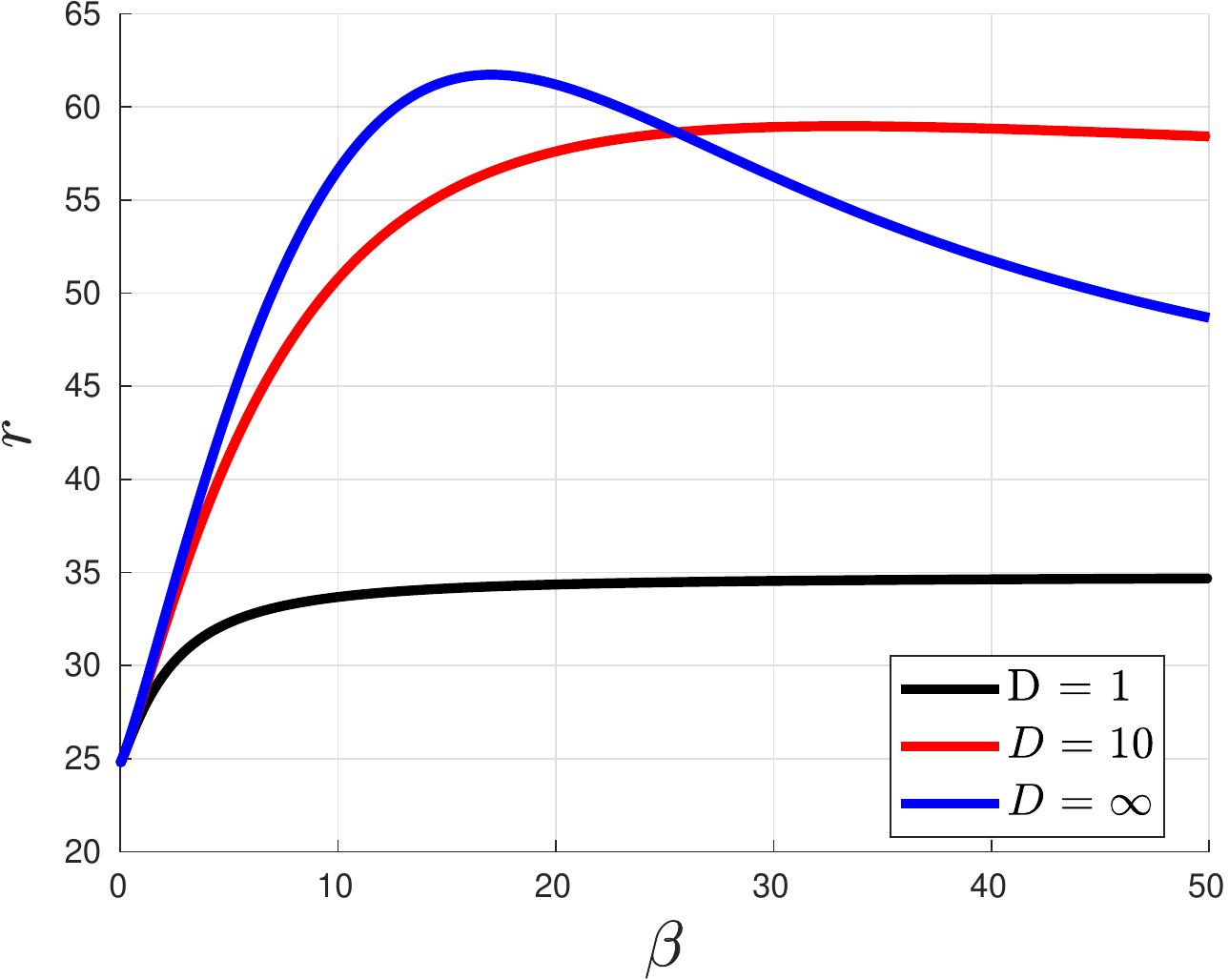}
\caption{In-phase mode.}
\end{subfigure}
\begin{subfigure}{0.32\linewidth}
\includegraphics[width=\linewidth]{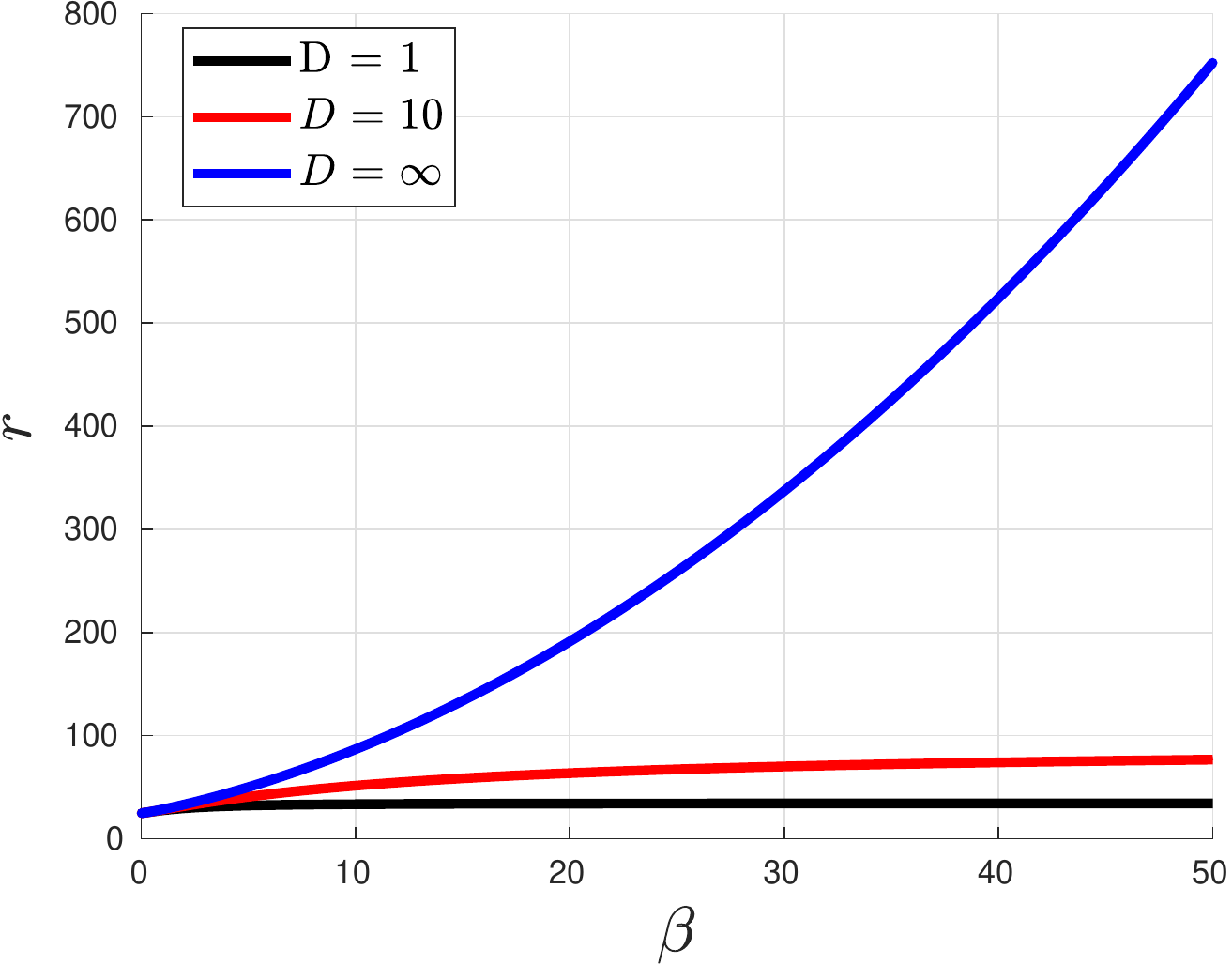}
\caption{Anti-phase mode.}
\end{subfigure}

\begin{subfigure}{0.32\linewidth}
\includegraphics[width=\linewidth]{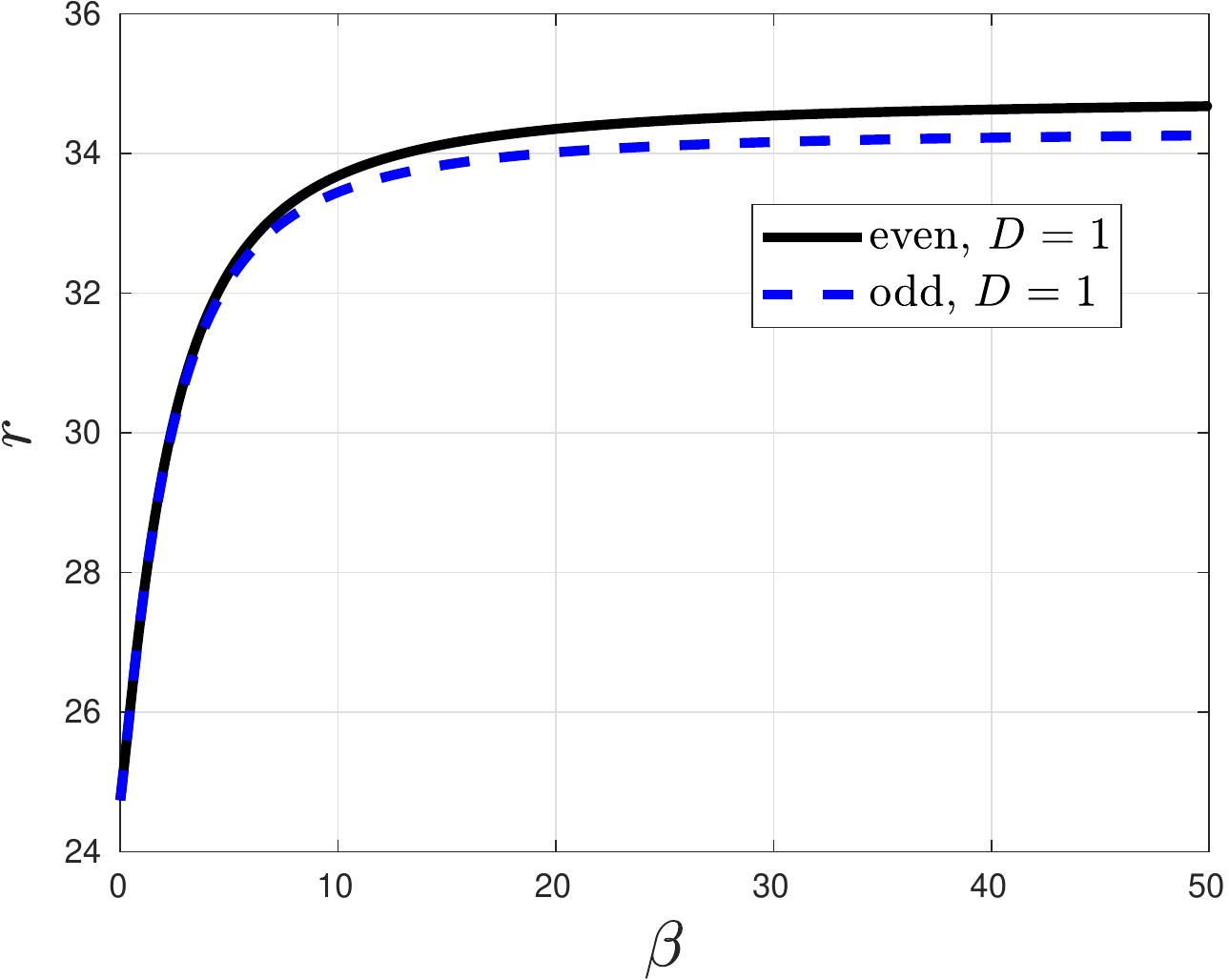}
\caption{$D=1$.}
\end{subfigure}
\begin{subfigure}{0.32\linewidth}
\includegraphics[width=\linewidth]{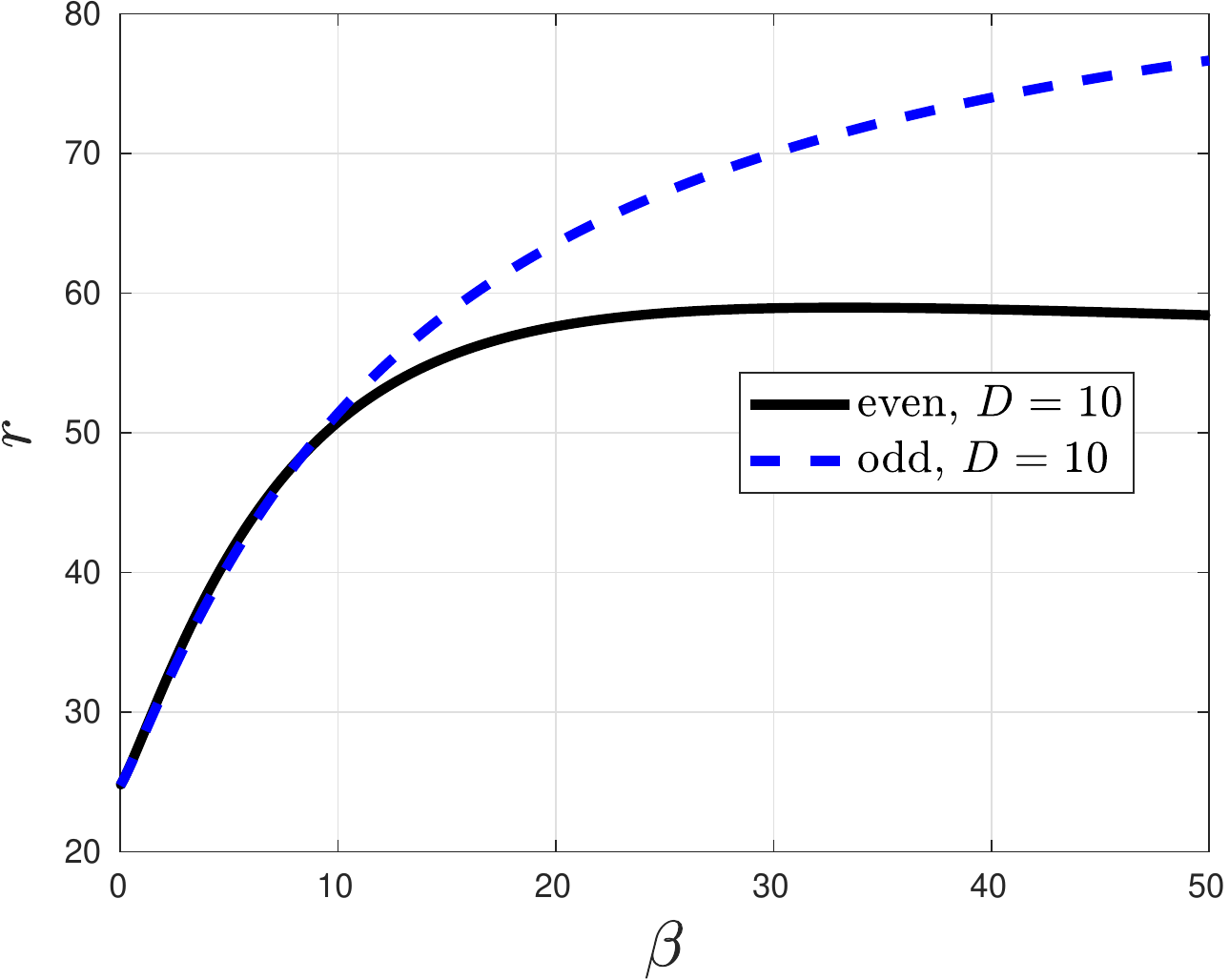}
\caption{$D=10$.}
\end{subfigure}
\begin{subfigure}{0.32\linewidth}
\includegraphics[width=\linewidth]{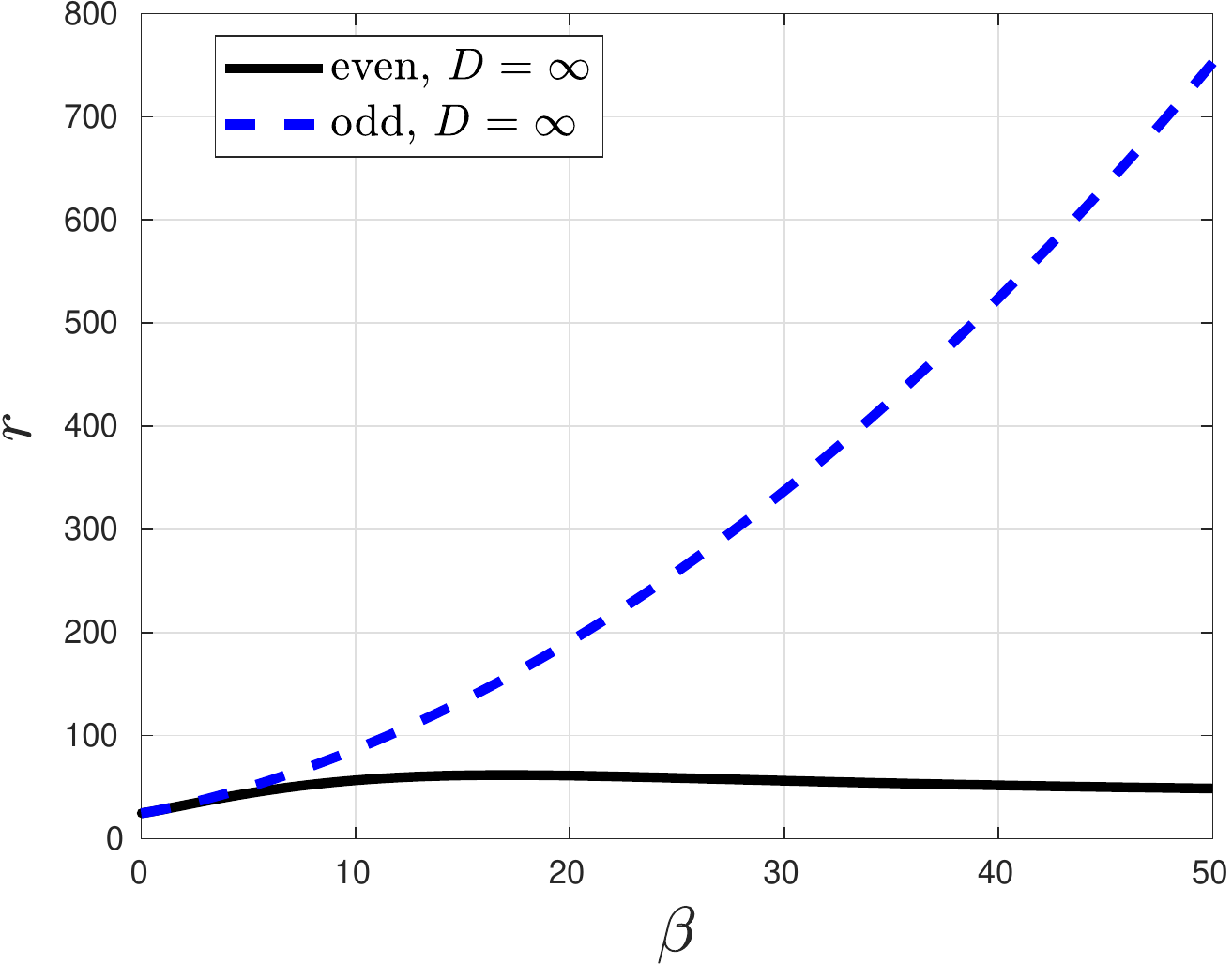}
\caption{$D = \infty$.}
\end{subfigure}
\caption{\label{fig:stabDiagLorenz_1} Numerically computed Hopf stability boundaries in the $r$ versus $\beta$ parameter plane for $D = 1$, $D=10$ and $D = \infty$. In panels (c), (d) and (e), the symmetric steady states are linearly stable below the lowermost stability boundary. The computation was performed with the software package \textsc{COCO} \cite{danko2013}. Other parameters are given by $L = 1,\, k = 1,\, \sigma = 10,\, b = 8/3\,.$}
\end{figure}

In Fig.~\ref{fig:stabDiagLorenz_1}, we investigate the effects of increasing the bulk diffusion level on stability boundaries in the parameter plane defined by the coupling strength $\beta$ and the Rayleigh number $r$. We distinguish between the even (panel (a)) and the odd (panel (b)) modes, with the two diagrams showing a significant increase in the critical Rayleigh number for Hopf bifurcations. Our linear stability results {therefore} suggest that a much higher $r$ value would be necessary for the emergence of chaotic dynamics when two identical Lorenz oscillators are coupled via a 1-D bulk diffusion field. This has been confirmed numerically, with simulations showing the stability of the symmetric steady states and giving no evidences of attracting chaos, when $r=28$ for a sufficiently large coupling strength (details not shown). Hence, in contrast to the preceding section, this special type of PDE-ODE coupling can also provide a stabilizing mechanism. When $D=1$, we remark in panel (c) of Fig.~\ref{fig:stabDiagLorenz_1} that the anti-phase mode becomes unstable first. {One possible interpretation for this observation is that synchronization with a common phase is harder when $D$ is small, since upon rescaling the spatial variable a small bulk diffusivity is equivalent to having the two oscillators located far from each other.}

\begin{figure}[htbp]
\centering
\begin{subfigure}{0.32\linewidth}
\includegraphics[width=\linewidth]{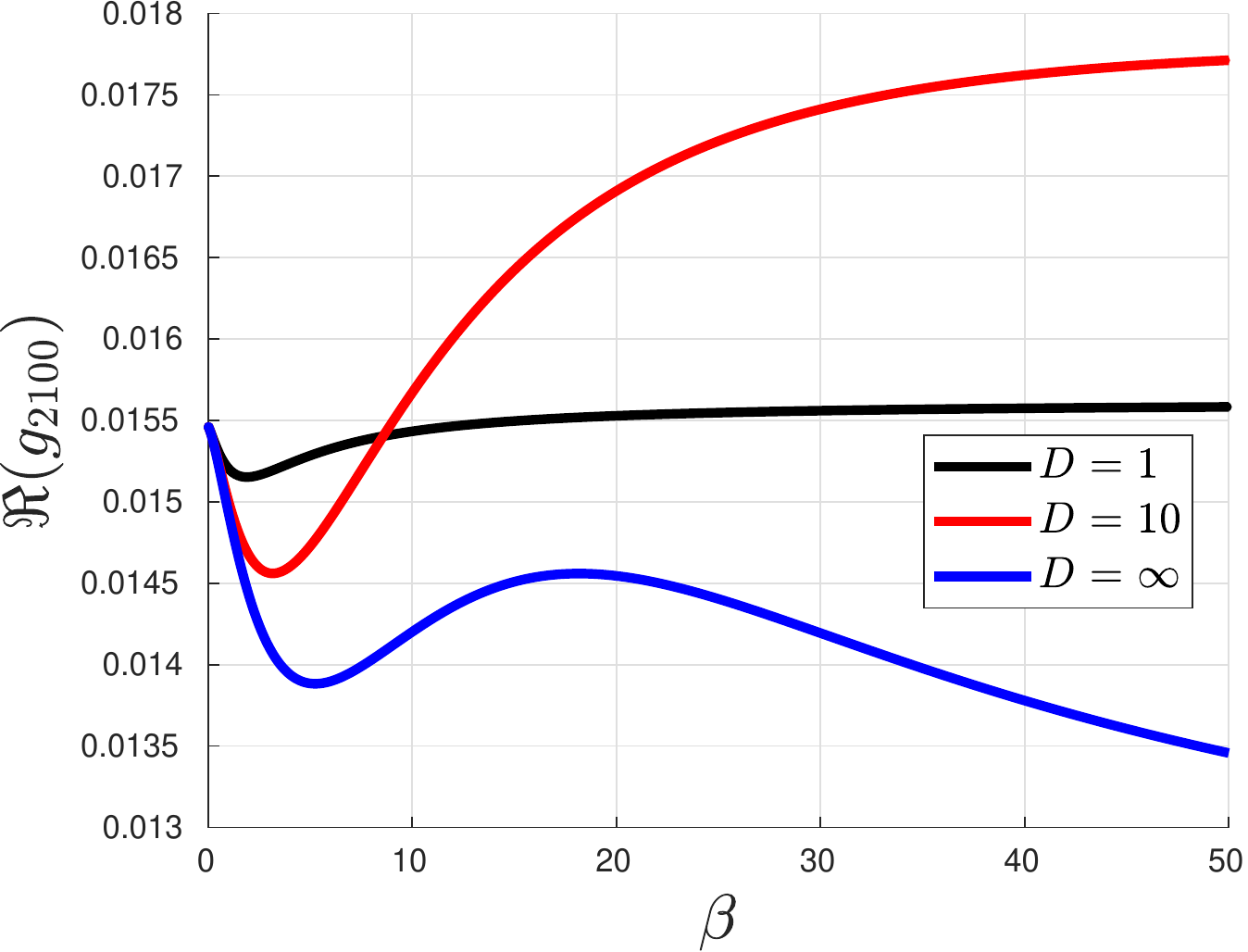}
\caption{In-phase mode}
\end{subfigure}
\begin{subfigure}{0.32\linewidth}
\includegraphics[width=\linewidth]{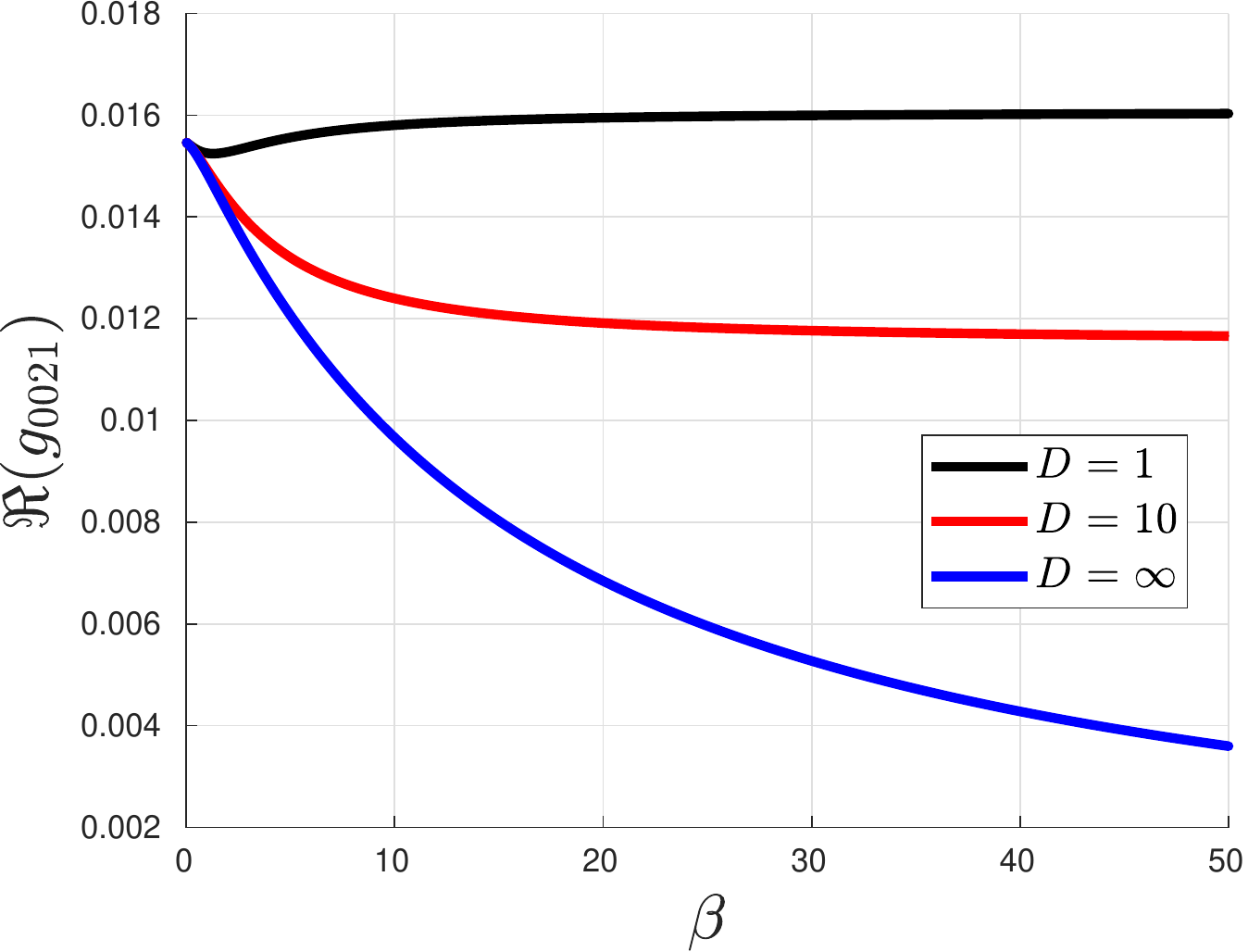}
\caption{Anti-phase mode}
\end{subfigure}
\caption{\label{fig:branching_behavior} Real parts of cubic normal form coefficients in \eqref{eq:amplitude_even} (panel (a)) and \eqref{eq:amplitude_odd} (panel (b)) along each Hopf stability boundary shown in Fig.~\ref{fig:stabDiagLorenz_1}. The Hopf bifurcations remain subcritical.}
\end{figure}

We then investigate the possible switch from a subcritical to a supercritical Hopf bifurcation as the strength of the coupling increases. This is shown in Fig.~\ref{fig:branching_behavior}, where the branching behavior near each Hopf stability boundary is deduced from a numerical evaluation of the cubic normal form coefficients in \eqref{eq:amplitude_even} and \eqref{eq:amplitude_odd}. As seen in this figure, our computations show that $g_{2100}$ and $g_{0021}$ each have positive real parts, an indication that the bifurcation remains subcritical over the range of $\beta$ and the values of $D$ considered for both even and odd modes.

\begin{figure}[htbp]
\centering
\begin{subfigure}{0.32\linewidth}
\includegraphics[width=\linewidth]{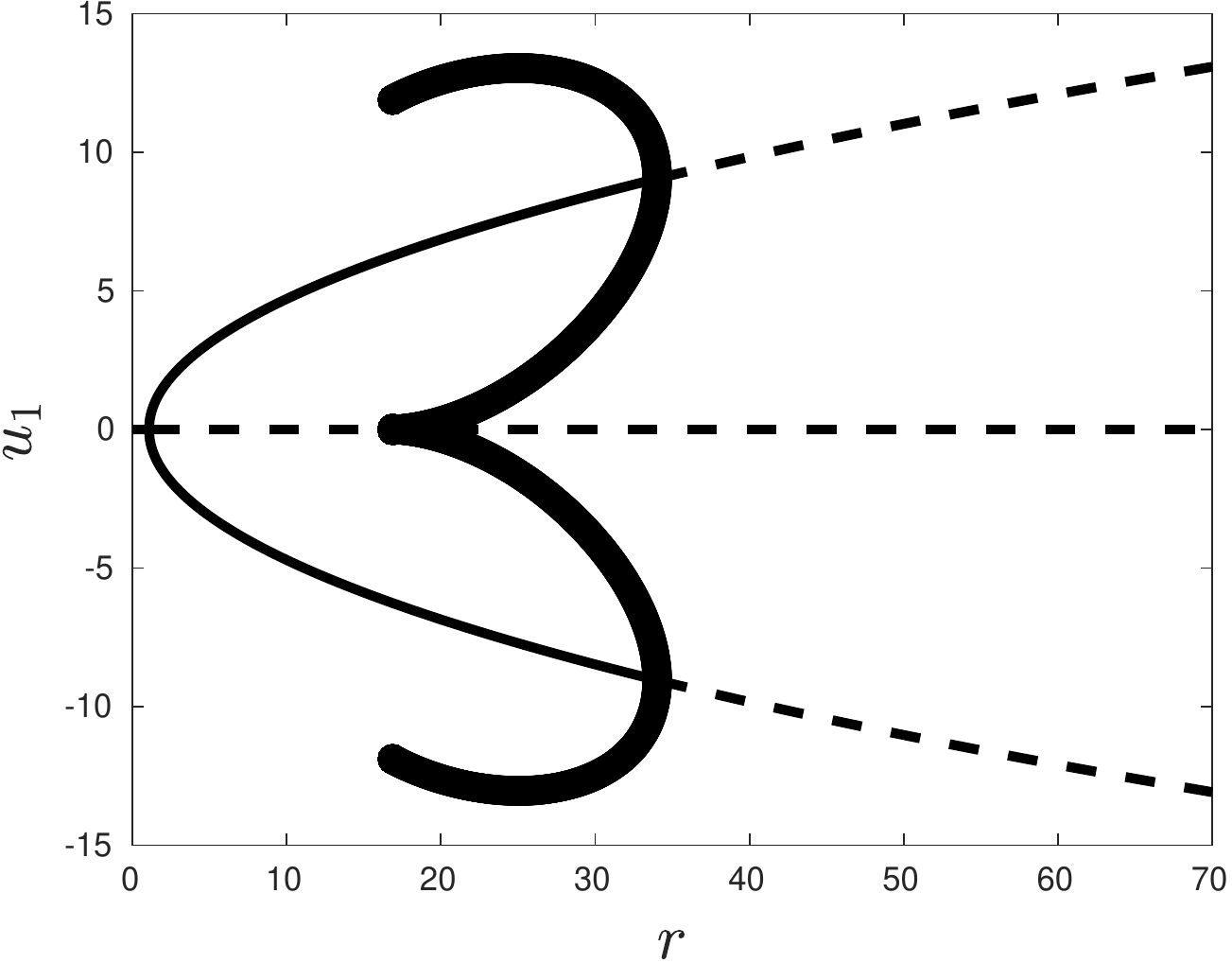}
\caption{$\beta=20,\, D=1$.}
\end{subfigure}
\begin{subfigure}{0.32\linewidth}
\includegraphics[width=\linewidth]{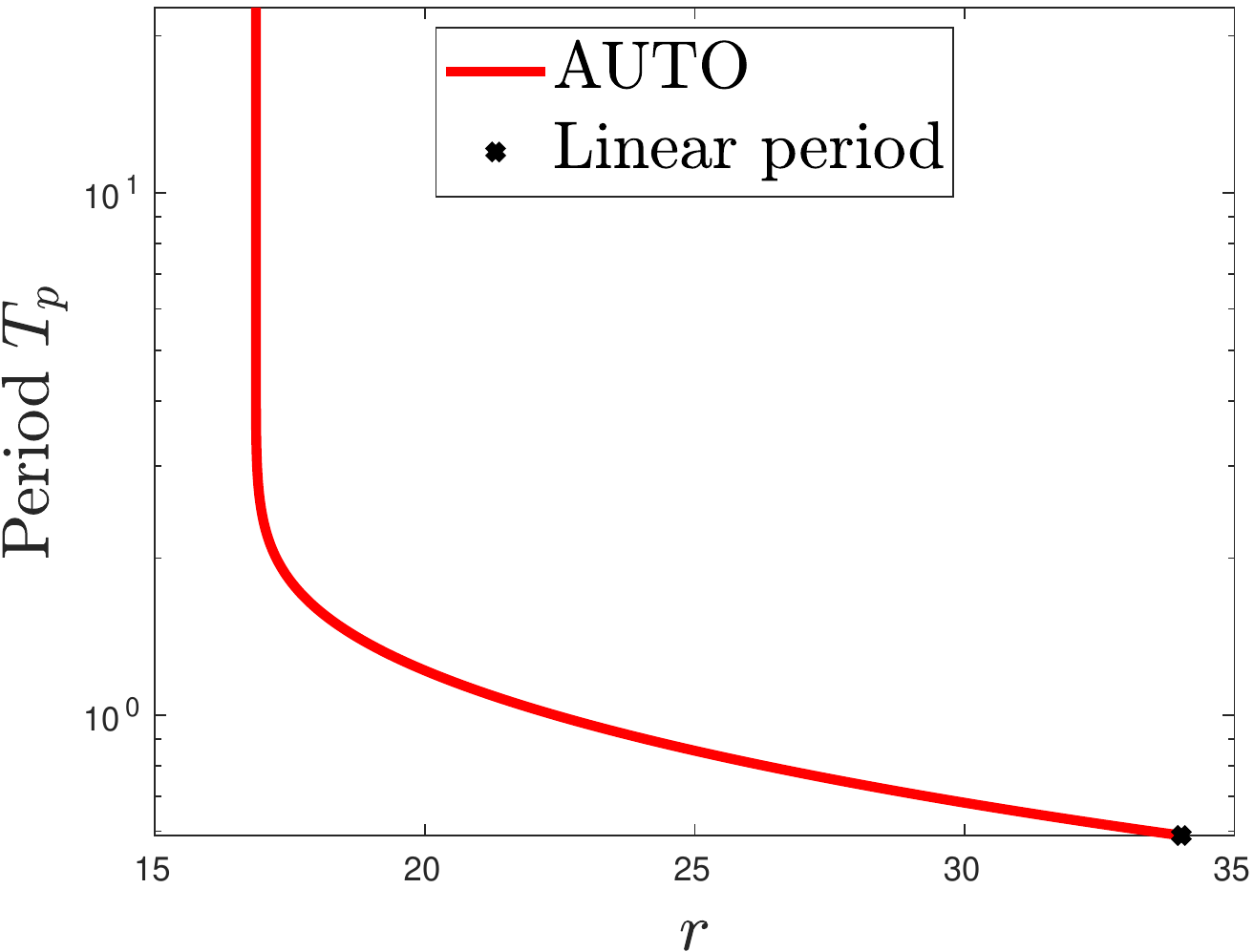}
\caption{$\beta=20,\, D=1$.}
\end{subfigure}
\begin{subfigure}{0.32\linewidth}
\includegraphics[width=\linewidth]{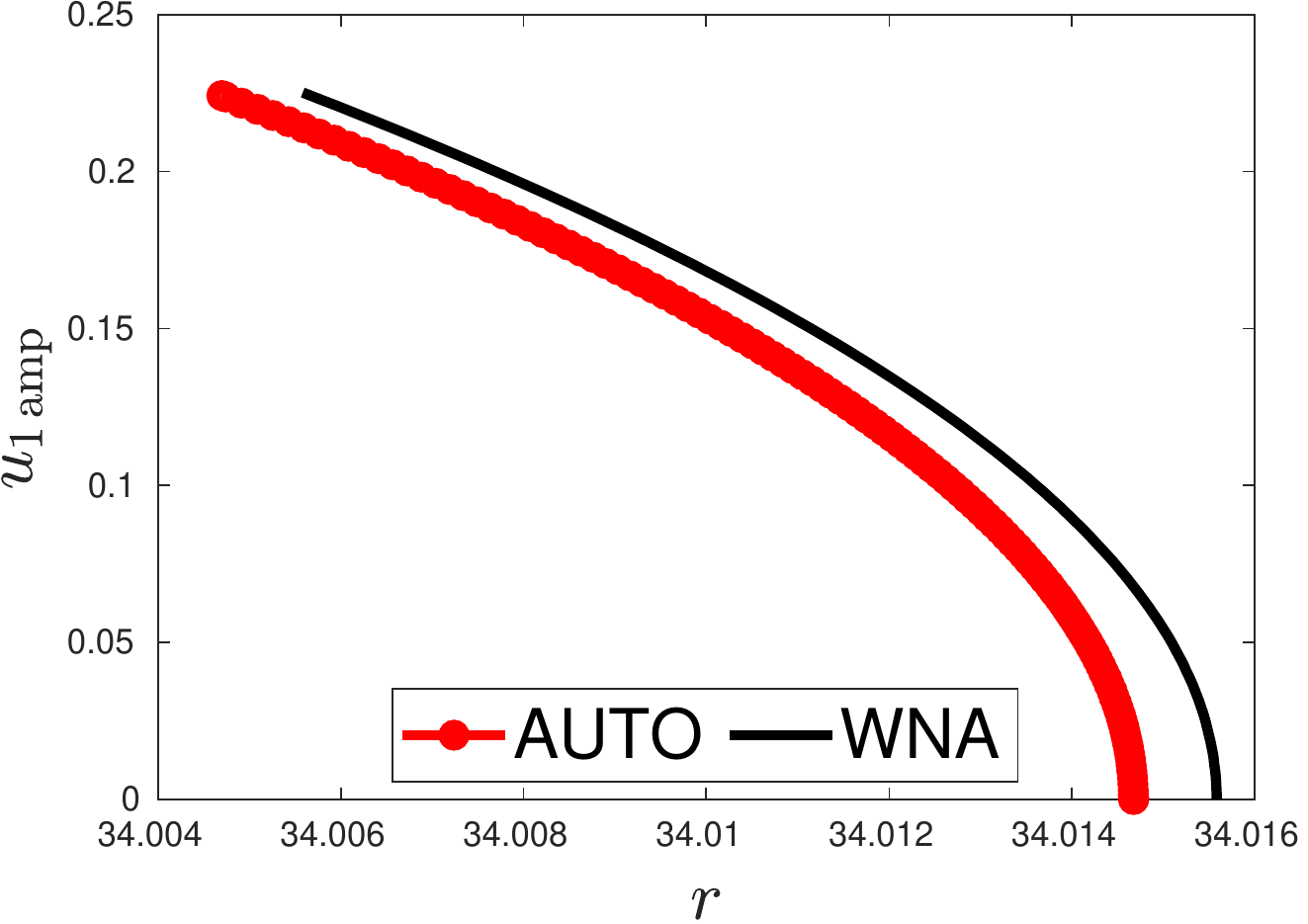}
\caption{$\beta=20,\, D=1$}
\end{subfigure}

\begin{subfigure}{0.32\linewidth}
\includegraphics[width=\linewidth]{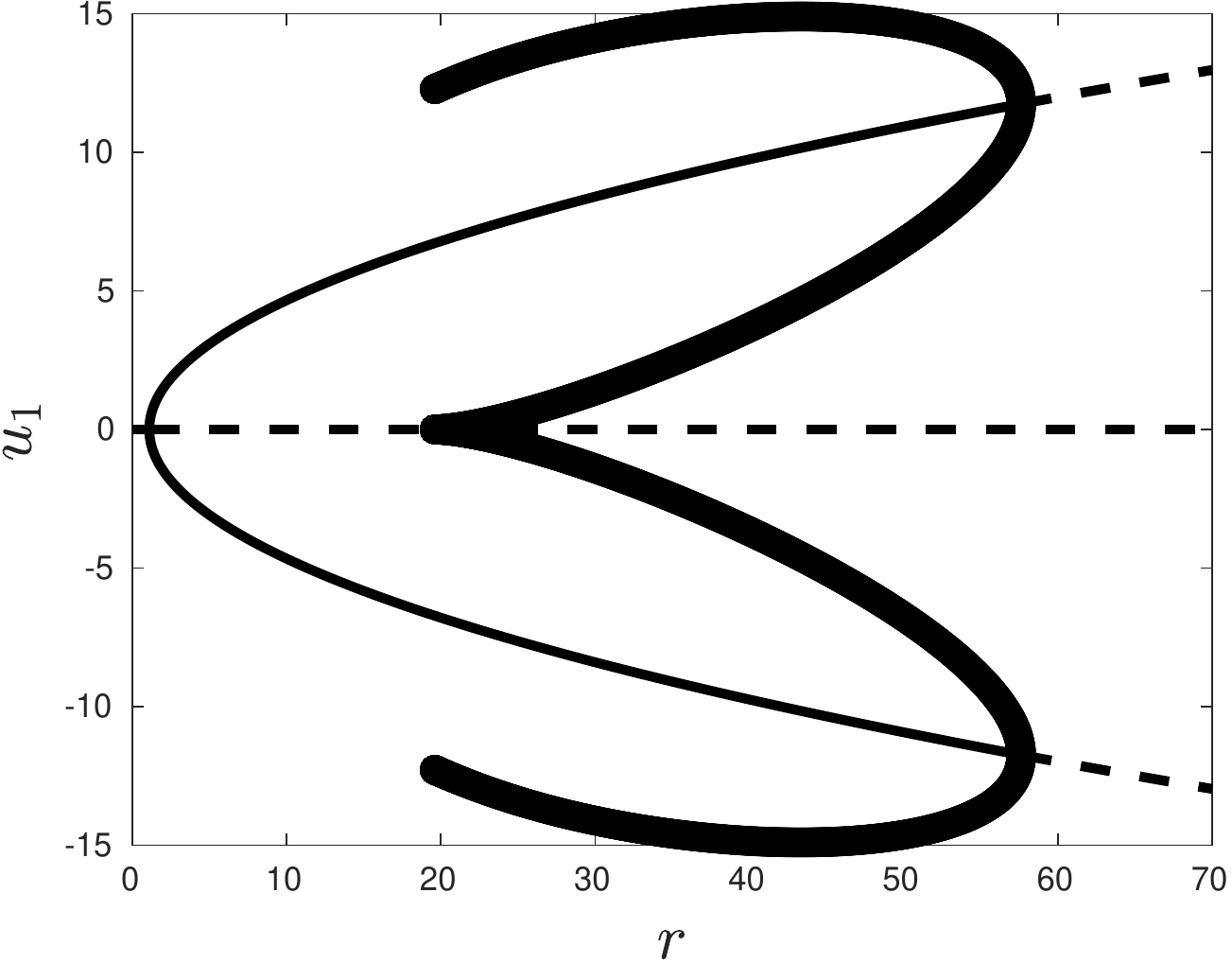}
\caption{$\beta=20,\, D=10$.}
\end{subfigure}
\begin{subfigure}{0.32\linewidth}
\includegraphics[width=\linewidth]{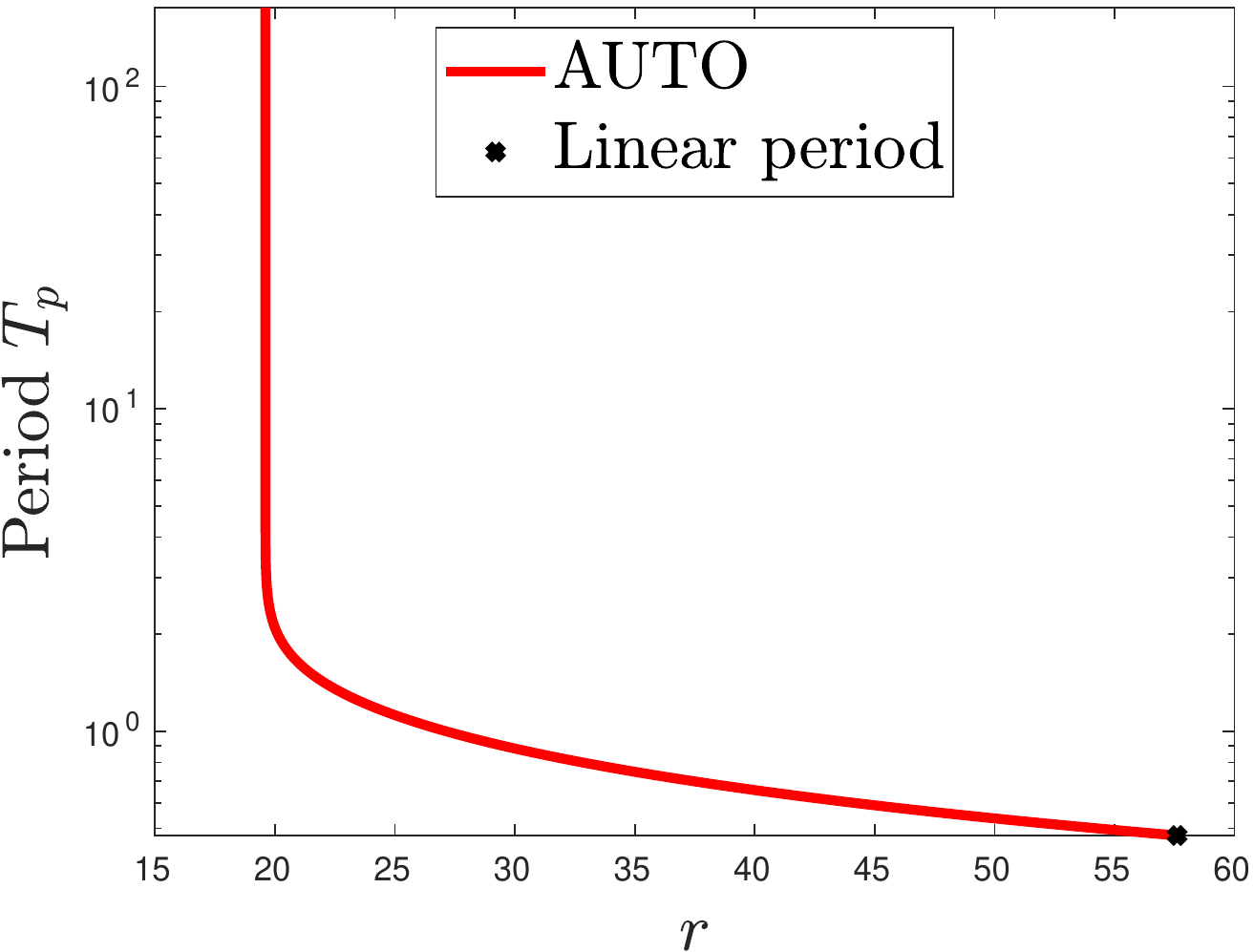}
\caption{$\beta=20,\, D=10$.}
\end{subfigure}
\begin{subfigure}{0.32\linewidth}
\includegraphics[width=\linewidth]{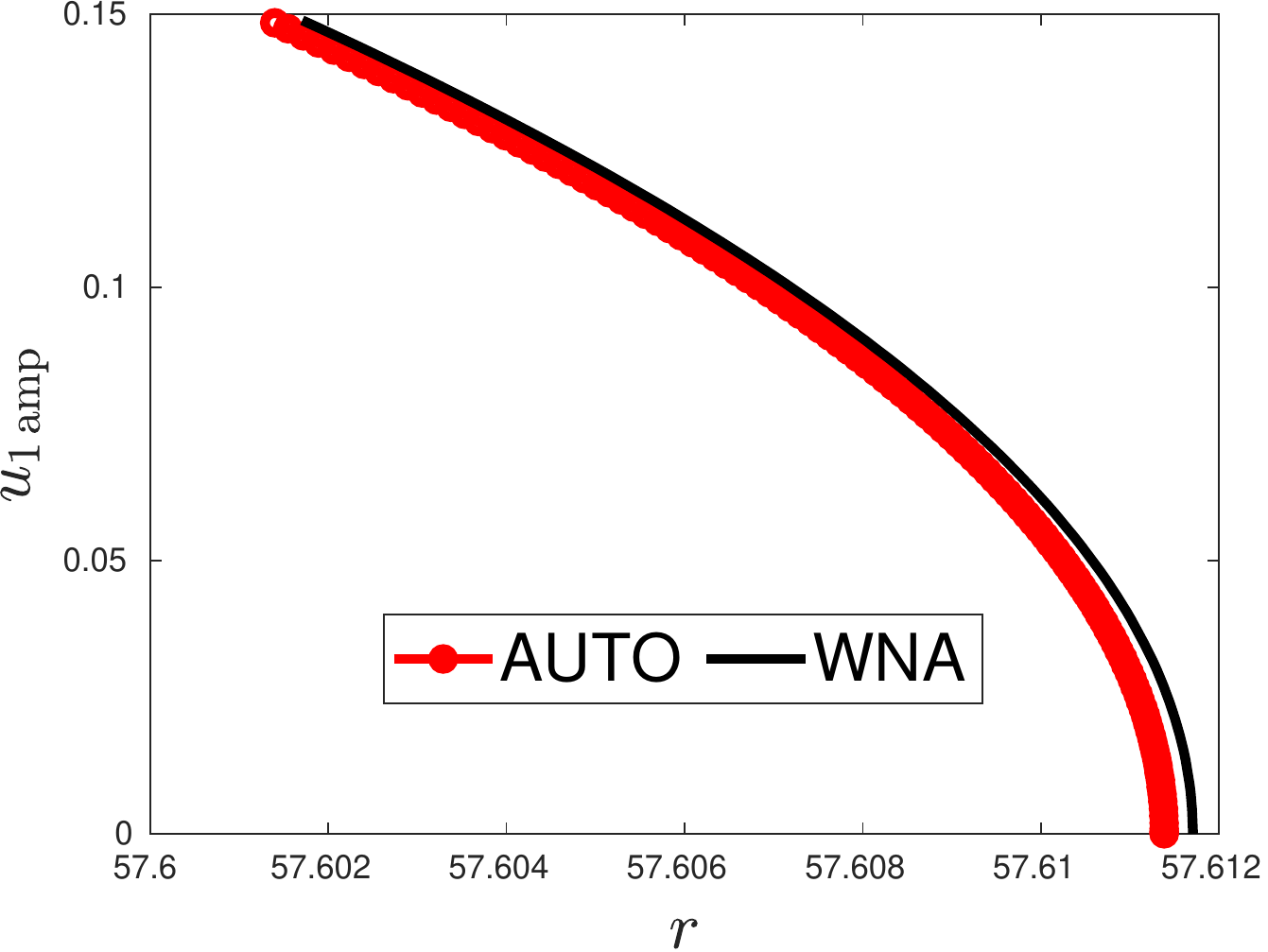}
\caption{$\beta=20,\, D=10$.}
\end{subfigure}
\caption{\label{fig:slice_r_finite_D} \textbf{Finite bulk diffusion.} Global and local bifurcation diagrams as a function of the Rayleigh number $r$, corresponding to {the $\beta=20$ vertical slice through the linear stability} diagrams shown in Fig.~\ref{fig:stabDiagLorenz_1} for $D=1$ (panels (a)-(c)) and $D=10$ (panels (d)-(f)). The sudden increase of the period seen in panels (b) and (e) suggests the presence of homoclinic orbits as the unstable branch collides with the origin. In panels (c) and (f), we observe a very small discrepancy between the bifurcation points as predicted by AUTO and as directly computed using the transcendental equation \eqref{eq:transcendental_beta}. This results from discretization errors. Here, $N=200$ grid points were employed to spatially discretize the coupled PDE-ODE system.}
\end{figure}

For the finite bulk diffusion regime, in Fig.~\ref{fig:slice_r_finite_D} we show global and local bifurcation diagrams as a function of the Rayleigh number on the vertical slice $\beta=20$. Both cases $D=1$ (panels (a)-(c)) and $D=10$ (panels (d)-(f)) are qualitatively similar, but most importantly they preserve the key features of the Lorenz ODE system, such as the symmetry of solutions and the destruction of the limit cycles via homoclinic bifurcations when the unstable periodic solution branches collide with the origin. However, we do remark a significant increase in the size of the bistability parameter regime, suggesting that the minimal Rayleigh number required for attracting chaos is much higher. In the weakly nonlinear regime (panels (c) and (f)), the amplitude of the unstable limit cycles as predicted by the weakly nonlinear theory is favorably compared with numerical bifurcation results. Note also that we only computed the branch of periodic solutions emerging from the primary Hopf bifurcation, corresponding to the anti-phase mode when $D=1$ and to the in-phase mode when $D=10$.

\begin{figure}[htbp]
\centering
\begin{subfigure}{0.32\linewidth}
\includegraphics[width=\linewidth]{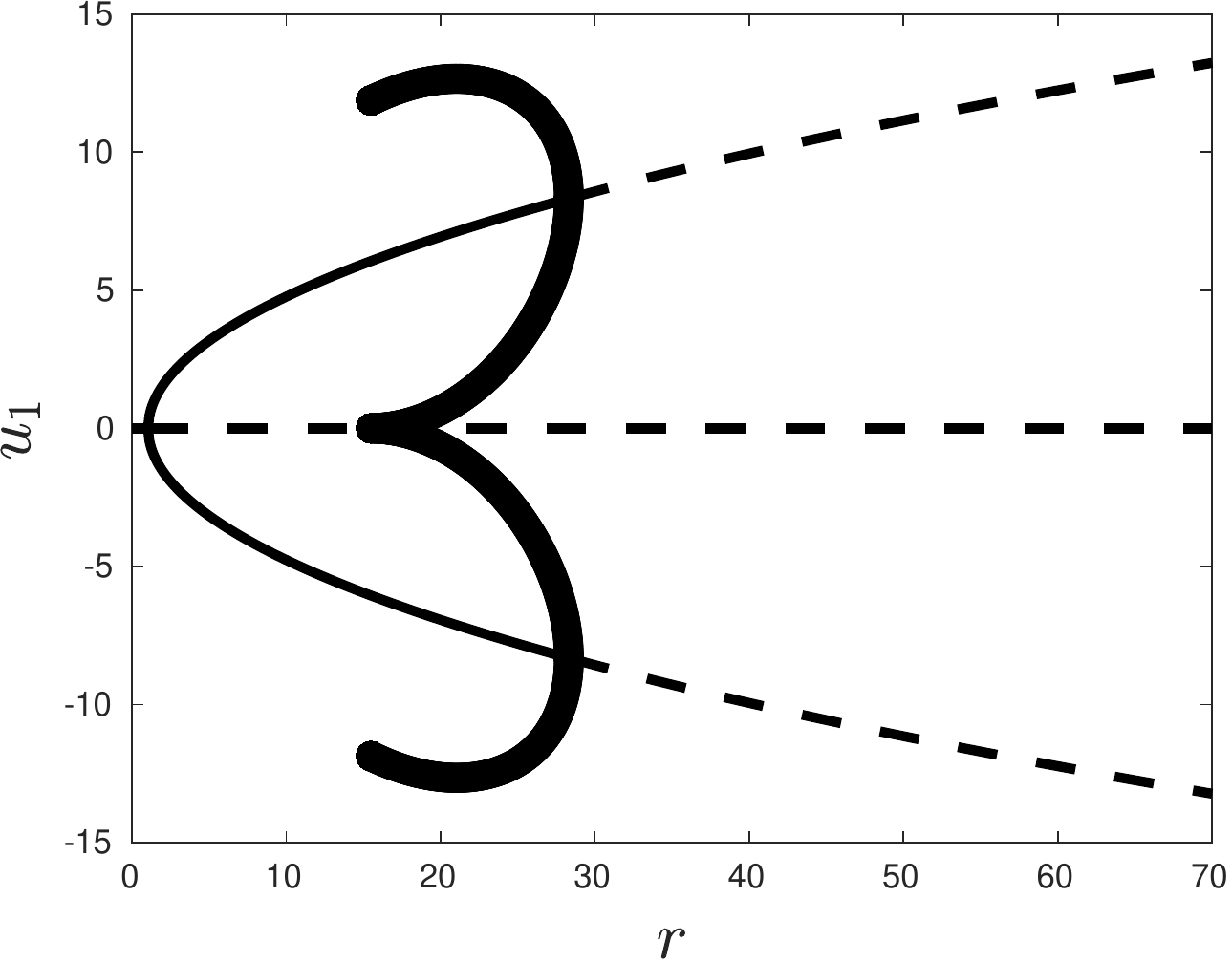}
\caption{$\beta=1,\, D = \infty$.}
\end{subfigure}
\begin{subfigure}{0.32\linewidth}
\includegraphics[width=\linewidth]{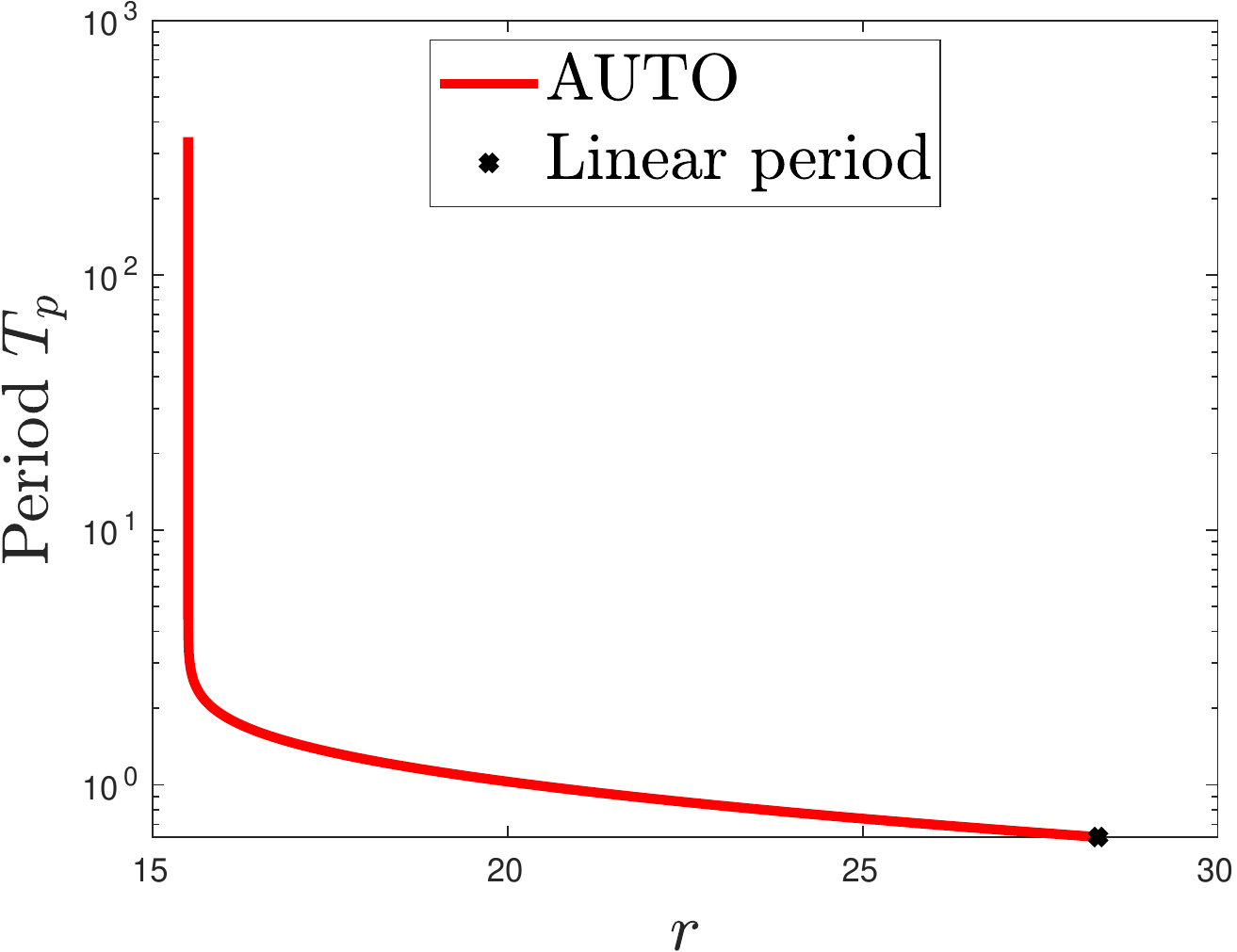}
\caption{$\beta=1,\, D = \infty$.}
\end{subfigure}
\begin{subfigure}{0.32\linewidth}
\includegraphics[width=\linewidth]{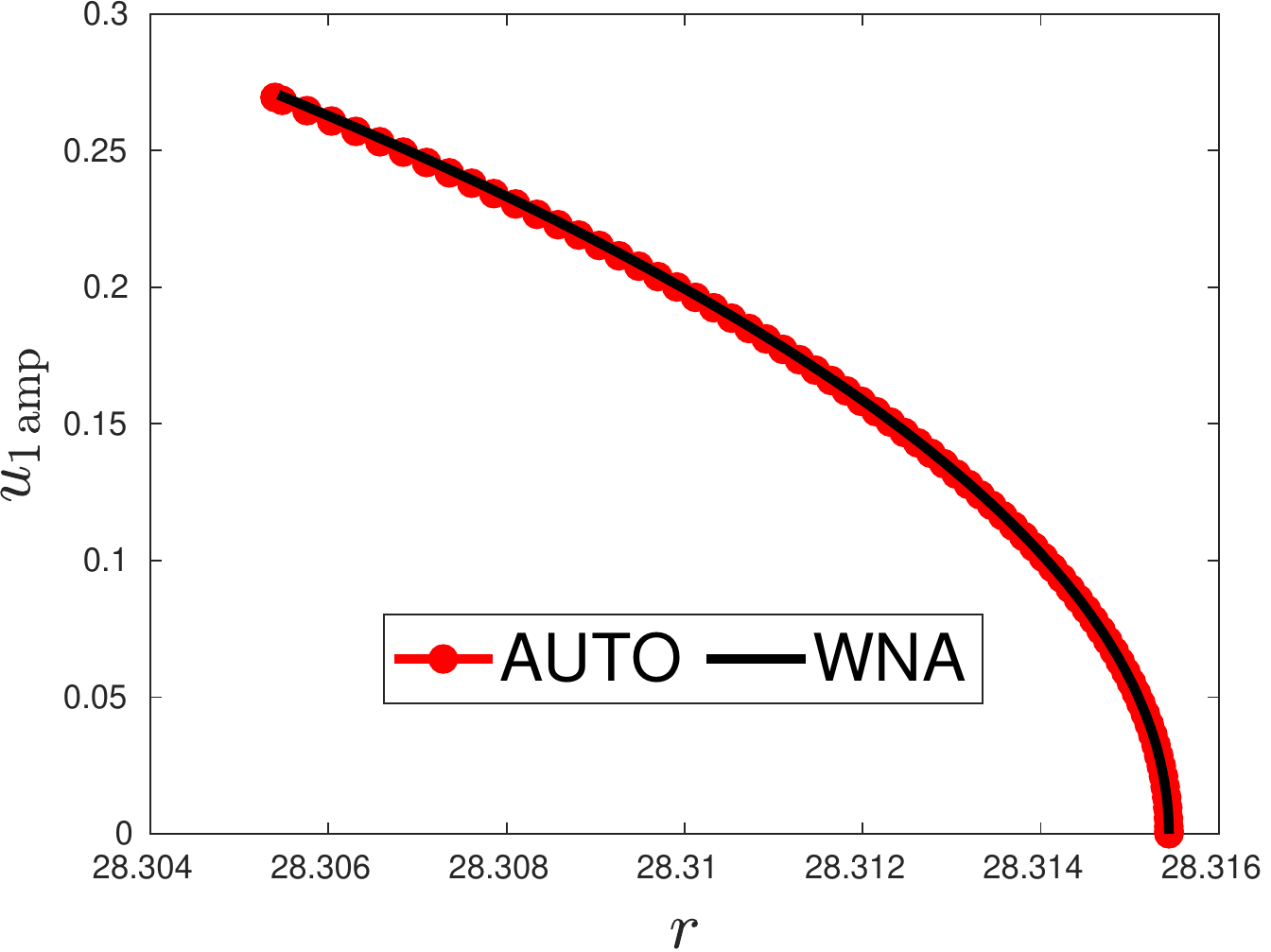}
\caption{$\beta=1,\, D = \infty$.}
\end{subfigure}

\begin{subfigure}{0.32\linewidth}
\includegraphics[width=\linewidth]{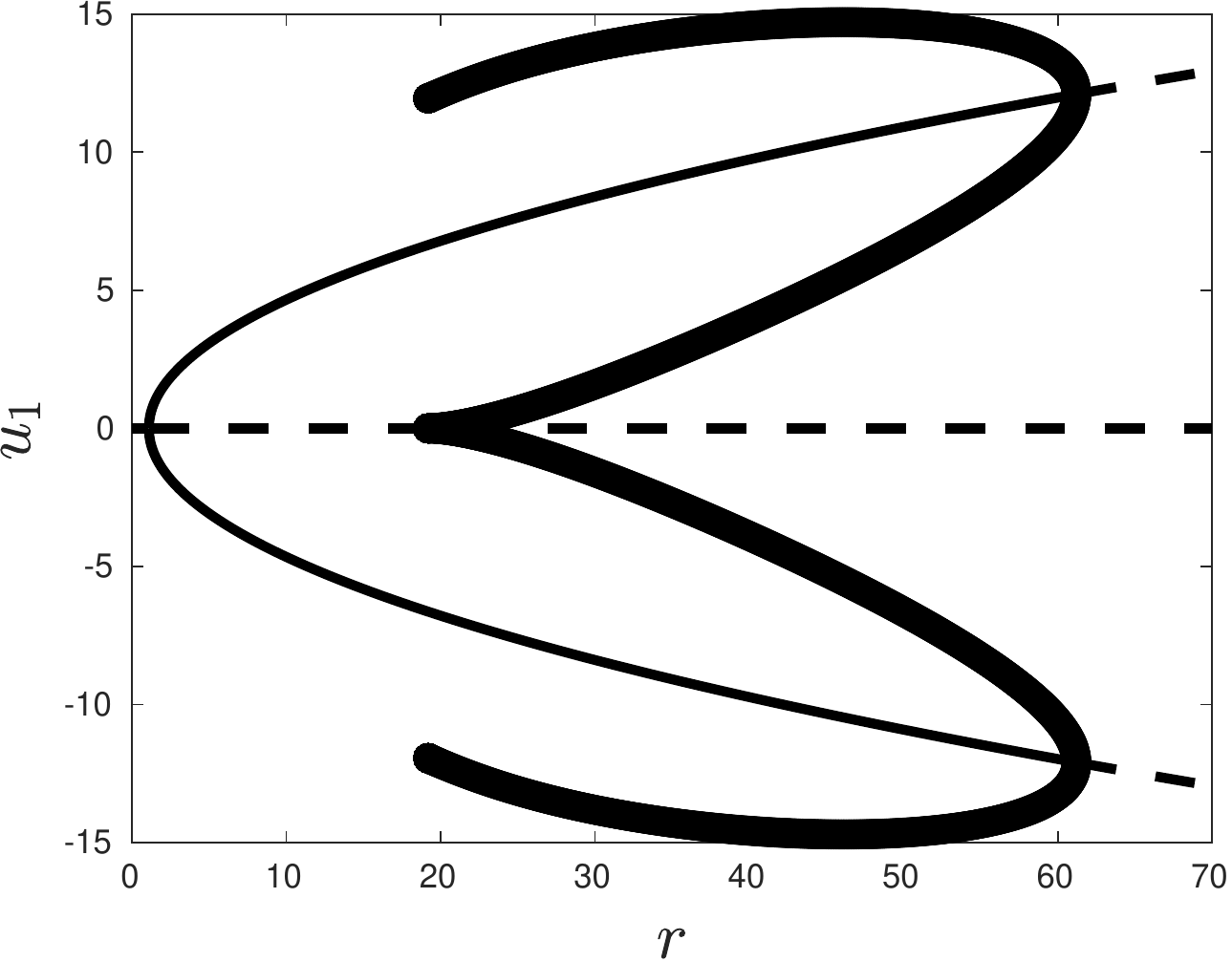}
\caption{$\beta=20,\, D = \infty$.}
\end{subfigure}
\begin{subfigure}{0.32\linewidth}
\includegraphics[width=\linewidth]{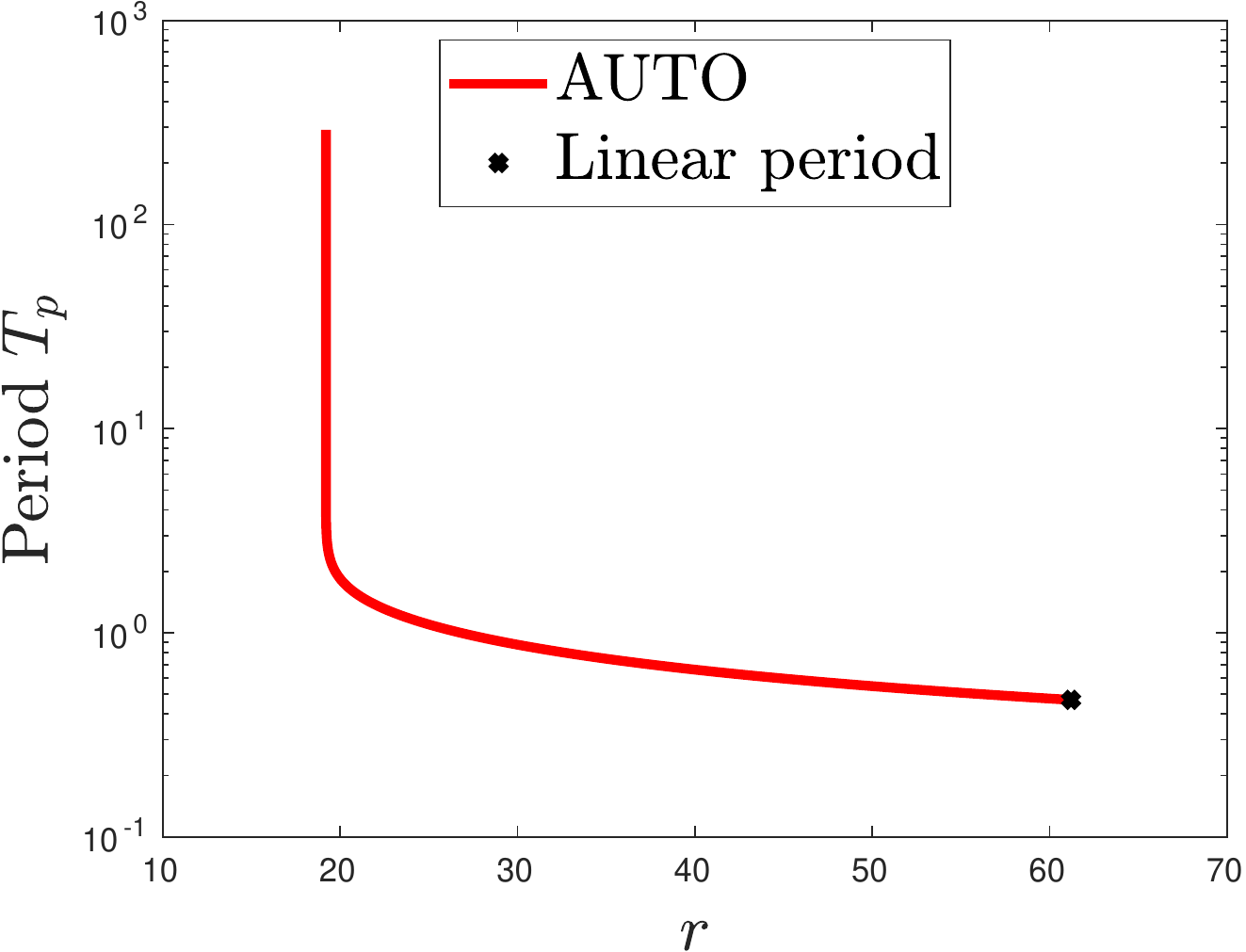}
\caption{$\beta=20,\, D = \infty$.}
\end{subfigure}
\begin{subfigure}{0.32\linewidth}
\includegraphics[width=\linewidth]{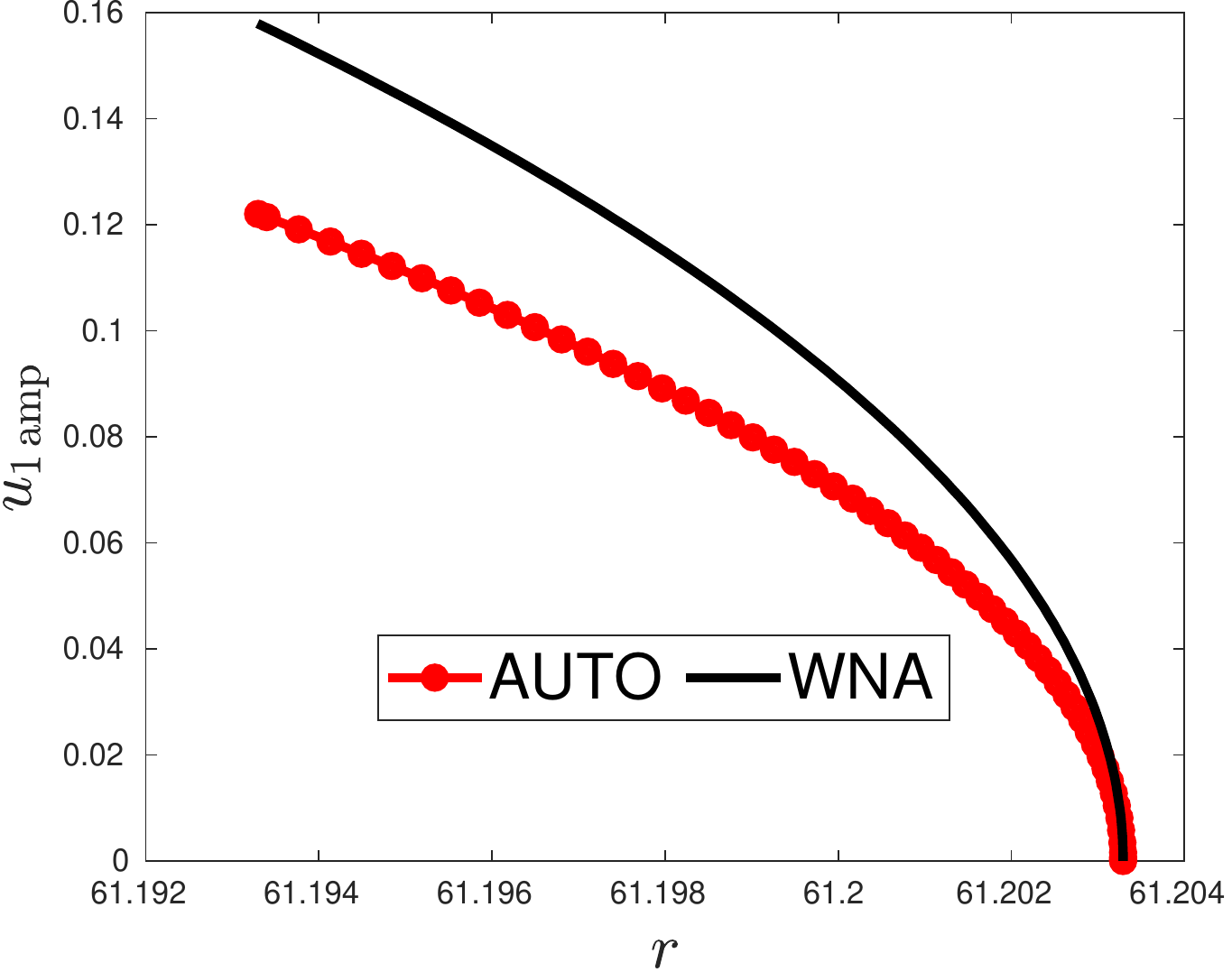}
\caption{$\beta=20,\, D = \infty$.}
\end{subfigure}
\caption{\label{fig:slice_r_infinite_D} \textbf{Well-mixed regime.} Global and local bifurcation diagrams for \eqref{eq:well_mixed} as a function of the Rayleigh number $r$, corresponding to the vertical slices $\beta=1$ (panels (a)-(c)) and $\beta=20$ (panels (d)-(f)).}
\end{figure}

Finally, {in Fig.~\ref{fig:slice_r_infinite_D} we show numerical results} of similar experiments performed in the infinite bulk diffusion case, as obtained with AUTO (cf.~\cite{doedel2007}) using the ODE system \eqref{eq:well_mixed}. They are consistent and qualitatively similar to their finite diffusion counterparts. Here also, attracting chaos likely occurs for significantly higher values of the Rayleigh number. In panel (f), the rather poor agreement between numerical and weakly nonlinear results at larger amplitudes is likely a result of the Hopf bifurcation being almost degenerate when $\beta$ becomes large.

\subsection{Synchronous chaos}\label{subsec:synchronous_chaos}

We now investigate the onset of synchronous chaos as the strength of the coupling $\beta$ and the bulk diffusion rate $D$ increase. For this purpose, we fix the Rayleigh number to be such that the symmetric steady states are linearly unstable for all values of $\beta$ and $D$. Hence, we choose $r=70$, which is above the linear stability boundary for the even mode in the well-mixed regime (see Fig.~\ref{fig:stabDiagLorenz_1}), where the dynamics is governed by \eqref{eq:well_mixed}. The stability of synchronous solutions is then determined from a computation of the largest Lyapunov exponent of a linearization of \eqref{eq:PDE_ODE_lorenz} around the synchronous manifold, where only transverse, or odd, perturbations are considered. The main result of this section is a phase diagram in the $D$ versus $\beta$ parameter plane that predicts the stability boundary for synchronous chaotic solutions.

{We recall from \S \ref{sec:intro} that synchronous chaos is the sensitivity to initial conditions on an invariant synchronous manifold $\bWW_s$, here defined as the subspace of solutions to \eqref{eq:PDE_ODE_lorenz} invariant under the action of reflection with respect to the midpoint $x=L$,}
\begin{equation}\label{eq:manifold}
\bWW_s = \left. \left\{ 
W_s = \begin{pmatrix}
C_s(x,t) \\
\bu_s(t) \\
\bu_s(t)
\end{pmatrix} 
\right|\, C_s(x,t) = C_s(2L-x,t)\, \right\}\,.
\end{equation}
{Reflection symmetry is readily obtained by imposing a
  no-flux boundary condition at the domain midpoint.} {In this
  way, $C_s(x,t)$ and $\bu_s(t)$ in \eqref{eq:manifold} satisfy the
  following reduced system:}
\begin{equation}\label{eq:synchronous}
\begin{split}
\frac{\partial C_s}{\partial t} &= D\frac{\partial^2 C_s}{\partial x^2} - k C_s\,, \quad 0 < x < L\,; \quad -D\partial_x C_s|_{x=0} = \beta (e_1^T\bu_s - C_s|_{x=0})\,, \quad  \partial_xC_s|_{x=L} = 0, \\
\frac{d\bu_s}{dt} &= \bFF(\bu_s) + \beta(C_s|_{x=0} - e_1^T\bu_s)e_1\,.
\end{split}
\end{equation}
Next, we introduce the following deviations from the synchronous manifold:
\begin{equation}
\eta(x,t) = C(x,t) - C_s(x,t),\, \quad \bphi(t) = \bu(t) - \bu_s(t)\,.
\end{equation}
Upon substituting this expression into the coupled PDE-ODE system and after linearizing, we obtain that $\eta(x,t)$ and $\bphi(t)$ satisfy the non-autonomous linear system
\begin{equation}\label{eq:non_autonomous}
\begin{split}
\frac{\partial \eta}{\partial t} &= D\frac{\partial^2\eta}{\partial x^2} - k \eta\,, \quad 0 < x < L\,; \quad -D\partial_x\eta|_{x=0} = \beta (e_1^T\bphi - \eta|_{x=0})\,, \quad  \eta(L,t) = 0\,, \\
\frac{d\bphi}{dt} &= J_s(t)\bphi + \beta(\eta|_{x=0} - e_1^T\bphi)e_1\,.
\end{split}
\end{equation}
Here, $J_s(t)$ is the Jacobian matrix of the nonlinear kinetics $\bFF(\bu)$ evaluated on the synchronous manifold. The central feature here is to impose an absorbing boundary condition at the domain midpoint in order to only select odd perturbations.

For the case of infinite bulk diffusion (system \eqref{eq:well_mixed}), the solutions on the synchronous manifold are spatially homogeneous. Therefore, we have that $C_{0s} \equiv C_{0s}(t)$ and $\bu_s(t)$ satisfy
\begin{equation}\label{eq:synchronous_WM}
\frac{dC_{0s}}{dt} = \frac{\beta}{L} e_1^T \bu_s - \left(k + \frac{\beta}{L}\right) C_{0s}\,, \quad \frac{d \bu_s}{dt} = \bFF(\bu_s) + \beta (C_{0s} - e_1^T\bu_s)e_1\,,
\end{equation}
and the corresponding non-autonomous linearization reduces to
\begin{equation}\label{eq:non_autonomous_WM}
\eta \equiv 0\,, \qquad \frac{d\bphi}{dt} = J_s(t)\bphi - \beta E \bphi\,.
\end{equation}

{We now provide some details on Lyapunov exponents and their
  computation (see \cite{meiss2007} for a more in-depth coverage). Let
  $\Lambda_{\max} \equiv \Lambda_{\max}(W_s; \beta, D)$ be the largest
  Lyapunov exponent of the non-autonomous linear system
  \eqref{eq:non_autonomous} (or \eqref{eq:non_autonomous_WM} if
  $D = \infty$). If $\Lambda_{\max} < 0$, then infinitesimal
  perturbations from the synchronous manifold decay exponentially and
  complete synchronization of both oscillators is
  expected. Conversely, when $\Lambda_{\max} > 0$ solutions on the
  synchronous manifold are unstable to any transverse perturbations.}
{ In order to obtain a} { numerical approximation to
  $\Lambda_{\max}$, we must solve simultaneously the coupled PDE-ODE
  system \eqref{eq:synchronous} and the odd linearization
  \eqref{eq:non_autonomous}, and then compute the following quantity:
\begin{equation}
\Lambda_{\max}(T) \approx \frac{1}{T} \log \frac{\|\WW(T)\|}{\|\WW(0)\|}\,, \quad \WW(T) = \begin{pmatrix} \eta(x,T) \\ \bphi(T) \end{pmatrix}\,,
\end{equation}
where $T$ is a sufficiently long integration time, chosen here to be
$10^4$.}  {Before implementing the time integration scheme a
spatial discretization of \eqref{eq:synchronous} and
\eqref{eq:non_autonomous} must be performed}, {and for this we use a
method of lines approach with $N=100$ equidistant grid
points. Finally, our algorithm to compute $\Lambda_{\max}$ follows
Appendix A.3 of \cite{meiss2007}, where the essential role of regular
renormalization of tangent vectors, in order to preserve accuracy, is
emphasized. Hence, we select the renormalization step to be
$\Delta t=1$, often used to compute Lyapunov exponents for a single
Lorenz system \cite{meiss2007}.}

\begin{figure}[htbp]
\centering
\begin{subfigure}{0.24\linewidth}
\includegraphics[width=\linewidth]{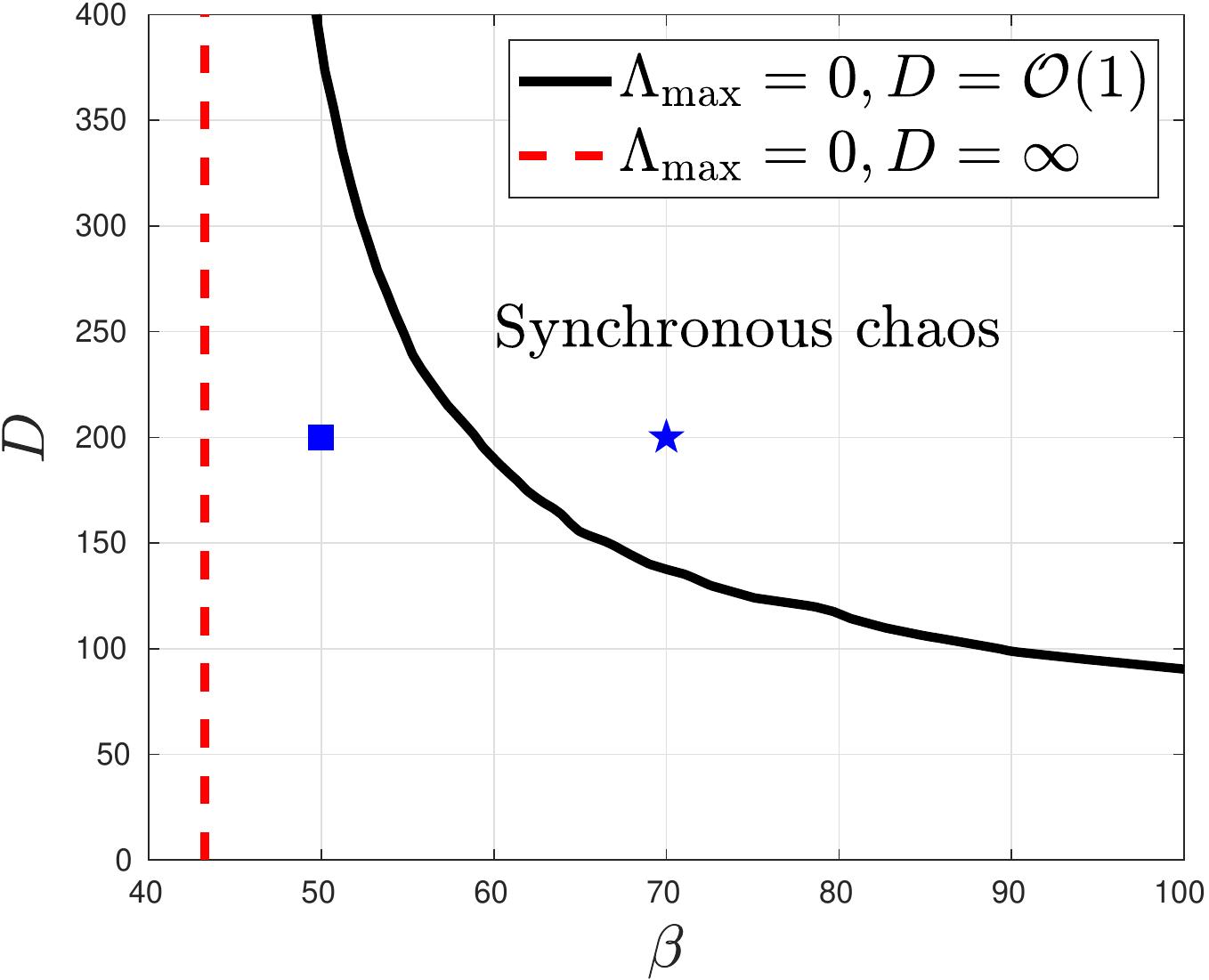}
\end{subfigure}
\begin{subfigure}{0.24\linewidth}
\includegraphics[width=\linewidth]{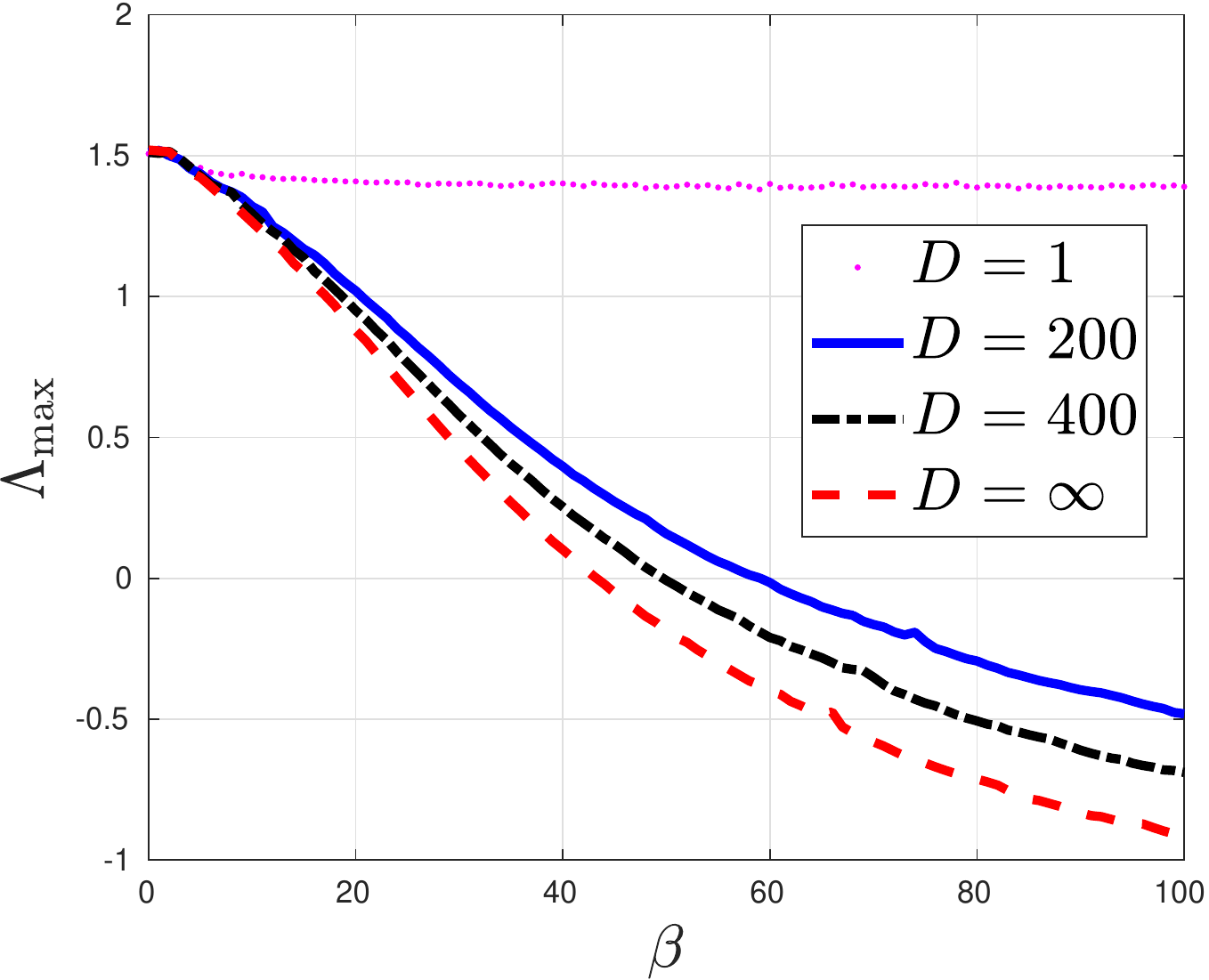}
\end{subfigure}
\begin{subfigure}{0.24\linewidth}
\includegraphics[width=\linewidth]{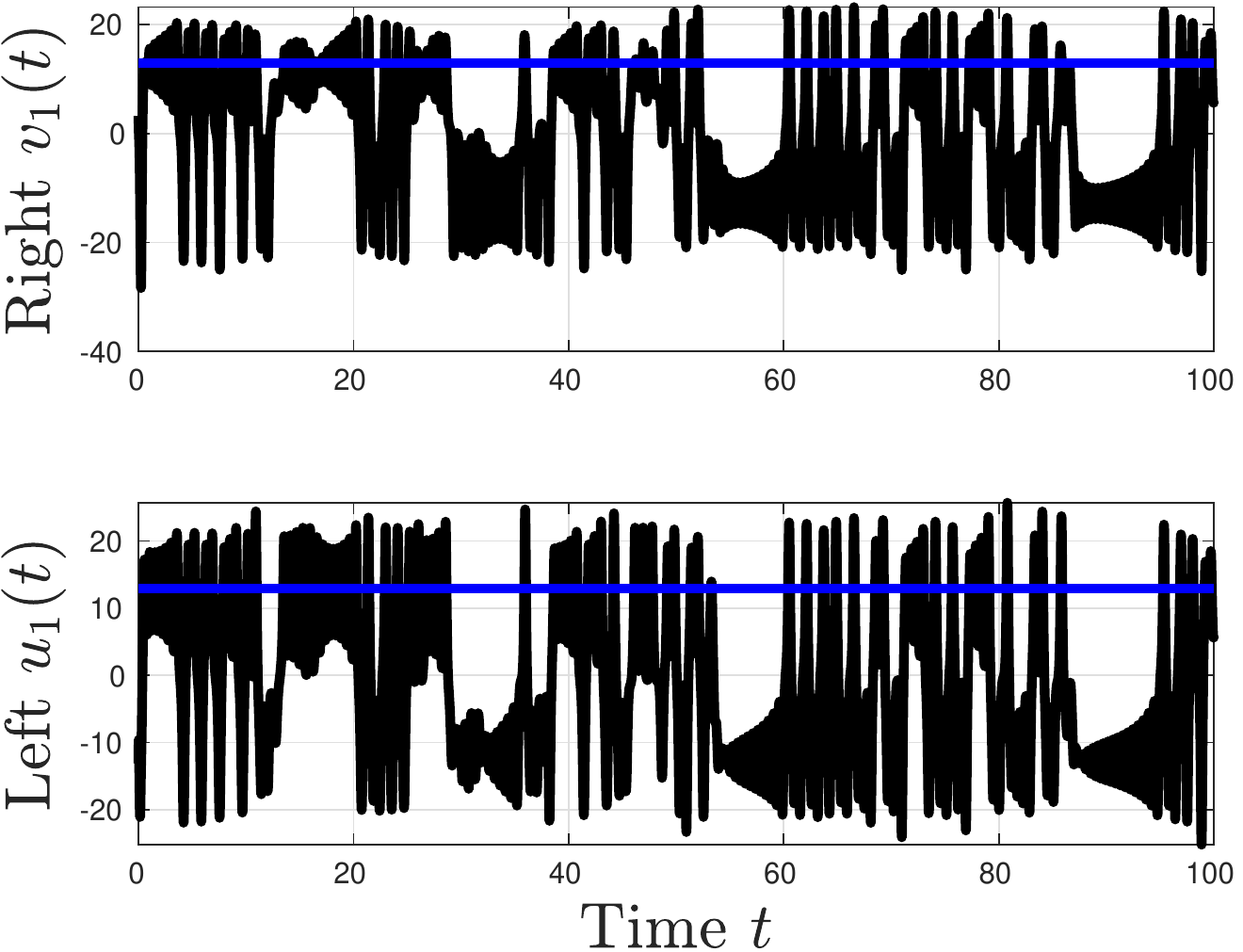}
\end{subfigure}
\begin{subfigure}{0.24\linewidth}
\includegraphics[width=\linewidth]{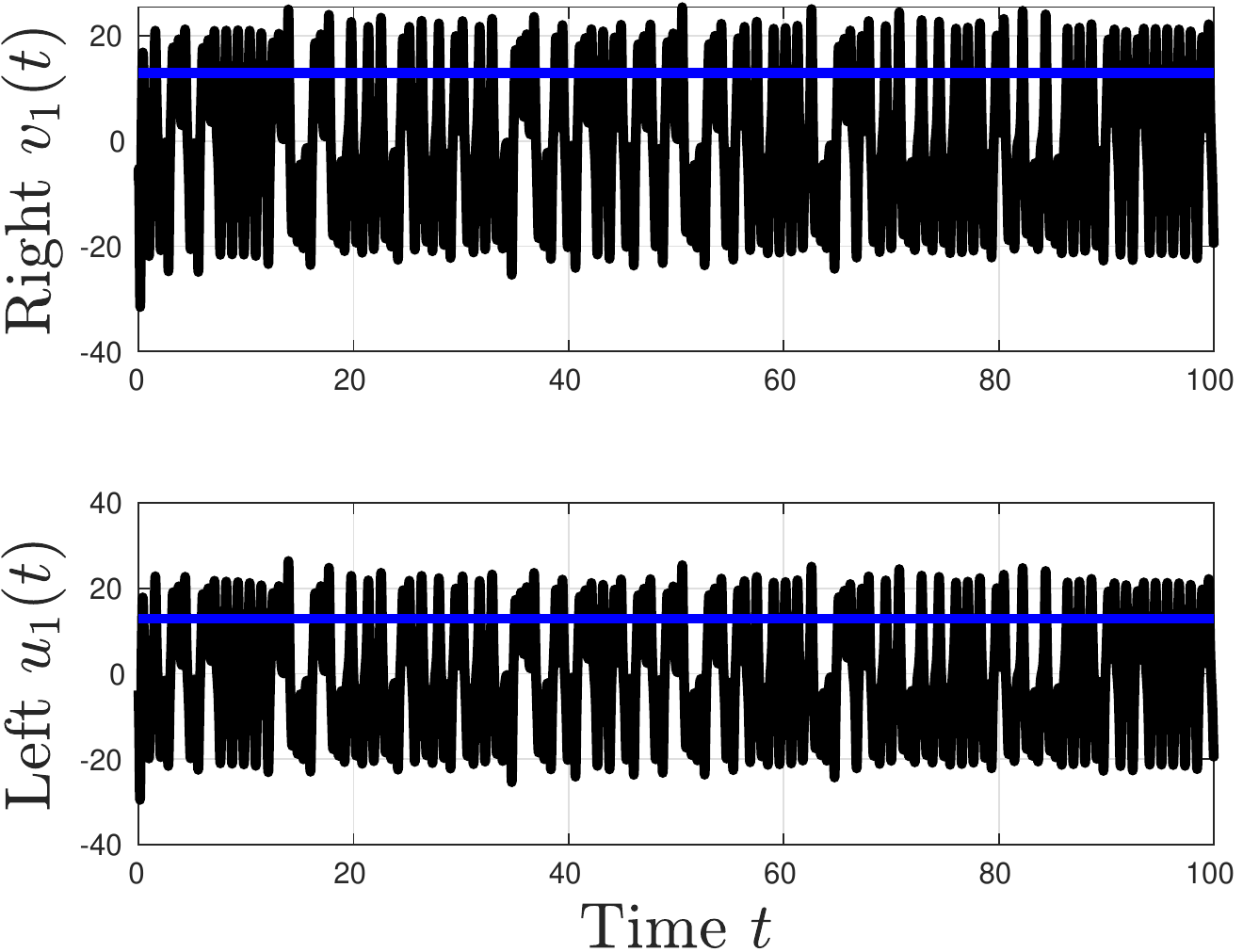}
\end{subfigure}
\caption{\label{fig:syncChaos} {Far left panel: Synchronous chaos stability boundary in the $D$ versus $\beta$ parameter plane. The red-dashed curve indicates $\Lambda_{\max}=0$ when $D=\infty$. Middle left panel: Plot of $\Lambda_{\max}$ as a function of the coupling strength $\beta$ for $D=1,\,200,\,400$ and $\infty$, indicating that $D$ must be large enough for $\Lambda_{\max}$ to become negative. Middle right panel: Numerically computed chaotic trajectories, with no synchronization, for $\beta=50$ and $D=200$ (indicated by a square in the far left panel). Far right panel: Synchronous chaotic oscillations for $\beta=70$ and $D=200$ (indicated by a star in the far left panel). Other parameters are $L=1,\,k=1,\,\sigma = 10,\, b = 8/3,\, r=70$.}}
\end{figure}

\begin{figure}[htbp]
\centering
\begin{subfigure}{0.32\linewidth}
\includegraphics[width=\linewidth]{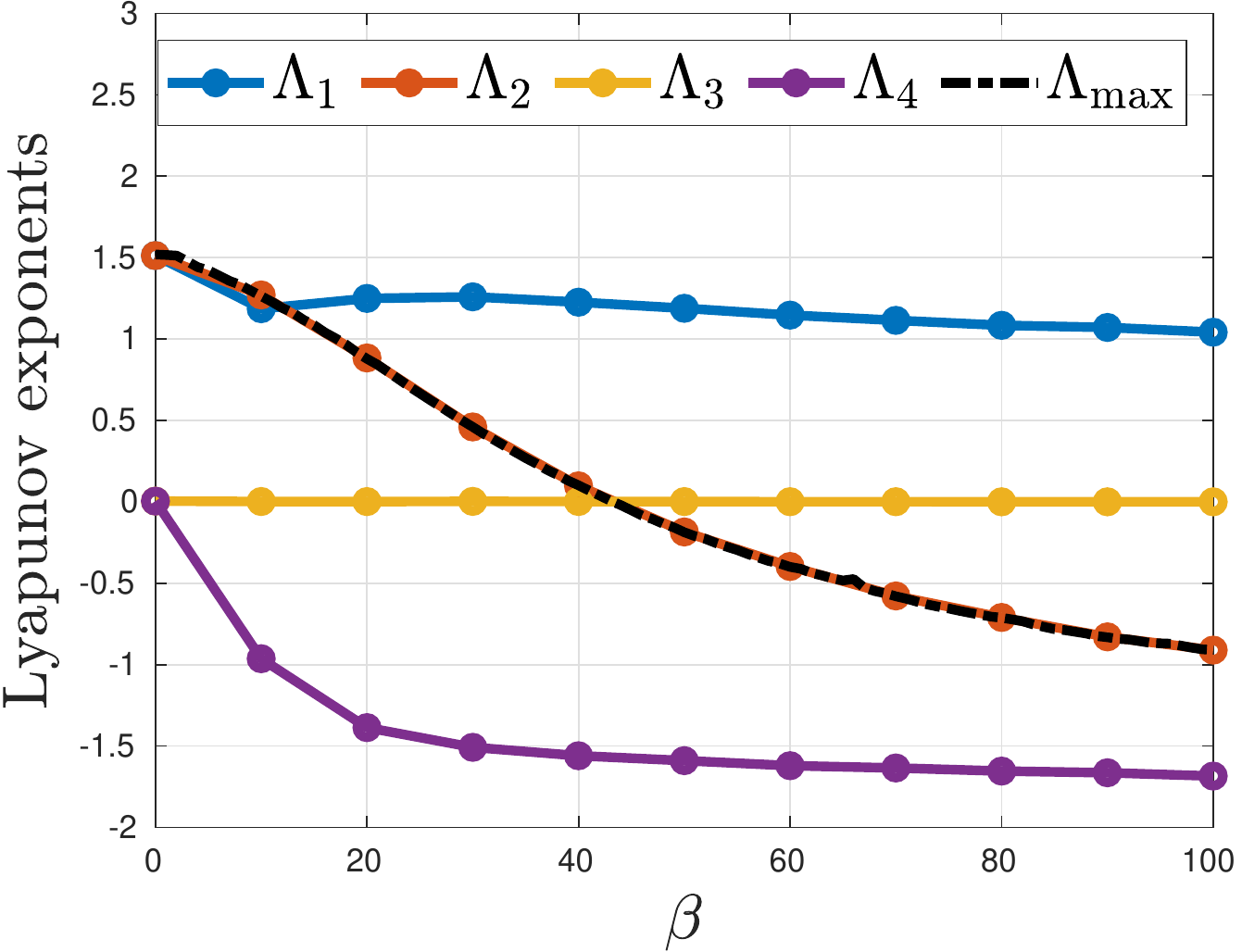}
\caption{$D = \infty$}
\end{subfigure}
\begin{subfigure}{0.32\linewidth}
\includegraphics[width=\linewidth]{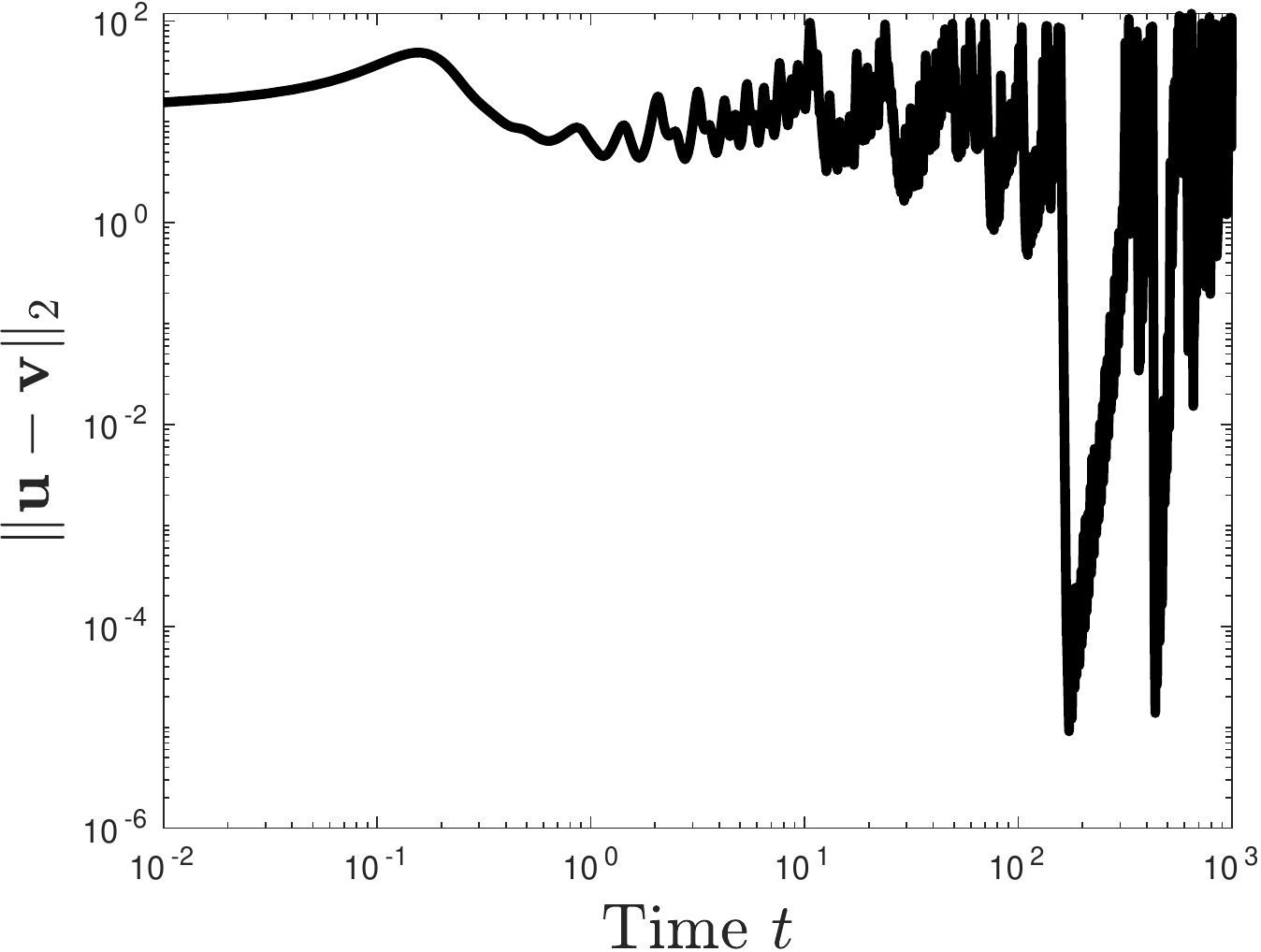}
\caption{$\beta = 40,\, D = \infty$}
\end{subfigure}
\begin{subfigure}{0.32\linewidth}
\includegraphics[width=\linewidth]{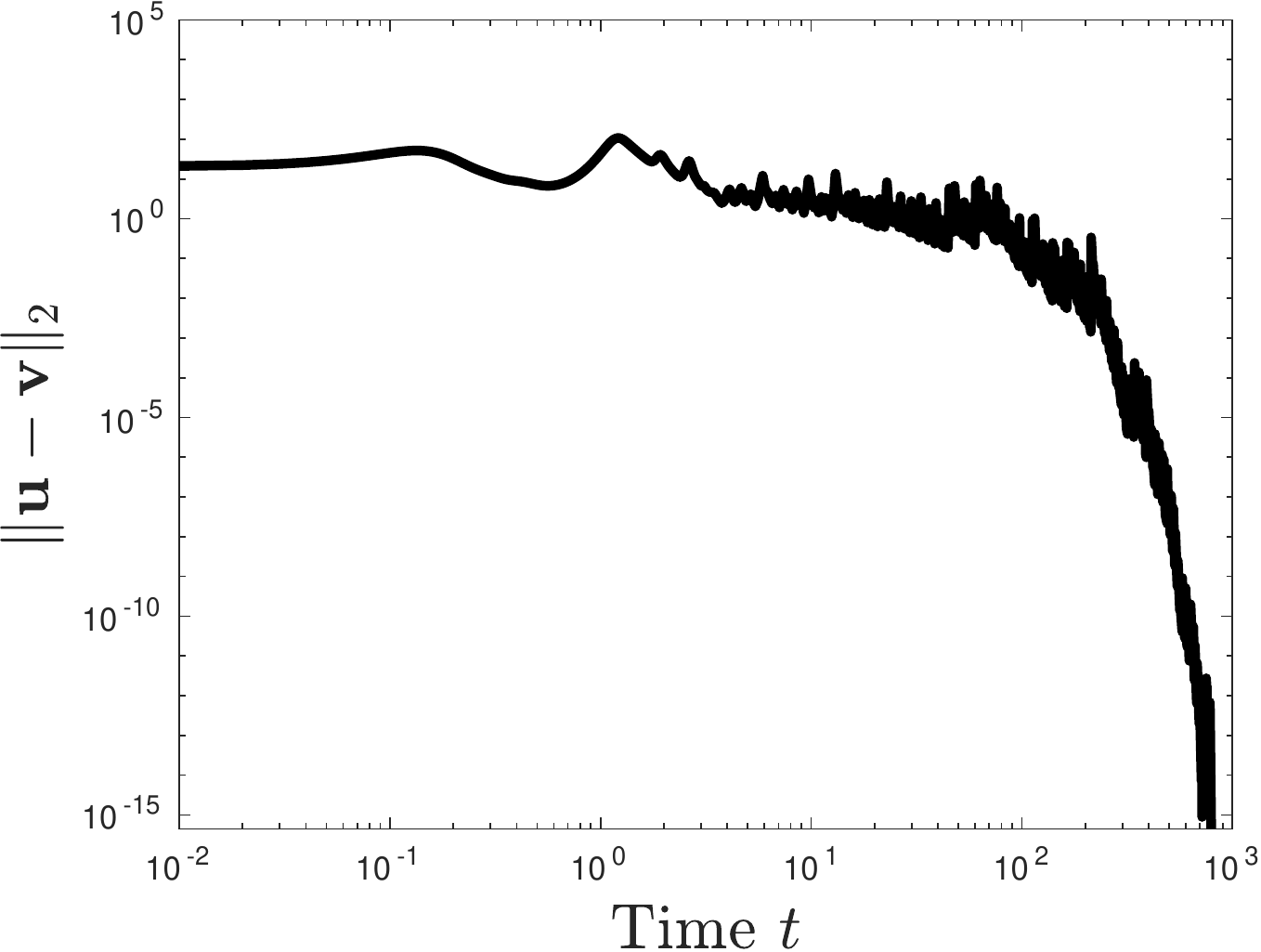}
\caption{$\beta = 45,\, D = \infty$}
\end{subfigure}
\caption{\label{fig:leading} {Transition to synchronous chaos in the infinite $D$ case. Panel (a): Largest four Lyapunov exponents numerically computed from the ODE \eqref{eq:well_mixed} and its linearization as a function of $\beta$. We observe that $\Lambda_2$ agrees with $\Lambda_{\max}$, the largest Lyapunov exponent computed when considering transverse perturbations to the synchronous manifold. At least one positive $\Lambda_j$ indicates chaos, while a negative $\Lambda_{max}$ indicates the synchronous manifold is attracting. Panels (b)-(c): Two simulation results, with random initial conditions. Synchronization is obtained in panel (c) for $\beta = 45$, where we expect the distance $\|\bu(t)-\bv(t)\|$ to decay to zero as time increases. Other parameters are the same as in Fig.~\ref{fig:syncChaos}.}}
\end{figure}

\begin{figure}[htbp]
\centering
\begin{subfigure}{0.32\linewidth}
\includegraphics[width=\linewidth]{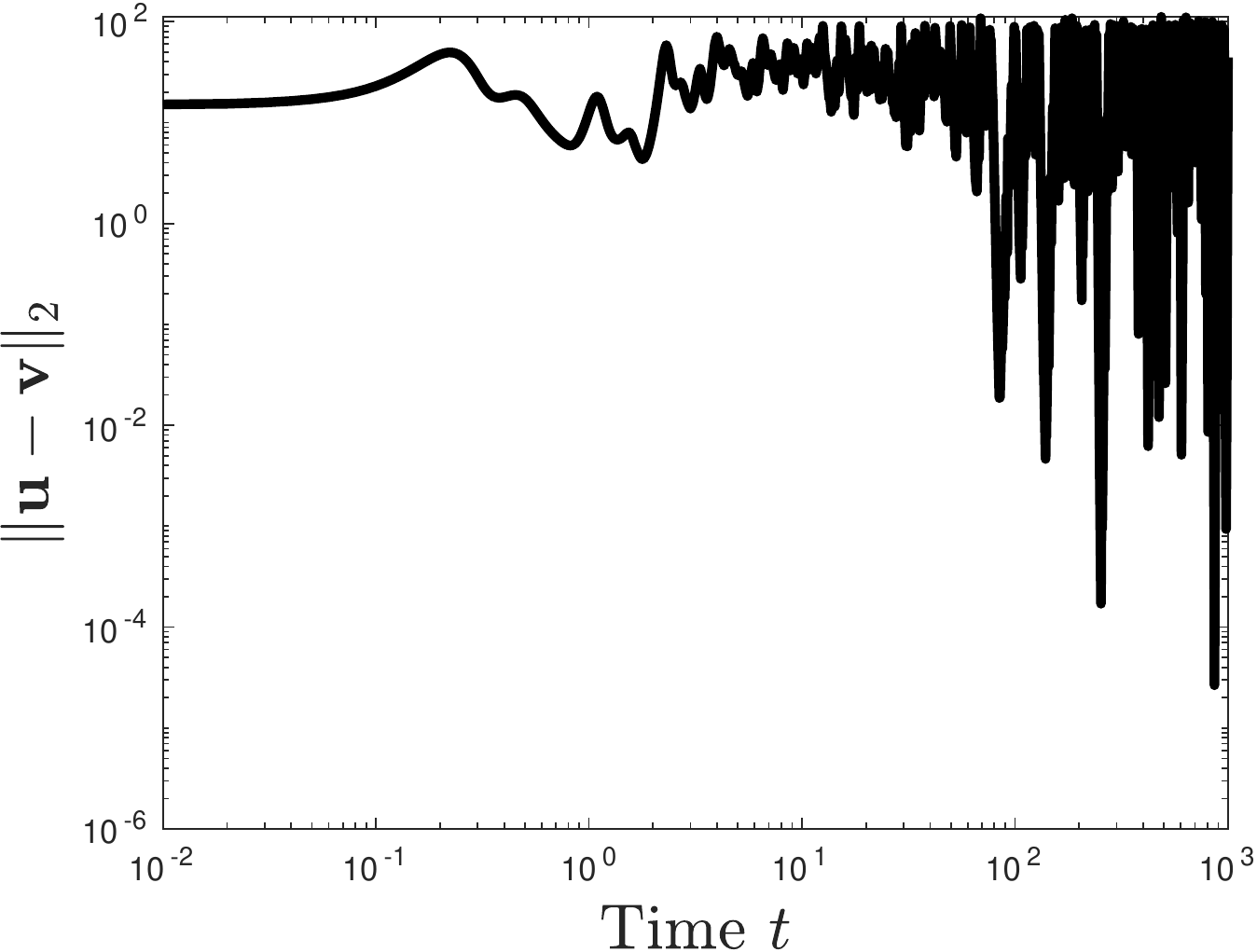}
\caption{$\beta = 100,\, D = 50$}
\end{subfigure}
\begin{subfigure}{0.32\linewidth}
\includegraphics[width=\linewidth]{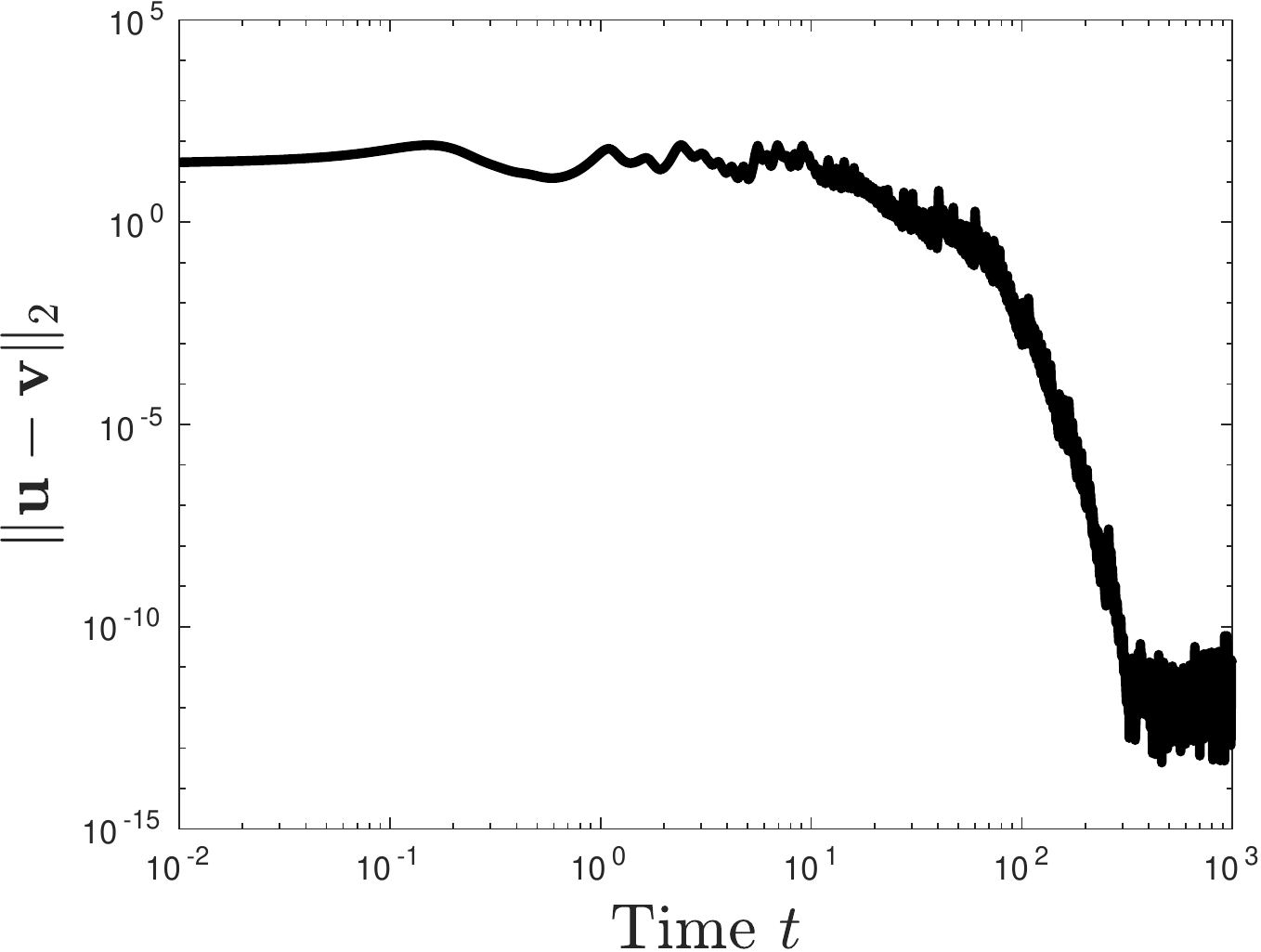}
\caption{$\beta = 100,\, D = 100$}
\end{subfigure}
\begin{subfigure}{0.32\linewidth}
\includegraphics[width=\linewidth]{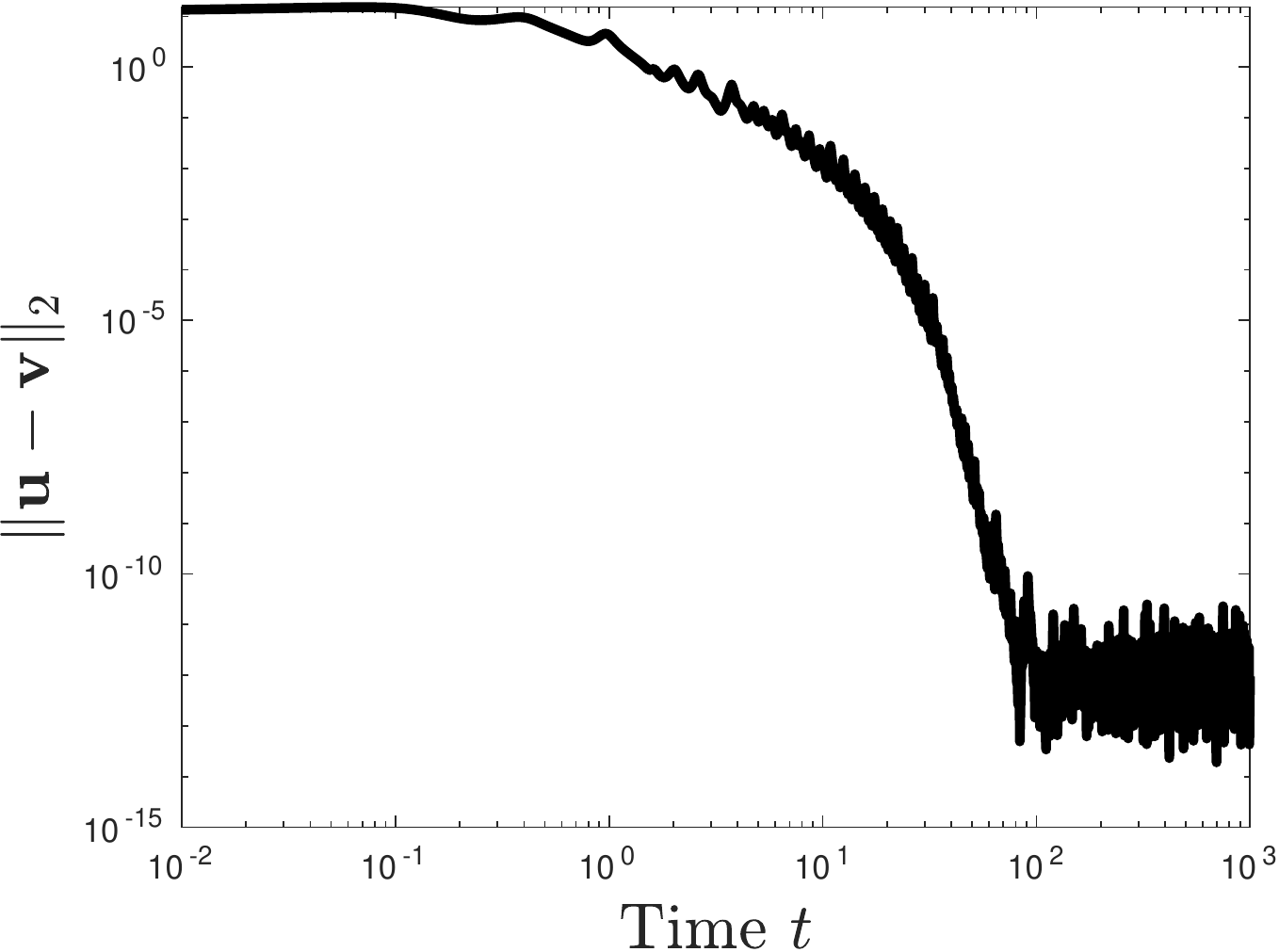}
\caption{$\beta = 100,\, D = 150$}
\end{subfigure}
\caption{\label{fig:onset_D_order_1} Onset of synchronous chaos in the finite bulk diffusion regime on the vertical slice $\beta = 100$, with other parameters the same as in Fig.~\ref{fig:syncChaos}. Each plot gives the Euclidean distance between the two oscillators as a function of time, starting from random initial conditions.}
\end{figure}

Next, we compute the largest Lyapunov exponent in the $D$ versus
$\beta$ parameter plane, with the aim of approximating the level curve
$\Lambda_{\max} = 0$. The result is shown in the left panel of
Fig.~\ref{fig:syncChaos}, where we find that synchronous chaos,
corresponding to where $\Lambda_{\max} < 0$, holds to the right of the
stability boundary. Not surprisingly, the critical diffusion level is
approximately inversely proportional to the coupling
strength. {This implies that a smaller diffusion level is
  necessary for complete synchronization to occur if $\beta$ gets
  larger.} {Moreover, as $D$ tends to infinity the stability boundary
  should approach an asymptote in $\beta \approx 43$, corresponding to
  $\Lambda_{\max} = 0$ as computed from \eqref{eq:synchronous_WM} and
  \eqref{eq:non_autonomous_WM}. At last, examples of numerically
  computed chaotic trajectories when $D=200$ are shown in the middle and far right panels of Fig.~\ref{fig:syncChaos} for random initial
  conditions. As expected from the stability diagram, complete
  synchronization fails for $\beta = 50$ while it succeeds for
  $\beta=70$. However, a more appropriate synchrony measure would be
  to compute the Euclidean distance $\|\bu(t)-\bv(t)\|$. This is done
  in Fig.~\ref{fig:leading} and \ref{fig:onset_D_order_1}.}

{We now briefly discuss the relationship between
  $\Lambda_{\max}$ and the spectrum of Lyapunov exponents directly
  computed from the full system, with no symmetry reduction.}
{To illustrate this relationship}, {and since the
size of the spectrum equals the dimension of the dynamical system, we
focus on the infinite $D$ case, for which there are only 7 Lyapunov
exponents (3 for each Lorenz oscillator and 1 for the coupling
variable; see \ref{eq:well_mixed}). In
Fig.~\ref{fig:leading}, the largest four exponents (denoted as
$\Lambda_1,\,\Lambda_2,\,\Lambda_3$ and $\Lambda_4$) are shown as a
function of the coupling strength $\beta$, where we conclude that
synchronous chaos is characterized by a single exponent being
positive.} {In contrast, chaos without synchronization
corresponds to having two exponents being positive.} {We also
remark that $\Lambda_2$ exactly corresponds to $\Lambda_{\max}$, thus
allowing us to recover the stability threshold $\beta \approx 43$
previously obtained for the infinite bulk diffusion case. This
threshold is confirmed from numerical simulations in the middle and
right panels of Fig.~\ref{fig:leading}.} Thus, we claim that our
computational approach, which is to compute the largest exponent of an
odd linearization around the synchronous manifold, is more accurate
and efficient (especially when $D$ is finite) than if we were to
consider the full spectrum of Lyapunov exponents.

{We conclude this section with Fig.~\ref{fig:onset_D_order_1}, which illustrates a transition to synchronous chaos as the diffusivity $D$ increases while the coupling strength $\beta$ is kept fixed. As the system goes further into the synchronous chaos stability regime, faster convergence onto the synchronous manifold is observed.}

\section{Discussion}\label{sec:discussion}

In this paper, we have developed a comprehensive weakly nonlinear theory for a class of PDE-ODE systems that couple 1-D bulk diffusion with arbitrary nonlinear kinetics at the two endpoints of the interval. From a multi-scale asymptotic expansion, in \S \ref{sec:weakly_nonlinear_theory} we derived amplitude equations characterizing the weakly nonlinear oscillations of in-phase and anti-phase spatio-temporal oscillations. In \S \ref{sec:selkov}, our analysis was shown to compare favorably with numerical bifurcation results for a coupled PDE-ODE model with Sel'kov kinetics. Our second example is given in \S \ref{sec:lorenz}, where we considered the diffusive coupling of two Lorenz oscillators. There we showed how this coupling mechanism can provide a stabilizing mechanism and suppress chaotic oscillations at parameter values that are well known to yield chaos in the isolated Lorenz ODE. We also considered the well-mixed regime, defined as the infinite bulk diffusion limit, for which the coupled PDE-ODE system is replaced by two globally coupled ODE systems. Finally, in \S \ref{subsec:synchronous_chaos} we predicted the transition to synchronous chaos as the coupling strength and the diffusivity increase, from a numerical computation of the largest Lyapunov exponent of an appropriate non-autonomous linearization around the synchronous manifold, where only odd (or transverse) perturbations are considered.

{In the formulation of our PDE-ODE model we have assumed a
  scalar coupling, so that only one variable from each of the two
  compartments is coupled with the bulk diffusion field. Qualitatively
  different dynamics is to be expected for other coupling schemes than
  the one considered here, that are obtained by replacing the basis
  vector $e_1$ with $e_j,\, j\neq 1$ in \eqref{eq:BC} and
  \eqref{eq:kinetics}. Our choice for the Sel'kov kinetics was partly
  motivated by earlier studies (cf.~\cite{gou2017,gou2016DH}) and by
  our own numerical experiments} {which concluded that there
  are no oscillatory dynamics for a scalar coupling} { via
  the inhibitor species $u_2$. Different coupling schemes were also
  explored for the Lorenz example. Although we have observed a similar
  stabilizing effect for each of the three possible coupling schemes,
  with a larger Rayleigh number necessary for the system to undergo a
  Hopf bifurcation, results from non-exhaustive numerical simulations
  suggested} {that synchronous chaos is possible}
{only for the specific coupling considered here in \S
  \ref{sec:lorenz}.}

{Finally, results from \S \ref{sec:selkov} and \S
  \ref{sec:lorenz} shed light on some of the key differences between
  the finite and infinite bulk diffusion regimes. From a modeling
  point of view, one effect of the finite spatial diffusion of a
  signaling chemical consists in introducing time delays into the
  spatially segregated system, which are well known to cause
  oscillatory dynamics. This mechanism is at play here for the example
  with Sel'kov kinetics, as we observe from the stability diagram in
  Fig.~\ref{fig:stabDiagSelkov_1} that no oscillations are possible as
  the bulk diffusivity gets too large, hence effectively suppressing
  time delays between the two localized ODE compartments. The role of
  diffusion induced delays on oscillations is discussed in a number of
  references, including \cite{glass2012} and \S 5 of
  \cite{gou2016DH}.} {However, for our second example based on
  the Lorenz model}, {both diffusion regimes yield
  qualitatively similar dynamics. Not surprisingly, we found the
  minimal coupling strength for synchronous chaos to be smaller in the
  infinite versus finite diffusion cases. }

Among the open problems related to bulk coupled PDE-ODE systems that warrant further investigation, it would be interesting to use global bifurcation software to numerically path-follow the solution branch originating from the torus bifurcation points detected in \S \ref{sec:selkov} for the Sel'kov model (see Fig.~\ref{fig:slice_2}). This would allow us to determine whether this model can provide a bifurcation cascade leading to spatio-temporal chaos.

It would also be interesting to extend our weakly nonlinear theory to analyze periodic ring spatio-temporal patterns in systems composed of several oscillators spatially segregated on a 1-D interval with periodic boundary conditions. The derivation of this novel class of models was given in \cite{gou2017}, where it was also shown how Floquet theory can be employed to study the linear stability of symmetric steady states. {Moreover, it would be interesting to perform a weakly nonlinear analysis for the quasi-steady state version of this modeling paradigm, whereby each ODE compartment acts as a localized source term within the diffusion equation. A model of this type is given in \cite{glass2012}, as well as in \cite{hess2} and \cite{hess3} with applications to the study of spatial effects in gene regulatory systems. A weakly nonlinear analysis for a specific such system was given in \cite{hess1}.}

%--------------------Bibliography------------------

\section*{Acknowledgments}
F.~Paquin-Lefebvre was partially supported by a NSERC Doctoral Award and a UBC Four-Year Doctoral Fellowship. W.~Nagata and M.~J.~Ward gratefully acknowledge the support of the NSERC Discovery Grant Program.

\begin{appendix}
\renewcommand{\theequation}{\Alph{section}.\arabic{equation}}
\setcounter{equation}{0}

\section{{Exact solution of the inhomogeneous linear systems arising at second order}}\label{sec:Wjklm}
{
In this appendix we briefly outline the computation of the coefficients $W_{jklm}$ at $\orderTwo$ of the multi-scale expansion. First, we have that $W_{0000}$, which arises from the perturbation of the bifurcation parameters within the symmetric steady state, satisfies a linear inhomogeneous equation,
\begin{equation}\label{eq:W_0000}
\LL(\mu_0; W_{0000}) + \begin{pmatrix} \omega^2 C_e D_1 \\ - p_0 E \bu_e \beta_1 \\ - p_0 E \bu_e \beta_1 \end{pmatrix} = \bm{0}\,,
\end{equation}
subject to inhomogeneous boundary conditions,
\begin{equation}\label{eq:BC_W_0000}
\begin{split}
& D_0\partial_x C_{0000} + \kappa\left(e_1^T\bu_{0000} - C_{0000}\right) = \frac{\kappa p_0}{D_0}e_1^T\bu_e D_1\,, \quad x = 0\,, \\
& D_0\partial_x C_{0000} - \kappa\left(e_1^T\bv_{0000} - C_{0000}\right) = -\frac{\kappa p_0}{D_0}e_1^T\bu_e D_1\,, \quad x = 2L\,.
\end{split}
\end{equation}
It is readily seen that the solution must be even and that $\bu_{0000} = \bv_{0000}$. As a result, a suitable ansatz to $C_{0000}(x)$ is given by
\begin{equation}\label{eq:ansatz_C_0000}
C_{0000}(x) = K_1\frac{\cosh(\omega(L-x))}{\cosh(\omega L)} + K_2(x-L)\frac{\sinh(\omega(L-x))}{\cosh(\omega L)}.
\end{equation}
By inserting \eqref{eq:ansatz_C_0000} within \eqref{eq:W_0000} and \eqref{eq:BC_W_0000}, we can readily establish that the unknown constants are given by
\begin{align}
K_1 &= (1-p_0)e_1^T\bu_{0000} + \frac{\kappa\omega\left( \tanh(\omega L)(\kappa L - D_0) + \kappa D_0\omega \right)}{2 D_0\left(D_0\omega\tanh(\omega L) + \kappa\right)^2}e_1^T\bu_e D_1, \\
K_2 &= \frac{\kappa\omega}{2 D_0\left(D_0\omega\tanh(\omega L) + \kappa\right)}e_1^T\bu_e D_1.
\end{align}
Next, the evaluation of $C_{0000}$ at the endpoints leads to
\begin{equation}\label{eq:C_0000_0}
C_{0000}|_{x=0,2L} = (1-p_0)e_1^T\bu_{0000} + e_1^T\bu_e \delta D_1, \quad \delta = \frac{\kappa \omega^2 L\sech^2(\omega L) - \kappa \omega \tanh(\omega L)}{2(D_0\omega \tanh(\omega L) + \kappa)^2}.
\end{equation}
Finally, the substitution of \eqref{eq:C_0000_0} within \eqref{eq:W_0000} leads to an $n\times n$ linear system for $\bu_{0000}$ given by,
\begin{equation}
 \left[\bPhi_+(0)\right]\bu_{0000} = \alpha^T\mu_1 E \bu_e \quad \Rightarrow \quad \bu_{0000} = \alpha^T\mu_1 \left[\bPhi_+(0)\right]^{-1} E \bu_e.
\end{equation}
Here, $\alpha$ is a two-dimensional vector defined by
\begin{equation}
 \alpha = p_0 \xi_1 - \beta_0 \delta \xi_2, \quad \xi_1 = \begin{pmatrix} 1 \\ 0 \end{pmatrix}, \quad \xi_1 = \begin{pmatrix} 0 \\ 1 \end{pmatrix}.
\end{equation}

The linear inhomogeneous systems satisfied by the other $W_{jklm}$ are listed as
\begin{gather*}
\LL\left(\mu_0; W_{2000}\right) - 2i\lambda_I^+ W_{2000} = - \BB(\WW_+,\WW_+)\,, \quad \LL\left(\mu_0; W_{0020}\right) - 2i\lambda_I^- W_{0020} = - \BB(\WW_-,\WW_-)\,, \\
\LL\left(\mu_0; W_{0200}\right) + 2i\lambda_I^+ W_{2000} = - \BB(\overline{\WW_+},\overline{\WW_+})\,, \quad \LL\left(\mu_0; W_{0002}\right) + 2i\lambda_I^- W_{0002} = - \BB(\overline{\WW_-},\overline{\WW_-})\,, \\
\LL\left(\mu_0; W_{1100}\right) = - 2\BB(\WW_+,\overline{\WW_+})\,, \quad \LL\left(\mu_0; W_{0011}\right) = - 2\BB(\WW_-,\overline{\WW_-})
\end{gather*}
from which it follows that $W_{0200} = \overline{W_{2000}}$ and $W_{0002} = \overline{W_{0020}}$. Explicit solutions for $W_{2000}$, $W_{1100}$, $W_{0020}$ and $W_{0011}$ are given by
\begin{align*}
W_{2000} &= \begin{pmatrix}
             (1-p_+(2i\lambda_I^+))\frac{\cosh\left(\Omega_{2I}^+(L-x)\right)}{\cosh\left(\Omega_{2I}^+L\right)} e_1^T \bu_{2000} \\
             \bu_{2000} \\
             \bu_{2000}
            \end{pmatrix}, \quad \bu_{2000} = -[\bPhi_+(2i\lambda_I^+)]^{-1}B(\bphi_+,\bphi_+), \\
W_{1100} &= \begin{pmatrix}
             (1-p_0)\frac{\cosh\left(\omega(L-x)\right)}{\cosh\left(\omega L\right)} e_1^T \bu_{1100} \\
             \bu_{1100} \\
             \bu_{1100}
            \end{pmatrix}, \quad \bu_{1100} = -2[\bPhi_+(0)]^{-1}B(\bphi_+,\overline{\bphi_+}), \\
W_{0020} &= \begin{pmatrix}
             (1-p_+(2i\lambda_I^-))\frac{\cosh\left(\Omega_{2I}^-(L-x)\right)}{\cosh\left(\Omega_{2I}^-L\right)} e_1^T \bu_{0020} \\
             \bu_{0020} \\
             \bu_{0020}
            \end{pmatrix}, \quad \bu_{0020} = -[\bPhi_+(2i\lambda_I^-)]^{-1}B(\bphi_-,\bphi_-), \\
W_{0011} &= \begin{pmatrix}
             (1-p_0)\frac{\cosh\left(\omega(L-x)\right)}{\cosh\left(\omega L\right)} e_1^T \bu_{1100} \\
             \bu_{0011} \\
             \bu_{0011}
            \end{pmatrix}, \quad \bu_{0011} = -2[\bPhi_+(0)]^{-1}B(\bphi_-,\overline{\bphi_-}),
\end{align*}
where $\Omega_{2I}^\pm$ is defined by
\begin{equation}
 \Omega_{2I}^\pm = \sqrt{\frac{k + 2i\lambda_I^\pm}{D}}.
\end{equation}
}

\section{{Derivation of the well-mixed ODE system}}\label{sec:well_mixed}

{
In this appendix, we derive the ODE system \eqref{eq:well_mixed} governing the dynamics in the $D = \infty$ case. For this purpose, we consider the intermediate case of a large (but finite) diffusivity $D \gg 1$ and expand the bulk variable $C(x,t)$ in a regular asymptotic power series of $\frac{1}{D} \ll 1$,
\begin{equation}
C = C_0 + \frac{1}{D}C_1 + \ldots\,,
\end{equation}
and upon inserting within
\begin{equation}
\begin{split}
& \frac{1}{D}C_t = C_{xx} - \frac{k}{D}C\,, \qquad 0 < x < 2L\,, \qquad t > 0\,, \\
& -C_x(0,t) = \frac{\beta}{D} (e_1^T\bu(t) - C(0,t))\,, \quad C_x(2L,t) = \frac{\beta}{D} (e_1^T\bv(t) - C(2L,t))\,,
\end{split}
\end{equation}
we find, at leading order, that $C_0$ satisfies $C_{0xx} = 0$ subject to $C_{0x} = 0$ in $x=0,\, 2L$. Hence, we effectively have that $C_0 \equiv C_0(t)$ is spatially uniform. At the next order, we have
\begin{equation}
C_{1xx} = \frac{dC_0}{dt} + k C_0\,,
\end{equation}
and upon integrating from $x=0$ to $x=2L$ and using the boundary conditions
\begin{equation}
-C_{1x}|_{x=0} = \beta (e_1^T\bu - C_0)\,, \quad C_{1x}|_{x=2L} = \beta (e_1^T\bv - C_0)\,,
\end{equation}
we obtain the following ODE for $C_0(t)$:
\begin{equation}
\frac{dC_0}{dt} = \frac{\beta}{2L}e_1^T\left(\bu + \bv\right) - \left(k + \frac{\beta}{L}\right)C_0\,.
\end{equation}
}

\end{appendix}

\bibliographystyle{plainnat}
\bibliography{references}

\end{document}